\documentclass[10pt, a4paper, twocolumn]{article} 
\usepackage{amsfonts,amssymb}
\usepackage{graphicx}
\usepackage{float}
\usepackage{amsmath}
\usepackage{cuted}

\usepackage{blindtext}

\newcommand{\ket}[1]{\left|#1\right\rangle}
\newcommand{\bra}[1]{\left\langle#1\right|}

\usepackage[english]{babel} 

\usepackage{microtype} 

\usepackage{amsmath,amsfonts,amsthm} 

\usepackage[svgnames]{xcolor} 

\usepackage[hang, small, labelfont=bf, up, textfont=it]{caption} 

\usepackage{booktabs} 

\usepackage{lastpage} 

\usepackage{graphicx} 

\usepackage{enumitem} 
\setlist{noitemsep} 

\usepackage{sectsty} 
\allsectionsfont{\usefont{OT1}{phv}{b}{n}} 

\usepackage{geometry} 

\usepackage[T1]{fontenc} 
\usepackage[utf8]{inputenc} 

\usepackage{XCharter} 

\usepackage{fancyhdr} 
\fancypagestyle{firstpage}{ 
	\fancyhf{}
}

\newcommand{\institution}[1]{{\footnotesize\usefont{OT1}{phv}{m}{sl}\color{Black}#1}} 

\usepackage{titling} 

\newcommand{\HorRule}{\color{DarkGoldenrod}\rule{\linewidth}{1pt}} 

\pretitle{
	\vspace{-30pt} 
	\HorRule\vspace{10pt} 
	\fontsize{32}{36}\usefont{OT1}{phv}{b}{n}\selectfont 
	\color{DarkRed} 
}

\posttitle{\par\vskip 15pt} 

\preauthor{} 

\postauthor{ 
	\vspace{10pt} 
	\par\HorRule 
	\vspace{20pt} 
}

\usepackage{lettrine} 
\usepackage{fix-cm}	

\newcommand{\initial}[1]{ 
	\lettrine[lines=3,findent=4pt,nindent=0pt]{
		\color{DarkGoldenrod}
		{#1}
	}{}%
}

\usepackage{xstring} 

\newcommand{\lettrineabstract}[1]{
	\StrLeft{#1}{1}[\firstletter] 
	\initial{\firstletter}\textbf{\StrGobbleLeft{#1}{1}} 
}

\usepackage[backend=bibtex,style=authoryear,natbib=true]{biblatex} 

\usepackage[autostyle=true]{csquotes} 

\title{Analytical view on tunnable electrostatic quantum swap gate in tight-binding model} 

\author{ Krzysztof Pomorski\textsuperscript{1,2,3}
	\newline\newline 
	\textsuperscript{1}\institution{University College Dublin, School of Computer Science, Ireland}\\ 
	\textsuperscript{2}\institution{University College Dublin, School of Electrical and Electronic Engineering, Ireland}\\ 
	\textsuperscript{3}\institution{Quantum Hardware Systems:\texttt{www.quantumhardwaresystems.com}} 
}

\date{\today} 

\begin{document}

\maketitle 

\thispagestyle{firstpage} 


\lettrineabstract{Generalized electrostatic quantum swap gate implemented on the chain of 2 double coupled quantum dots using single electron in semiconductor is presented in tight-binding simplistic model specifying both analytic and numerical results. The anticorrelation principle coming from Coulomb electrostatic repulsion is exploited. The formation of quantum entanglement is specified and supported by analytical results. The difference between classical and quantum picture is given. The corraltions between geometry of quantum structures and entanglement dynamics are shown.  The presented results have its significance in cryogenic CMOS quantum technologies that gives perspective of implementation of semiconductor quantum computer on massive scale.}


\section{Description of position based-qubit in tight-binding model}
\begin{figure}[htb]
\centering
	\includegraphics[scale=0.6]{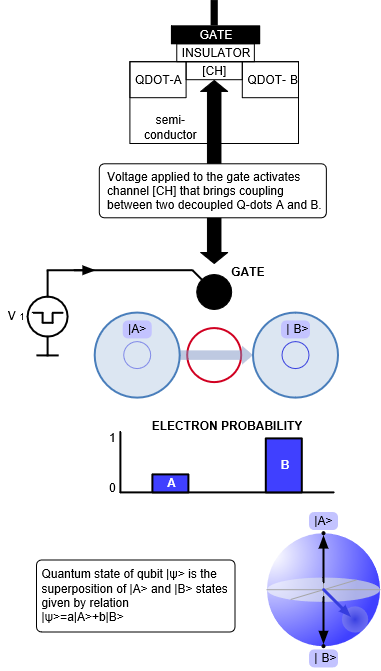} 
	\caption{Scheme of electrostatic quantum gate as two interacting qubits [3-8] referring to the previous figure showing implementation of the single qubit. } 
	\label{PositionQubit} 
	\includegraphics[scale=0.6]{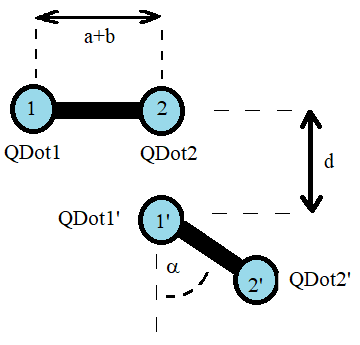} 
	\caption{Geometrical parametrization of generalized electrostatic quantum gate} 
	\label{GQSwapGate} 
\end{figure}

We refer to the physical situation from Fig.\ref{PositionQubit} as given by [1-5] 
and we consider position based-qubit in tight-binding model \cite{SEL} and its the Hamiltonian of this system is given as
\begin{eqnarray}
\label{simplematrix}
\hat{H}(t)=
\begin{pmatrix}
E_{p1}(t) & t_{s12}(t) \\
t_{s12}^{\dag}(t) & E_{p2}(t)
\end{pmatrix}_{[x=(x_1,x_2)]}= \nonumber \\
(E_1(t)\ket{E_1}_t \bra{E_1}_t+E_2(t)\ket{E_2}\bra{E_2})_{[E=(E_1,E_2)]}.
\end{eqnarray}
The Hamiltonian $\hat{H}(t)$ eigenenergies $E_1(t)$ and $E_2(t)$ with $E_2(t)>E_1(t)$ are given as
\begin{eqnarray}
E_1(t)= \left(-\sqrt{\frac{(E_{p1}(t)-E_{p2}(t))^2}{4}+|t_{s12}(t)|^2}+\frac{E_{p1}(t)+E_{p2}(t)}{2}\right), \nonumber \\
E_2(t)= \left(+\sqrt{\frac{(E_{p1}(t)-E_{p2}(t))^2}{4}+|t_{s12}(t)|^2}+\frac{E_{p1}(t)+E_{p2}(t)}{2}\right),
\end{eqnarray}
and energy eigenstates $\ket{E_1(t)}$ and $\ket{E_2(t)}$ have the following form
\begin{eqnarray}
\ket{E_1,t}=
\begin{pmatrix}
\frac{(E_{p2}(t)-E_{p1}(t))+\sqrt{\frac{(E_{p2}(t)-E_{p1}(t))^2}{2}+|t_{s12}(t)|^2}}{-i t_{sr}(t)+t_{si}(t)} \\
-1
\end{pmatrix},  \nonumber \\
\ket{E_2,t}=
\begin{pmatrix}
\frac{-(E_{p2}(t)-E_{p1}(t))+\sqrt{\frac{(E_{p2}(t)-E_{p1}(t))^2}{2}+|t_{s12}(t)|^2}}{t_{sr}(t) - i t_{si}(t)} \\
1
\end{pmatrix}.
\end{eqnarray}
This Hamiltonian gives a description of two coupled quantum wells as depicted in Fig.2.
In such situation we have real-valued functions $E_{p1}(t)$, $E_{p2}(t)$ and complex-valued functions $t_{s12}(t)=t_s(t)=t_{sr}(t)+i t_{si}(t)$ and $t_{s21}(t)=t_{s12}^{*}(t)$, what is equivalent to the knowledge of four real valued time-dependent continuous or discontinues functions $E_{p1}(t)$, $E_{p1}(2)$ , $t_{sr}(t)$ and $t_{si}(t)$. The quantum state is a superposition of state localized at node 1 and 2 and therefore is given as
\begin{equation}
\ket{\psi}_{[x]}=\alpha(t)\ket{1,0}_x+\beta(t)\ket{0,1}_x=
\alpha(t)
\begin{pmatrix}
1 \\
0 \\
\end{pmatrix}
+
\beta(t)
\begin{pmatrix}
0 \\
1 \\
\end{pmatrix} ,
\end{equation}
where $|\alpha(t)|^2$ ($|\beta(t)|^2$) is probability of finding particle at node 1(2) respectively, which brings $|\alpha(t)|^2+|\beta(t)|^2=1$ and obviously $\bra{1,0}_x\ket{|1,0}_x=1=\bra{0,1}_x\ket{|0,1}_x$ and $\bra{1,0}_x\ket{|0,1}_x=0=\bra{0,1}_x\ket{|1,0}_x$. In Schr\"odinger formalism, states $\ket{1,0}_x$ and $\ket{0,1}_x$ are Wannier functions that are parameterized by position $x$. We work in tight-binding approximation and quantum state evolution with time as given by
\begin{equation}
 i \hbar \frac{d}{dt}\ket{\psi(t)}=\hat{H}(t)\ket{\psi(t)}=E(t)\ket{\psi(t)}.
\end{equation}
The last equation has an analytic solution
\begin{equation}
\ket{\psi(t)}=e^{\frac{1}{i \hbar}\int_{t_0}^{t}\hat{H}(t_1)dt_1}\ket{\psi(t_0)}=e^{\frac{1}{i \hbar}\int_{t_0}^{t}\hat{H}(t_1)dt_1}
\begin{pmatrix}
\alpha(0) \\
\beta(0) \\
\end{pmatrix}
\end{equation}
and in quantum density matrix theory we obtain
\begin{eqnarray*}
\hat{\rho}(t)=\hat{\rho}^{\dag}(t)=\ket{\psi(t)}\bra{\psi(t)}=\nonumber \\ =\hat{U}(t,t_0)\hat{\rho}(t_0)\hat{U}(t,t_0)^{-1}= \nonumber \\
=e^{\frac{1}{i \hbar}\int_{t_0}^{t}\hat{H}(t_1)dt_1}(\ket{\psi(t_0)}\bra{\psi(t_0)})e^{-\frac{1}{i \hbar}\int_{t_0}^{t}\hat{H}(t_1)dt_1}= \nonumber \\
=e^{\frac{1}{i \hbar}\int_{t_0}^{t}\hat{H}(t_1)dt_1}
\bigg(
\begin{pmatrix}
\alpha(0) \\
\beta(0) \\
\end{pmatrix} 
\begin{pmatrix}
\alpha^{*}(0) & \beta^{*}(0) \\
\end{pmatrix}
\bigg)e^{-\frac{\int_{t_0}^{t}\hat{H}(t_1)dt_1}{i \hbar}} \nonumber \\
=\hat{U}(t,t_0)  
\begin{pmatrix}
|\alpha(0)|^2 & \alpha(0)\beta^{*}(0)  \\
\beta(0)\alpha(0)^{*} &  |\beta(0)|^2 \\
\end{pmatrix} \hat{U}(t,t_0)^{\dag}.\nonumber \\
\end{eqnarray*}

Having Hermitian matrix $\hat{A}$ with real-valued coefficients $a_{11}(t)$, $a_{22}(t)$, $a_{12r}(t)$, $a_{12i}(t)$  and Pauli matrices $\sigma_1$, $\sigma_2$, $\sigma_3$, $\sigma_0=\hat{I}_{2 by 2}$ we observe that
\begin{eqnarray*}
\hat{A}_{2 \times 2}=
\begin{pmatrix}
 a_{11} & a_{12r}+ia_{12i} \\
 a_{12r}-ia_{12i} & a_{22}
\end{pmatrix}, = \nonumber \\
=a_{12r}\sigma_1 -a_{12i}\sigma_2+\frac{1}{2}(a_{11}-a_{22})\sigma_3+\frac{1}{2}(a_{11}+a_{22})\sigma_0.
\end{eqnarray*}
and for $\hat{A}_{2N \times 2N}=\Sigma_{k_1,k_2,..,k_N} b_{k_1,k_2,..,k_N}(\sigma_{k_1}\times \sigma_{k_2} \times .. \times  \sigma_{k_N})$ we obtain the unique matrix decomposition in terms of Pauli matrix tensor products, where $k_i=0,..,3$.
Using the above property for matrix of size 2$\times$2 we obtain $e^{\frac{1}{i\hbar}\int_{t_0}^{t}\hat{H}(t_1)dt_1}=\hat{U}(t,t_0), $ and assuming $E_{p1}(t)=E_{p2}(t)=E_{p}(t)$ and we  are given matrix $e^{\frac{1}{i\hbar}\int_{t_0}^{t}\hat{H}(t_1)dt_1}= 
 $ 
\begin{eqnarray*}
\begin{pmatrix}
e^{\frac{-i \int_{t_0}^{t}E_p(t')dt'}{\hbar}}ch\left(\frac{\sqrt{-\int_{t_0}^{t}(|t_{s}(t')|^2)dt'}}{\hbar}\right) & \frac{e^{\frac{-i \int_{t_0}^{t}E_{p}(t')dt'}{\hbar}} (\int_{t_0}^{t}(t_{s}^{*}(t'))dt')sh\left(\frac{\sqrt{-\int_{t_0}^{t}|t_{s}(t')|^2)}}{\hbar}\right)}{\sqrt{-\int_{t_0}^{t}((t_{si}(t')^2+t_{sr}(t'))^2)dt'}}  \\
\frac{e^{\frac{-i \int_{t_0}^{t}E_{p}(t')dt'}{\hbar}} (\int_{t_0}^{t}(-t_{s}(t'))dt')sh\left(\frac{\sqrt{-\int_{t_0}^{t}|t_{s}(t')|^2dt'}}{\hbar}\right)}{\sqrt{-\int_{t_0}^{t}((t_{si}(t')^2+t_{sr}(t'))^2)dt'}} & e^{\frac{-i\int_{t_0}^{t}E_p(t')dt'}{\hbar}}ch\left(\frac{\sqrt{-\int_{t_0}^{t}(|t_{s}(t')|^2)dt'}}{\hbar}\right)  \end{pmatrix}, \nonumber \\
\end{eqnarray*}
where $sh$(.) and $ch$(.) are $\sinh$ and $\cosh$ hyperbolic functions, where $|t_s(t)|^2=|t_{sr}(t)|^2+|t_{si}(t)|^2$. This matrix is unitary so $\hat{U}^{\dag}(t,t_0)=\hat{U}^{-1}(t,t_0)$.
At the very end we will also consider more general case when $E_{p1}(t) \neq E_{p2}(t)$. At first let us consider the case of two localized states in the left and right quantum well so there is no hopping which implies $t_s=0$.
In such case the evolution matrix $\hat{U}(t,t_0)$ is unitarian and has the following form 
\begin{eqnarray}
 \hat{U}(t,t_0)=\nonumber \\
e^{\frac{1}{i\hbar}\int_{t_0}^{t}\hat{H}(t_1)dt_1}= \nonumber \\
\begin{pmatrix}
e^{\frac{-i \int_{t_0}^{t}E_{p1}(t')dt'}{\hbar}} & 0 \\
0 & e^{\frac{-i \int_{t_0}^{t}E_{p2}(t')dt'}{\hbar}}
\end{pmatrix}= \nonumber \\
\frac{(e^{\frac{-i \int_{t_0}^{t}E_{p1}(t')dt'}{\hbar}}+e^{\frac{-i \int_{t_0}^{t}E_{p2}(t')dt'}{\hbar}})}{2} 
\begin{pmatrix}
1 & 0 \\
0 & 1 \\
\end{pmatrix}- \nonumber \\
\frac{(e^{\frac{-i \int_{t_0}^{t}E_{p1}(t')dt'}{\hbar}}-e^{\frac{-i \int_{t_0}^{t}E_{p2}(t')dt'}{\hbar}})}{2}
\begin{pmatrix}
1 & 0 \\
0 & -1 \\
\end{pmatrix}\nonumber \\ =
\frac{(e^{\frac{-i \int_{t_0}^{t}E_{p1}(t')dt'}{\hbar}}+e^{\frac{-i \int_{t_0}^{t}E_{p2}(t')dt'}{\hbar}})}{2} \sigma_0 + \nonumber \\
\frac{(e^{\frac{-i \int_{t_0}^{t}E_{p1}(t')dt'}{\hbar}}-e^{\frac{-i \int_{t_0}^{t}E_{p2}(t')dt'}{\hbar}})}{2} \sigma_3,
\end{eqnarray}
what implies that left and right quantum dot are two disconnected physical systems subjected to its own evolution with time. However since one electron is distributed between those physical systems the measurement conducted on the left quantum dot will have its immediate effect on the right quantum dot. Another extreme example is the situation when hopping energy is considerably bigger than localization energy. In such case we set $E_{p1}=E_{p2}=0$ and in case of non-zero hopping terms we obtain $\hat{U}(t,t_0)=$ given below 
\tiny 
\begin{eqnarray}
\hat{U}(t,t_0)=e^{\frac{1}{i\hbar}\int_{t_0}^{t}\hat{H}(t_1)dt_1}= \nonumber \\
\begin{pmatrix}
ch\left(\frac{\sqrt{-\int_{t_0}^{t}(|t_{s}(t')|^2)dt'}}{\hbar}\right) & \frac{ (\int_{t_0}^{t}(t_{s}^{*}(t'))dt')sh\left(\frac{\sqrt{-\int_{t_0}^{t}|t_{s}(t')|^2)}}{\hbar}\right)}{\sqrt{-\int_{t_0}^{t}((t_{si}(t')^2+t_{sr}(t'))^2)dt'}}\\
\frac{ (\int_{t_0}^{t}(-t_{s}(t'))dt')sh\left(\frac{\sqrt{-\int_{t_0}^{t}|t_{s}(t')|^2dt'}}{\hbar}\right)}{\sqrt{-\int_{t_0}^{t}((t_{si}(t')^2+t_{sr}(t'))^2)dt'}} & ch\left(\frac{\sqrt{-\int_{t_0}^{t}(|t_{s}(t')|^2)dt'}}{\hbar}\right)
\end{pmatrix}, \nonumber \\
\end{eqnarray}
\normalsize
$ $ \newline \newline \newline \newline $ $
Now it is time to move to most general situation of $E_{p1} \neq E_{p2}$, $t_{sr}, t_{si} \neq 0$. We have 4 elements of evolution matrix given as
\begin{eqnarray}
 \hat{U}(t,t_0)=e^{\frac{1}{i\hbar}\int_{t_0}^{t}\hat{H}(t_1)dt_1}= \nonumber \\
\begin{pmatrix}
 U(t,t_0)_{1,1} & U(t,t_0)_{1,2} \nonumber \\
 U(t,t_0)_{2,1}= U(t,t_0)_{1,2}^{*} & U(t,t_0)_{2,2}
\end{pmatrix}.
\end{eqnarray}
\begin{eqnarray}
 U(t,t_0)_{1,1}= \nonumber \\
\frac{\exp \left(-  \frac{\sqrt{-\hbar^2 \left(|\int_{t_0}^{t}dt'(E_{p1}(t')-E_{p2}(t'))|^2+4 \left(|\int_{t_0}^{t}dt't_{si}(t')|^2+|\int_{t_0}^{t}dt't_{sr}(t')|^2\right)\right)}+i \hbar \int_{t_0}^{t}dt'(E_{p1}(t')+E_{p2}(t'))}{2 \hbar^2}\right)}{2 \hbar
   \left( ( \int_{t_0}^{t}dt'(E_{p1}(t')-E_{p2}(t')))^2+4 \left(|\int_{t_0}^{t}dt't_{si}(t')|^2+|\int_{t_0}^{t}dt't_{sr}(t')|^2\right)\right)} \times \nonumber \\
\times  \Bigg[-i (\int_{t_0}^{t}dt'E_{p1}(t')) \sqrt{-\hbar^2 \left(|\int_{t_0}^{t}dt'(E_{p1}(t')-E_{p2}(t'))|^2+ 4 \left(|\int_{t_0}^{t}dt't_{si}(t')|^2+|\int_{t_0}^{t}dt't_{sr}(t')|^2\right)\right)}+ \nonumber \\  +\hbar \left(|\int_{t_0}^{t}dt'(E_{p1}(t')-E_{p2}(t'))|^2+4
   \left(|\int_{t_0}^{t}dt't_{si}(t')|^2+|\int_{t_0}^{t}dt't_{sr}(t')|^2\right)\right) \times \nonumber \\ e^{\frac{\sqrt{-\hbar^2 \left(|\int_{t_0}^{t}dt'(E_{p1}(t')-E_{p2}(t'))|^2+4 \left(|\int_{t_0}^{t}dt't_{si}(t')|^2+|\int_{t_0}^{t}dt't_{sr}(t')|^2\right)\right)}}{\hbar^2}}+ \nonumber\\  + \left(
   \left(( \int_{t_0}^{t}dt'(E_{p1}(t')-E_{p2}(t')))^2+4 \left(|\int_{t_0}^{t}dt't_{si}(t')|^2+|\int_{t_0}^{t}dt't_{sr}(t')|^2\right)\right) \right) + \nonumber \\
+i (\int_{t_0}^{t}dt'E_{p1}(t')) e^{\frac{\sqrt{-h^2 \left(|\int_{t_0}^{t}dt'(E_{p1}(t')-E_{p2}(t'))|^2+4 \left(|\int_{t_0}^{t}dt't_{si}(t')|^2+|\int_{t_0}^{t}dt't_{sr}(t')|^2\right)\right)}}{\hbar^2}} \times \nonumber \\ \sqrt{-\hbar^2 \left(|\int_{t_0}^{t}dt'(E_{p1}(t')-E_{p2}(t'))|^2+4
   \left(|\int_{t_0}^{t}dt't_{si}(t')|^2+|\int_{t_0}^{t}dt't_{sr}(t')|^2\right)\right)} \nonumber \\ -i (\int_{t_0}^{t}dt'E_{p2}(t')) e^{\frac{\sqrt{-\hbar^2 \left(|\int_{t_0}^{t}dt'(E_{p1}(t')-E_{p2}(t'))|^2+4 \left(|\int_{t_0}^{t}dt't_{si}(t')|^2+|\int_{t_0}^{t}dt't_{sr}(t')|^2\right)\right)}}{\hbar^2}}\times \nonumber \\
   \sqrt{-\hbar^2 \left(|\int_{t_0}^{t}dt'(E_{p1}(t')-E_{p2}(t'))|^2+4 \left(|\int_{t_0}^{t}dt't_{si}(t')|^2+|\int_{t_0}^{t}dt't_{sr}(t')|^2\right)\right)}+ \nonumber \\ + i (\int_{t_0}^{t}dt' E_{p2}(t')) \sqrt{-\hbar^2 \left(|\int_{t_0}^{t}dt'(E_{p1}(t')-E_{p2}(t'))|^2+4
   \left(|\int_{t_0}^{t}dt't_{si}(t')|^2+|\int_{t_0}^{t}dt't_{sr}(t')|^2\right) \right)} \Bigg].
\end{eqnarray}

\begin{eqnarray*}
U(t,t_0)_{1,2}=\nonumber \\
2 \hbar (\int_{t_0}^{t}dt'(t_{si}(t')-i t_{sr}(t'))) e^{-\frac{i \int_{t_0}^{t}dt'(E_{p1}(t')+E_{p2}(t'))}{2 \hbar}} \times \\
 \sinh \left(\frac{\sqrt{-\hbar^2 \left(|\int_{t_0}^{t}dt'(E_{p1}(t')-E_{p2}(t'))|^2+4
   \left(|\int_{t_0}^{t}dt't_{si}(t')|^2+|\int_{t_0}^{t}dt't_{sr}(t')|^2\right)\right)}}{2 h^2}\right) \times \\
\frac{1}{\sqrt{-\hbar^2 \left(|\int_{t_0}^{t}dt'(E_{p1}(t')-E_{p2}(t'))|^2+4 \left(|\int_{t_0}^{t}dt't_{si}(t')|^2+|\int_{t_0}^{t}dt't_{sr}(t')|^2\right)\right)}}= \nonumber \\
   =U(t,t_0)_{2,1}^{*}. \nonumber \\
\end{eqnarray*}

\onecolumn
\section{Description of 2 qubit interaction in general static case}
We consider most minimalist model of electrostatically interacting two position-based qubits that are double quantum dots A (with nodes 1 and 2  and named as U-upper qubit) and B (with nodes 1' and 2' and named as L-lower qubit) with local confinement potentials as given in the right side of Fig.1.
By introducing notation $\ket{1,0}_x=\ket{1},\ket{0,1}_x=\ket{2},\ket{1',0'}_x=\ket{1'},\ket{0',1'}_x=\ket{1'}$ the minimalistic Hamiltonian of the system of electrostatically interacting position based qubits can be written as
\begin{eqnarray}
\hat{H}=(t_{s21}(t)\ket{2}\bra{1}+t_{s12}(t)\ket{1}\bra{2})\hat{I}_{b})+(\hat{I}_a(t_{s2'1'}(t)\ket{2'}\bra{1'}+t_{s1'2'}(t)\ket{2'}\bra{1'})+ \nonumber \\
+(E_{p1}(t)\ket{1}\bra{1}+E_{p2}(t)\ket{2}\bra{2})\hat{I}_b+ \hat{I}_a(E_{p1'}(t)\ket{1'}\bra{1'}+E_{p2'}(t)\ket{2'}\bra{2'})+ \nonumber \\ 
+\frac{q^2}{d_{11'}}\ket{1,1'}\bra{1,1'}+\frac{q^2}{d_{22'}}\ket{2,2'}\bra{2,2'}+
\frac{q^2}{d_{12'}}\ket{1,2'}\bra{1,2'}+\frac{q^2}{d_{21'}}\ket{2,1'}\bra{2,1'}= \nonumber \\
H_{kinetic1}+H_{pot1}+H_{kinetic2}+H_{pot2}+H_{A-B}
\end{eqnarray} described by parameters
 $E_{p1}(t)$,$E_{p2}(t)$,$E_{p1'}(t)$,$E_{p2'}(t)$, $t_{s12}(t)$, $t_{s1'2'}(t)$ and distances between nodes k and l': $d_{11'}$,$d_{22'}$,$d_{21'}$,$d_{12'}$.
In such case q-state of the system is given as
\begin{eqnarray}
\ket{\psi}_t=\gamma_1(t)\ket{1,0}_U\ket{1,0}_L+\gamma_2(t)\ket{1,0}_U\ket{0,1}_L+\gamma_3(t)\ket{0,1}_U\ket{1,0}_L+\gamma_4(t)\ket{0,1}_U\ket{0,1}_L, \nonumber \\
\end{eqnarray}
where normalization condition gives $|\gamma_1(t)|^2+..|\gamma_4(t)|^2$. Probability of finding electron in upper system at node 1 is by action of projector $\hat{P}_{1U}=\bra{1,0}_U\bra{1,0}_L+\bra{1,0}_U\bra{0,1}_L$ on q-state $\hat{P}_{1U} \ket{\psi}$ so
it gives probability amplitude $|\gamma_1(t)+\gamma_3(t)|^2$ . On the other hand probability of finding electron from qubit A (U) at node 2 and electron from qubit B(L) at node 1 is obtained by projection $\hat{P}_{2U,1L}=\bra{0,1}_U\bra{1,0}_L$ acting on q-state
giving $(\bra{0,1}_U\bra{1,0}_L)\ket{\psi}$ that gives probability amplitude $|\gamma_3(t)|^2$.
Referring to picture from Fig.2
we set distances between nodes as $d_{11'}=d_{22'}=d_1$,$d_{12'}=d_{21'}=\sqrt{(a+b)^2+d_1^2}$ and assume Coulomb electrostatic energy to
 be of the form $E_c(k,l)=\frac{q^2}{d_{kl'}}$ and hence we obtain the matrix  Hamiltonian given as $ \hat{H}(t)= $ \tiny
\begin{eqnarray*}
\label{2bodies}
\begin{pmatrix}
E_{p1}(t)+E_{p1'}(t) + \frac{q^2}{d_1} & t_{s1'2'}(t) & t_{s12}(t) & 0 \\
t_{s1'2'}(t)^{*} & E_{p1}(t)+E_{p2'}(t)+\frac{q^2}{\sqrt{(d1)^2+(b+a)^2}} & 0 & t_{s12}(t) \\
t_{s12}^{*}(t) & 0 & E_{p2}(t)+E_{p1'}(t)+ \frac{q^2}{\sqrt{(d1)^2+(b+a)^2}} & t_{s1'2'}(t) \\
0 & t_{s12}^{*}(t) & t_{s1'2'}(t)^{*} & E_{p2}(t)+E_{p2'}(t)+ \frac{q^2}{d1} \\
\end{pmatrix} \nonumber \\
\end{eqnarray*}
\normalsize
We can introduce notation $E_{c1}=\frac{q^2}{d_1}$ and $E_{c2}=\frac{q^2}{\sqrt{d_1^2+(b+a)^2}}$. In most general case of 2 qubit electrostatic interaction one of which has 4 different Coulomb terms on matrix diagonal $E_{c1}=\frac{q^2}{d_{11'}}$, $E_{c2}\frac{q^2}{d_{12'}}$, $E_{c3}=\frac{q^2}{d_{21'}}$, $E_{c4}=\frac{q^2}{d_{22'}}$ and $\ket{\psi,t}=\hat{U}(t,t_0)\ket{\psi,t_0}$.
We introduce $q_1=E_{p1}(t)+E_{p1'}(t)+E_{c11'}$,$q_2=E_{p1}(t)+E_{p2'}(t)+E_{c12'}$, $q_3=E_{p2}(t)+E_{p1'}(t)+E_{c21'}$,$q_4=E_{p2}(t)+E_{p2'}(t)+E_{c22'}$ and in such case by using formula 8 one can decompose 2 particle Hamiltonian \ref{2bodies} as
\begin{eqnarray}
\hat{H}=\Big[\frac{(q_1+q_2+q_3+q_4)}{4}\sigma_0 \times \sigma_0 +
\frac{(q_1-q_2+q_3-q_4)}{4}\sigma_0 \times \sigma_3 + \nonumber \\
\frac{(q_1+q_2-q_3-q_4)}{4}\sigma_3 \times \sigma_0 +
\frac{(q_1-q_2-q_3+q_4)}{4}\sigma_3 \times \sigma_3 + \nonumber \\
+t_{sr1}(t)\sigma_0 \times \sigma_1 -t_{si1}(t) \sigma_0 \times \sigma_2+t_{sr2}(t)\sigma_1 \times \sigma_0 \nonumber \\
 - t_{si2}(t) \sigma_2 \times \sigma_0 \Big.
\end{eqnarray}
A very similar procedure is for the case of 3 or N interacting particles so one deals with tensor product of 3 or N Pauli matrices.
In order to simplify representation of unitary matrix describing physical system of 2 particles evolution with time it is helpful to define  $Q_1(t)=\int_{t_0}^{t}(E_{p1}(t')+E_{p1'}(t')+E_{c11'})dt'$,$Q_2(t)=\int_{t_0}^{t}(E_{p1}(t')+E_{p2'}(t')+E_{c12'})dt'$, $Q_3(t)=\int_{t_0}^{t}(E_{p2}(t')+E_{p1'}(t')+E_{c21'})dt'$,$Q_4(t)=\int_{t_0}^{t}(E_{p2}(t')+E_{p2'}(t')+E_{c22'})dt'$ and $TR1(t)=\int_{t_0}^{t}dt't_{s1r}(t')$ , $TI1(t)=\int_{t_0}^{t}dt't_{s1i}(t')$. We consider the situation when there is no hopping between q-wells $t_{s2}=0$ so, the second particle is localized among two quantum wells and first particle can move freely among 2 q-wells.
We obtain the following unitary matrix evolution with time with following $\hat{U}(t,t_0)_{1,2}=\hat{U}(t,t_0)_{1,4}=0=\hat{U}(t,t_0)_{2,3}=\hat{U}_{3,4}$ and

\tiny
\begin{eqnarray}
\hat{U}(t,t_0)_{1,1}=   \frac{1}{2 \sqrt{
(Q_1(t)-Q_3(t))^2+4 \left(TR_1(t)^2+TI_1(t)^2\right)}} \times
 \Bigg[Q_1(t)
\left(-e^{i \hbar \sqrt{
(Q_1(t)-Q_3(t))^2+4 \left(TR_1(t)^2+TI_1(t)^2\right)
}}\right) \nonumber \\ +
\left(\sqrt{ |Q_1(t)-Q_3(t)|^2+4(TR_1(t)^2+TI_1(t)^2)}
+Q_3(t) \right) \times  \left(-e^{i \hbar \sqrt{
(Q_1(t)-Q_3(t))^2+4 \left(TR_1(t)^2+TI_1(t)^2\right)
}}\right)+ \nonumber \\ \sqrt{(Q_1(t)-Q_3(t))^2+4 \left(TR_1(t)^2+TI_1(t)^2\right)} +(Q_1(t)-Q_3(t)))
e^{-\frac{1}{2} i \hbar \left(\sqrt{|\int_{t_0}^{t}dt'(q_{1}(t')-q_{3}(t'))|^2+4
 \left(t_{s1r}^2+t_{si1}^2\right)}+(Q_1(t)+Q_3(t))\right)} \Bigg] \nonumber \\
\end{eqnarray}
\normalsize
\begin{eqnarray}
\hat{U}(t,t_0)_{1,3}= \frac{2 (TI_1(t)-iTR_1(t)) e^{-\frac{1}{2}(Q_1(t)+Q_3(t)) i \hbar } \sin \left(\frac{1}{2} \hbar
   \sqrt{|Q_1(t)-Q_3(t)|^2+4(TR_1(t)^2+TI_1(t)^2)}\right)}{\sqrt{|Q_1(t)-Q_3(t)|^2+4(TR_1(t)^2+TI_1(t)^2)}}, \nonumber \\
\end{eqnarray}
\\,
\begin{eqnarray}
\hat{U}(t,t_0)_{2,2}=\Bigg[e^{(\frac{1}{2} i \hbar \left(\sqrt{  (Q_2(t)-Q_4(t))^2 +4(TR_1(t)^2+TI_1(t)^2)}-(Q_2(t)+Q_4(t))\right))} \times \nonumber \\ \times
 \frac{\left(\sqrt{(Q_2(t)-Q_4(t))^2 +4(TR_1(t)^2+TI_1(t)^2)}-Q_2(t)+Q_4(t)\right) }{2
   \sqrt{(Q_2-Q_4)^2+4(TR_1(t)^2+TI_1(t)^2)}} \nonumber \\
   -e^{\left(\frac{1}{2} i \hbar
   \left(-\sqrt{(Q_2(t)-Q_4(t))^2+ 4(TR_1(t)^2+TI_1(t)^2)}-(Q_2(t)+Q_4(t))\right)\right)} \times \nonumber \\ \times
\frac{\left(-\sqrt{ (Q_2(t)-Q_4(t))^2+ 4(TR_1(t)^2+TI_1(t)^2) }-Q_2(t)+Q_4(t)\right) }{2\sqrt{(Q_2-Q_4)^2+4(TR_1(t)^2+TI_1(t)^2)}}
\end{eqnarray}
$ $
\\
\\
$ $
\begin{eqnarray}
\hat{U}(t,t_0)_{3,3}=\frac{\exp \left(-\frac{1}{2} i \hbar \left(\sqrt{(Q_1(t)^2-Q_3(t))^2+4
   \left(TR_1(t)^2+TI_1(t)^2\right)}+Q_1(t)+Q_3(t)\right)
   \right)}{2 \sqrt{(Q_1(t)^2-Q_3(t))^2+4
   \left( TR_1(t)^2+TI_1(t)^2 \right)}} \times \nonumber \\
\Bigg[Q_1(t) \left(-1+e^{i \hbar \sqrt{(Q_1(t)-Q_3(t))^2+4
   \left(TR_1(t)^2+TI_1(t)^2 \right)}}\right)+ \nonumber \\
   \left(\sqrt{(Q_1(t)-Q_3(t))^2+4
   \left(TR_1(t)^2+TI_1(t)^2\right)}-\text{q3}\right) e^{i \hbar
   \sqrt{(Q_1(t)-Q_3(t))^2+4
   \left(TR_1(t)^2+TI_1(t)^2\right)}}+\nonumber \\+\sqrt{(Q_1(t)-Q_3(t))^2+4 \left(TR_1(t)^2+TI_1(t)^2\right)}+Q_3(t)\Bigg]
\end{eqnarray}

\begin{eqnarray*}
 \hat{U}(t,t_0)_{4,4}=\frac{\exp \left(-\frac{1}{2} i \hbar \left(\sqrt{(Q_2(t)-Q_4(t))^2+4
   \left(TR_1(t)^2+TI_1(t)^2\right)}+Q_2(t)+Q_4(t)\right)
   \right)}{2 \sqrt{(Q_2(t)-Q_4(t))^2+4
   \left(TR_1(t)^2+TI_1(t)^2\right)}} \times \nonumber \\
\times \Bigg[Q_2(t) \left(-1+e^{i \hbar \sqrt{(Q_2(t)-Q_4(t))^2+4
   \left(TR_1(t)^2+TI_1(t)^2\right)}}\right)+\nonumber \\ +\left(\sqrt{(Q_2(t)-Q_4(t)^2)^2 +4
   \left( TR_1(t)^2+TI_1(t)^2\right)}-Q_4(t)\right) e^{i \hbar
   \sqrt{(Q_2(t)-Q_4(t))^2+ 4
   \left(TR_1(t)^2+TI_1(t)^2\right)}}+ \nonumber \\
   \sqrt{(Q_2(t)-Q_4(t))^2 +4 \left(TR_1(t)^2+TI_1(t)^2\right)}+Q_4(t)\Bigg]
\end{eqnarray*}
\begin{eqnarray}
\hat{U}(t,t_0)_{2,4}=\frac{2 (TI_1(t)-i TR_1(t)) e^{-\frac{1}{2} i \hbar
   (Q_2(t)+Q_4(t))} \sin \left(\frac{1}{2} \hbar
   \sqrt{(Q_2(t)-Q_4(t))^2+4
   \left(TR_1(t)^2+TI_1(t)^2\right)}\right)}{\sqrt{(Q_2(t)-Q_4(t))^2+4 \left(TR_1(t)^2+TI_1(t)^2\right)}} \nonumber \\
\end{eqnarray}

\begin{figure}
\centering
\includegraphics[scale=0.4]{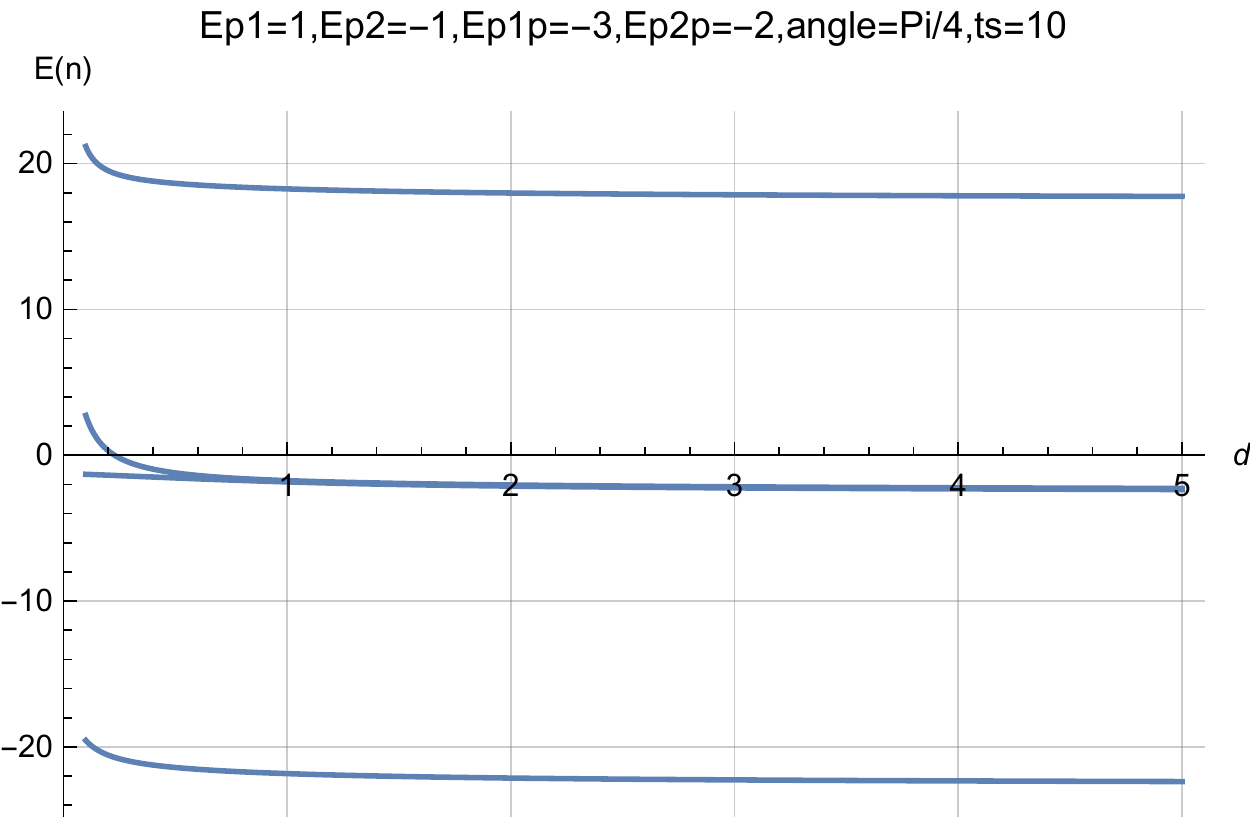}
\includegraphics[scale=0.4]{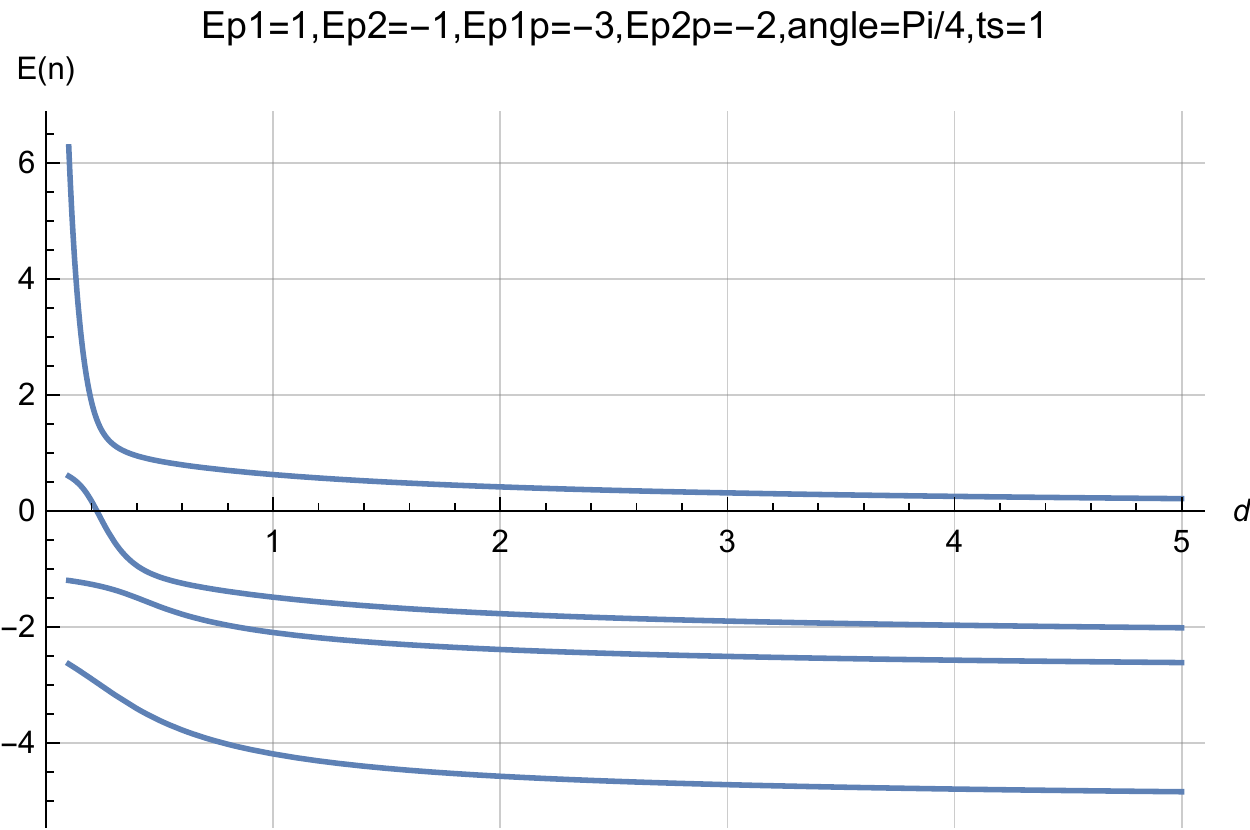}
\includegraphics[scale=0.4]{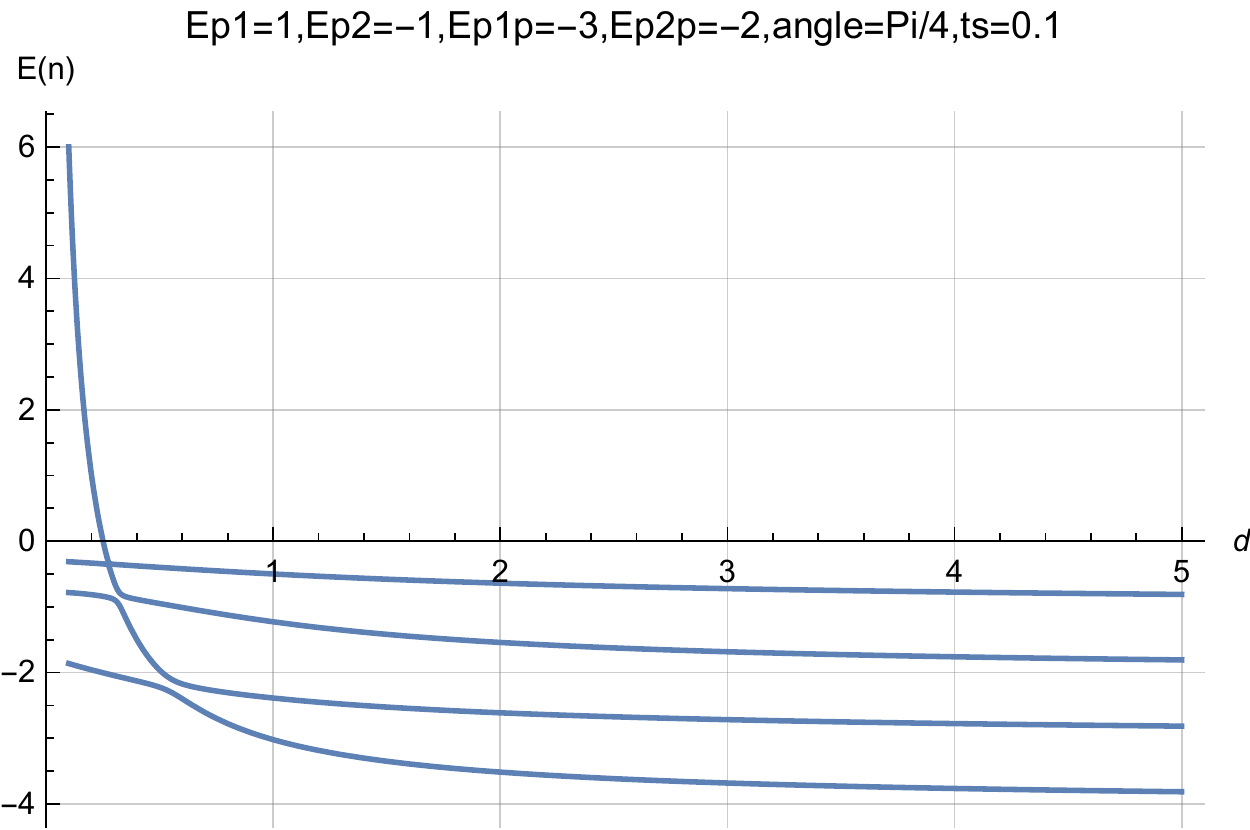}
\includegraphics[scale=0.4]{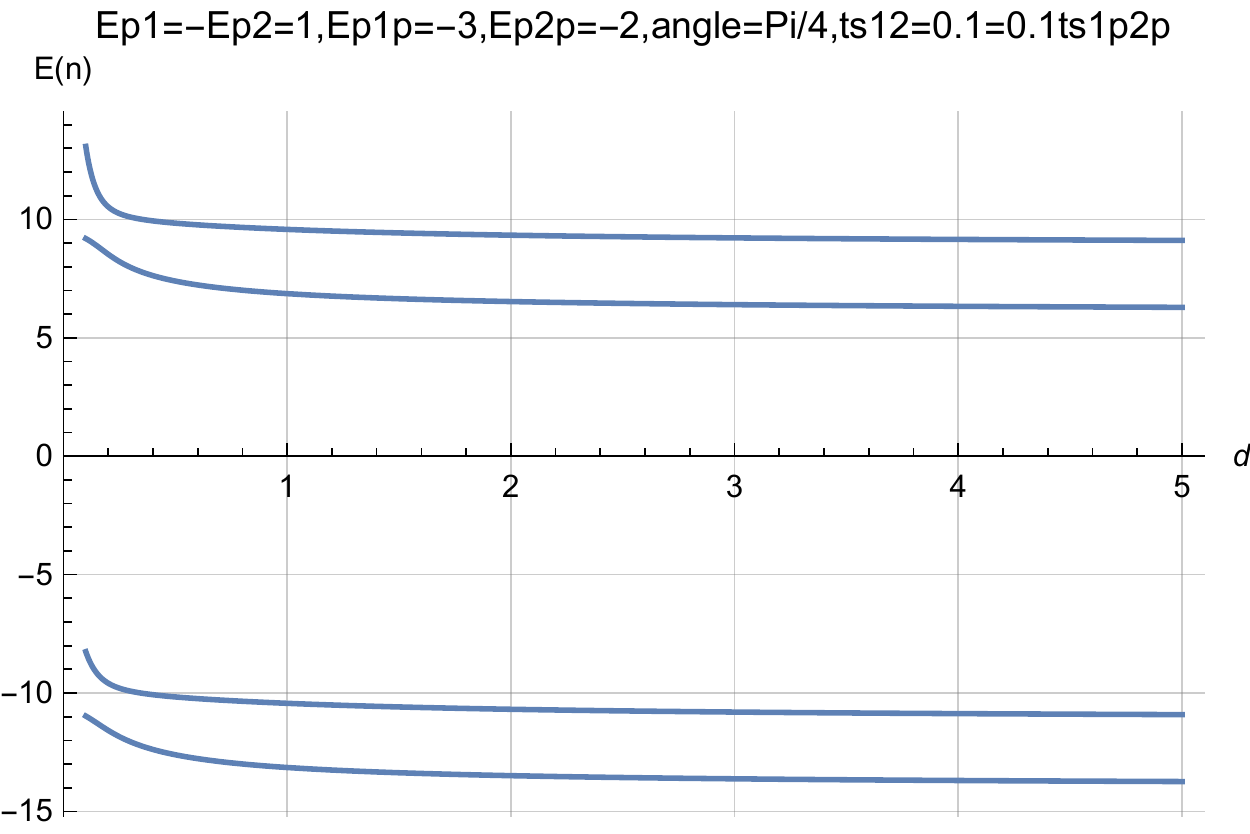}
\includegraphics[scale=0.4]{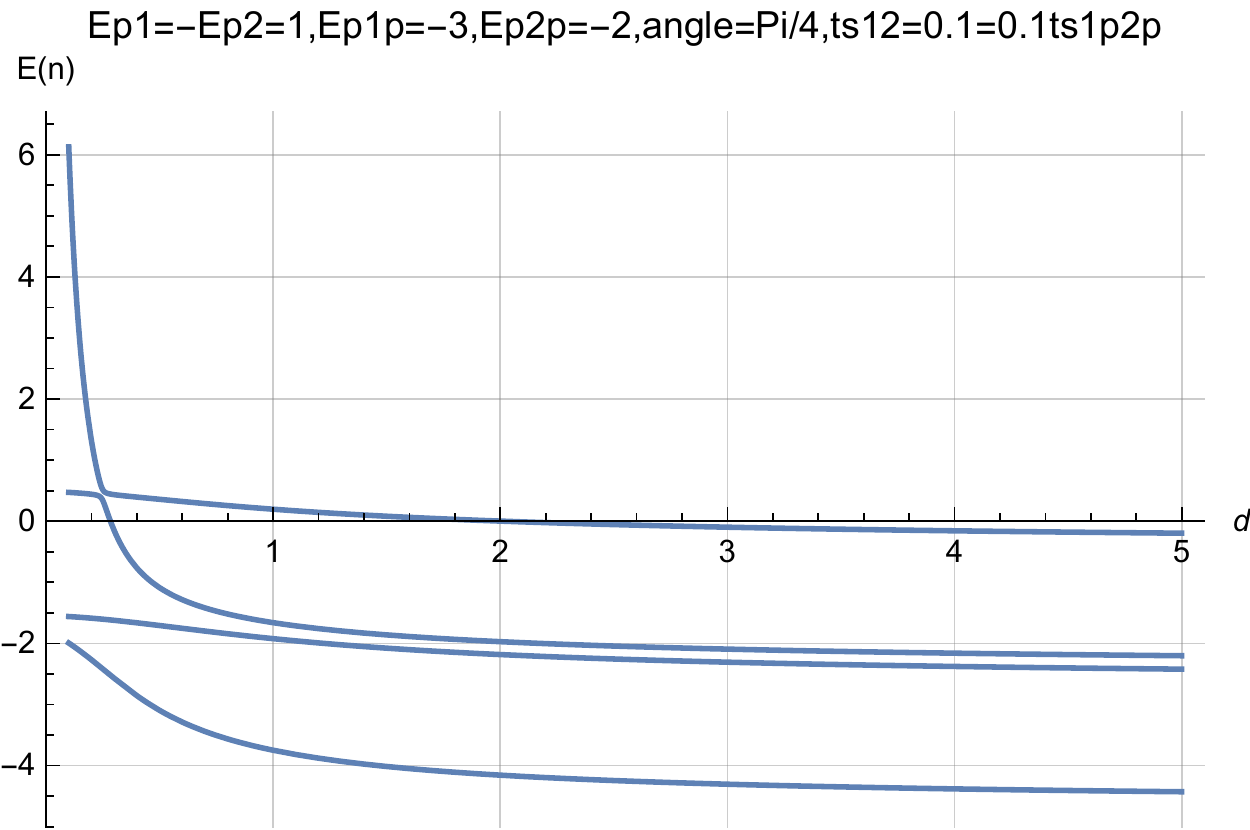}
\includegraphics[scale=0.4]{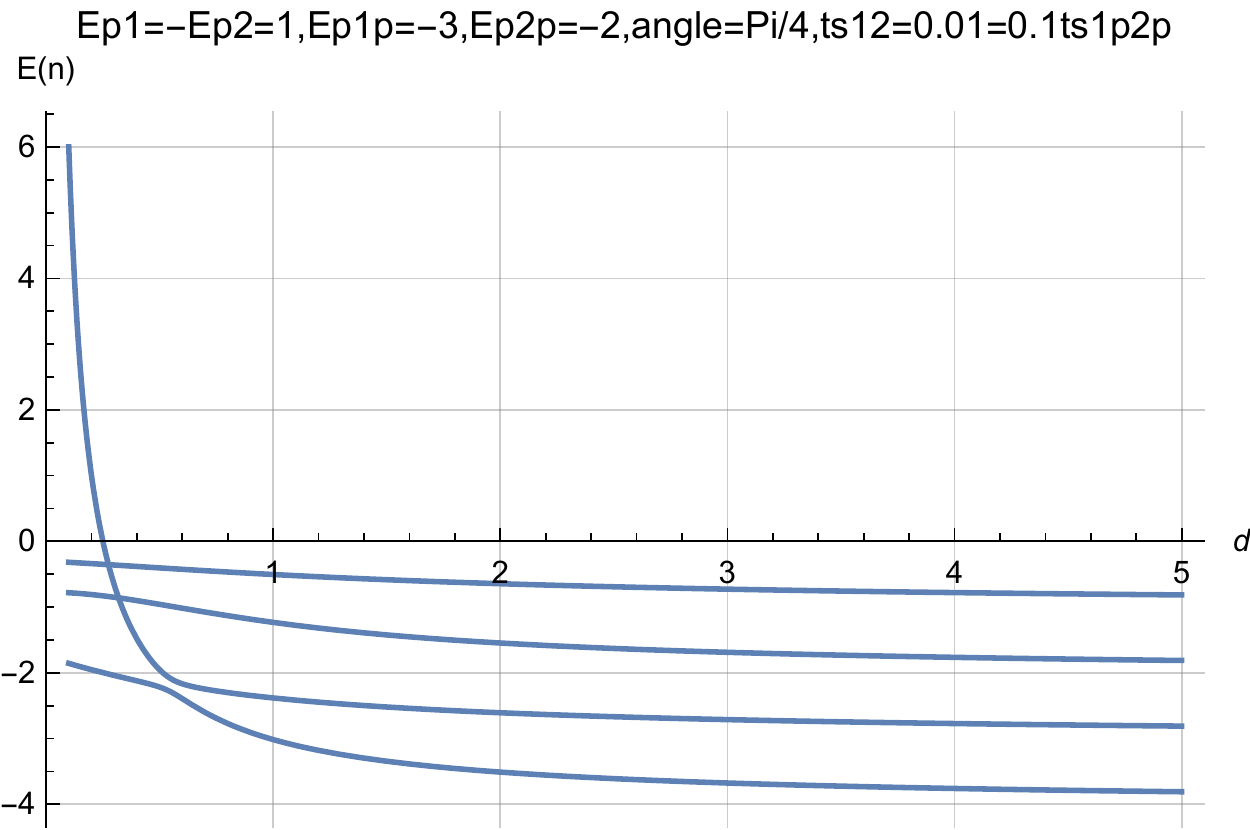}
\includegraphics[scale=0.4]{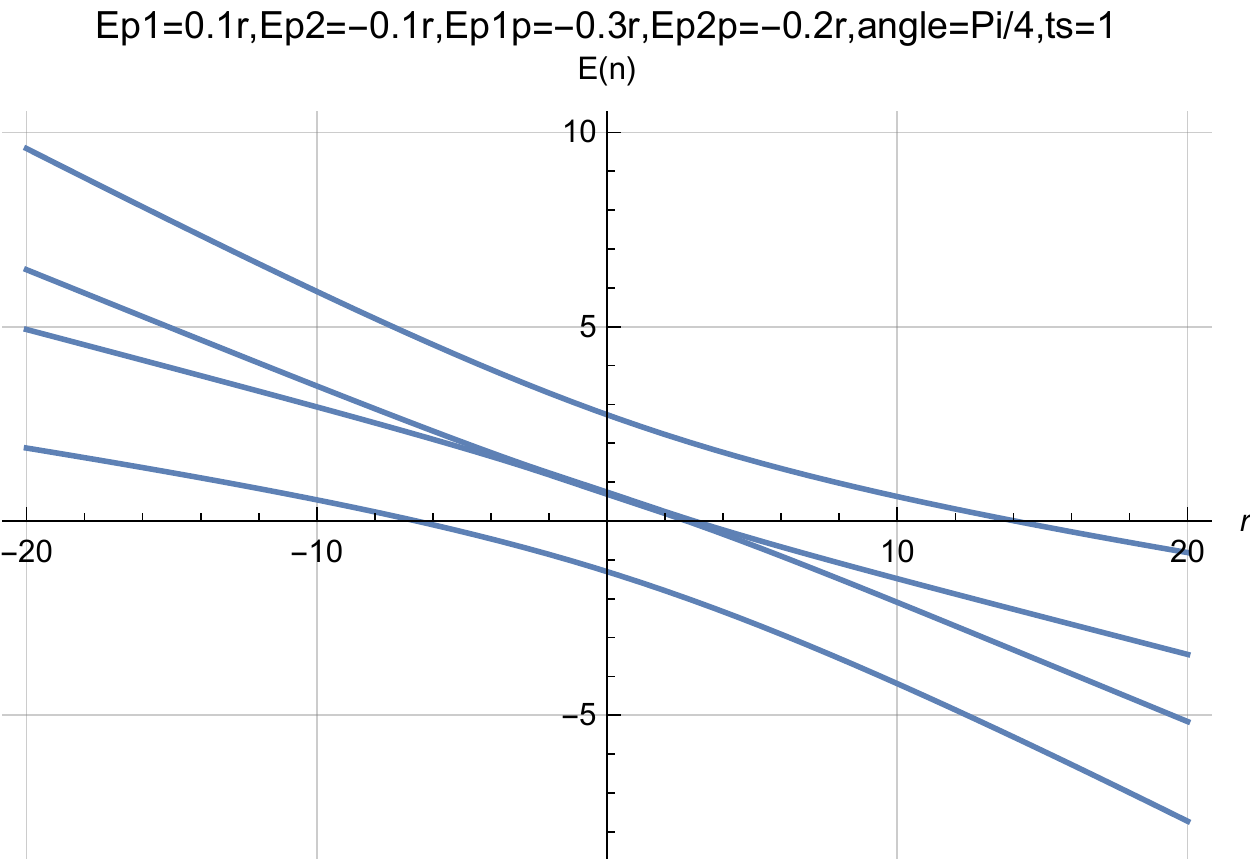}
\includegraphics[scale=0.4]{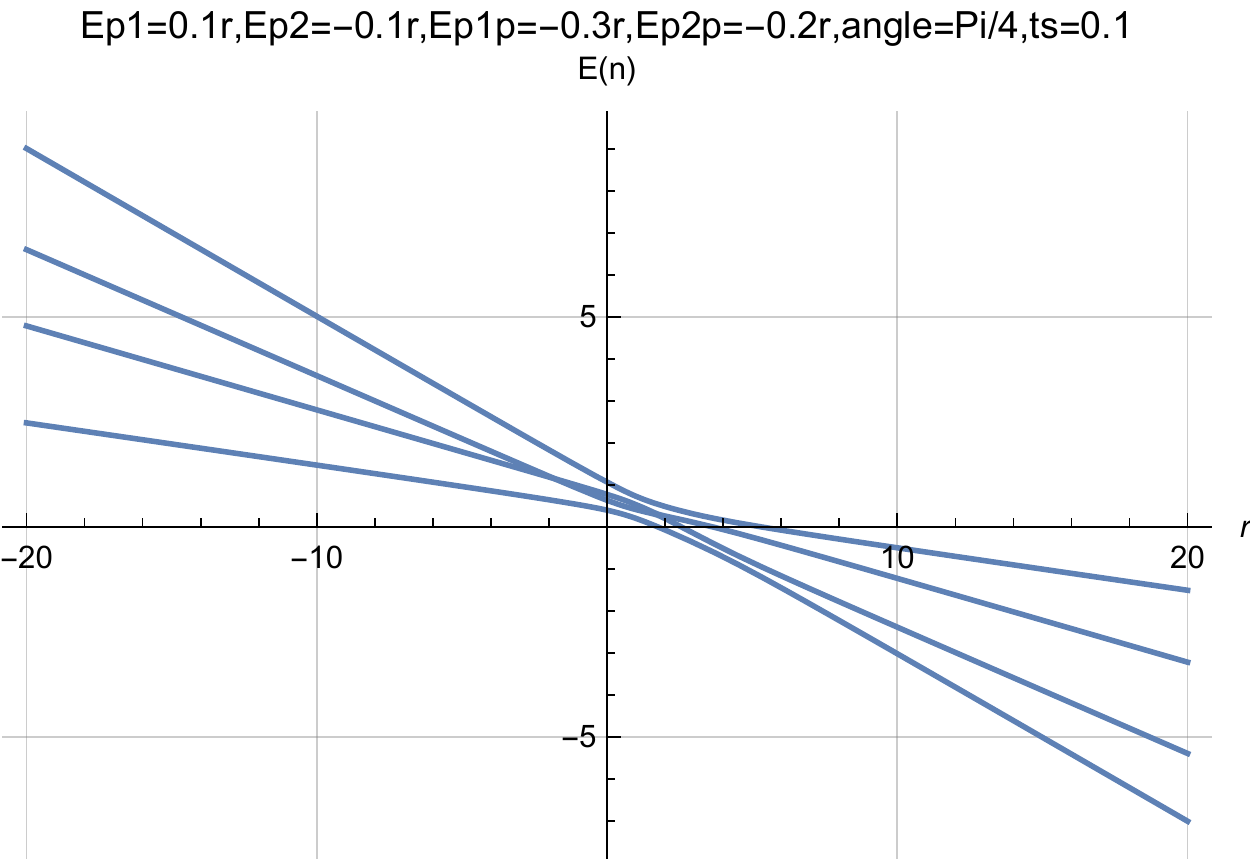}
\includegraphics[scale=0.4]{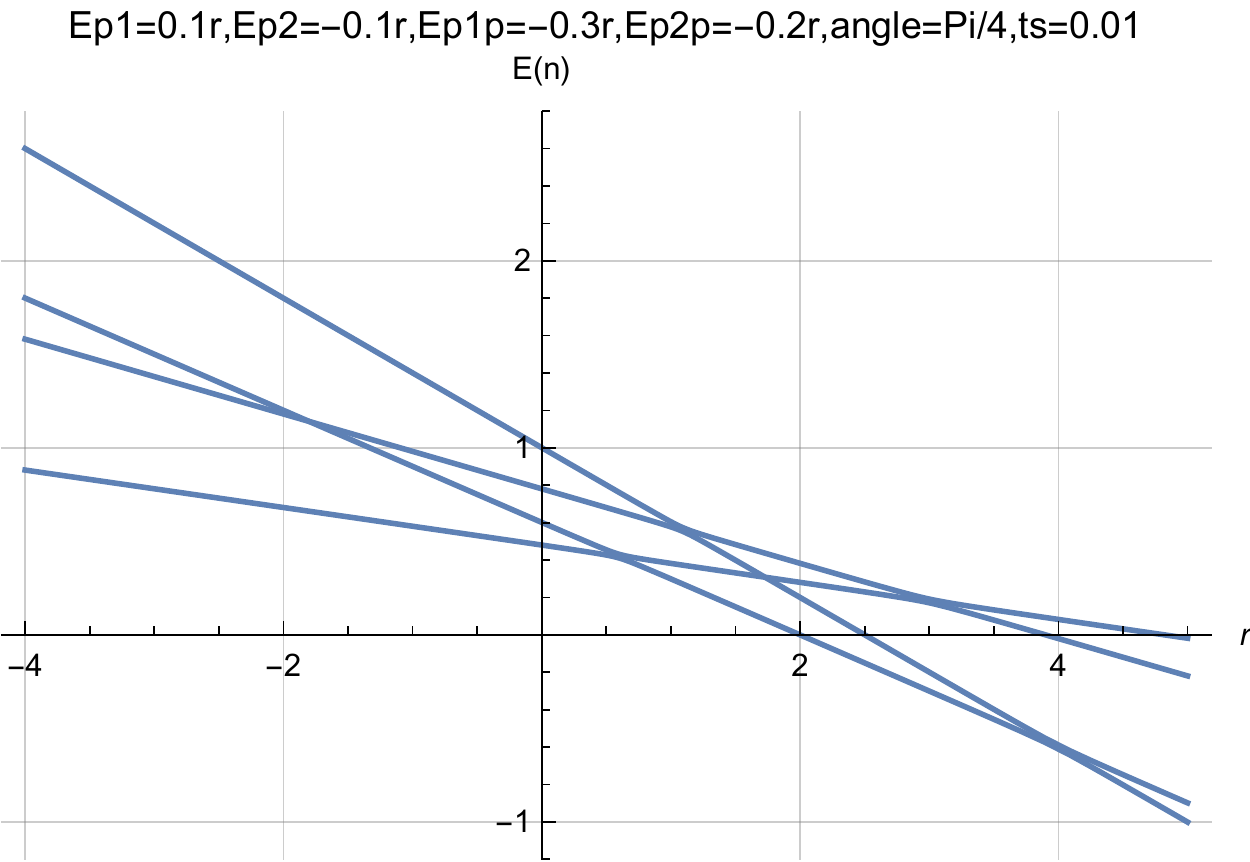}
\includegraphics[scale=0.4]{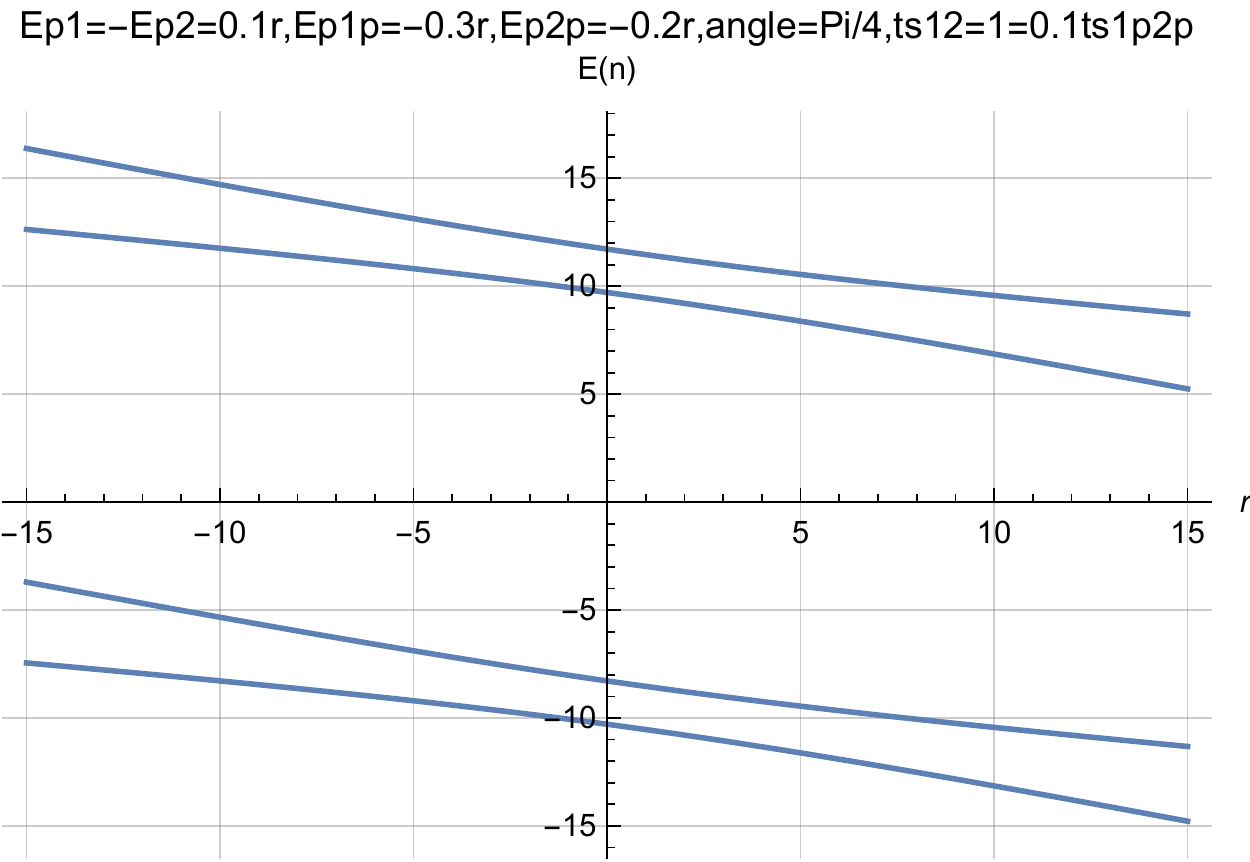}
\includegraphics[scale=0.4]{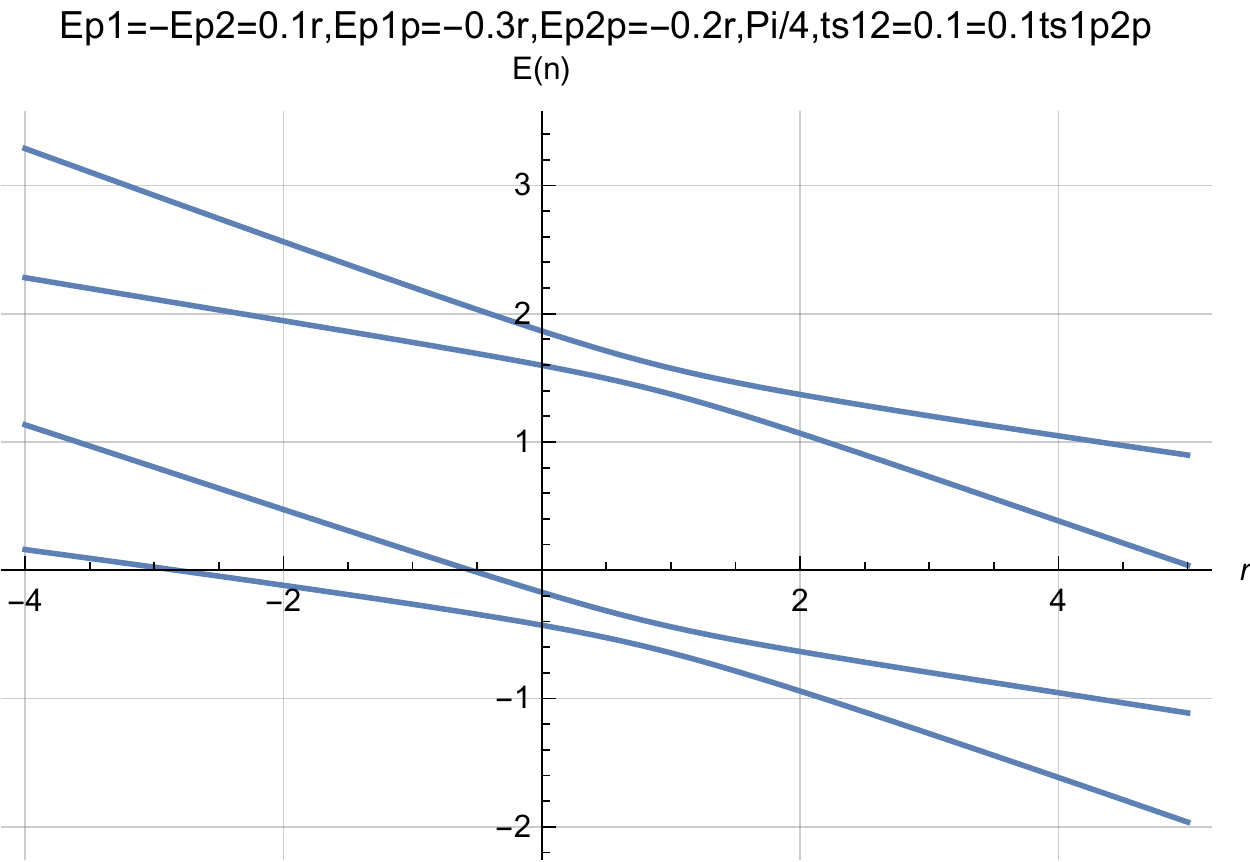}
\includegraphics[scale=0.4]{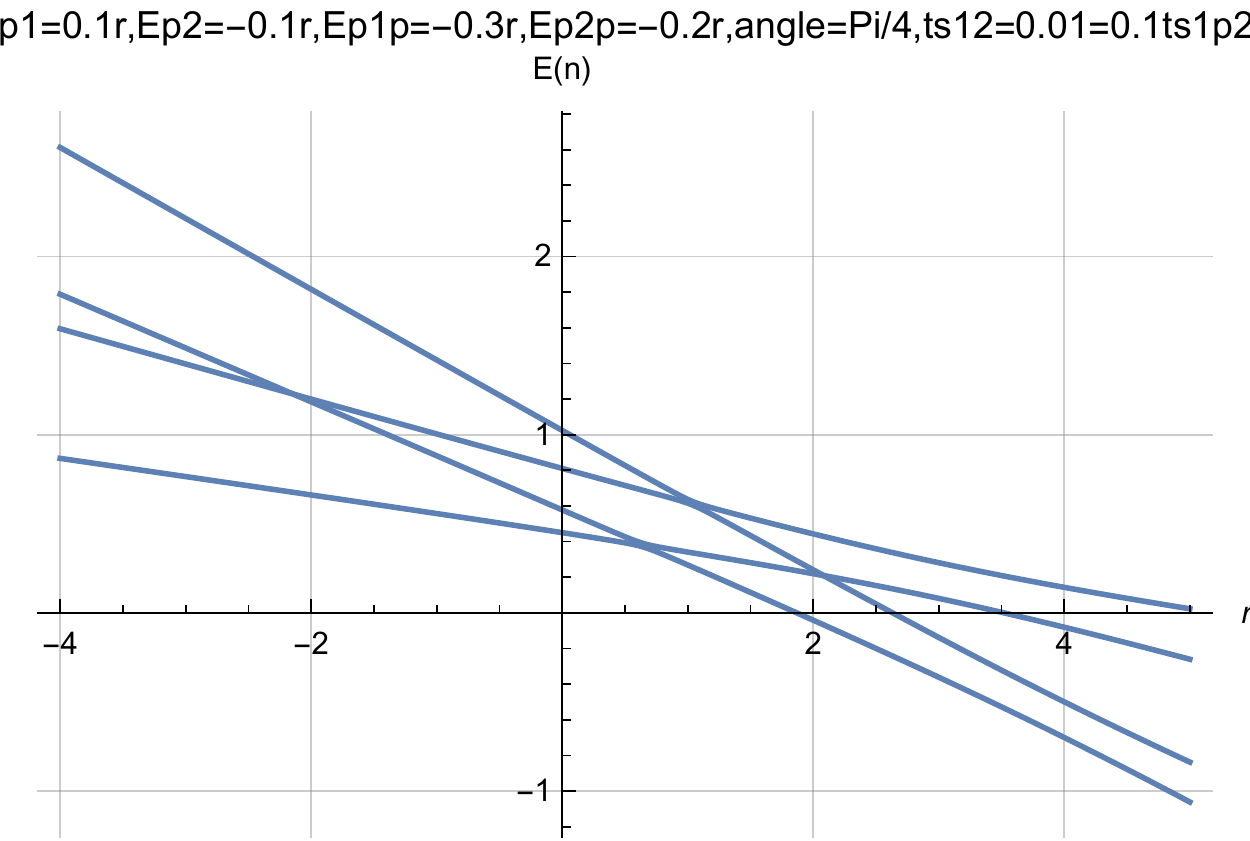}
\caption{Case of metallic-insulator transition seen from energy spectra of quantum swap gate in dependence on the distance $d$ (case of 2 electrostatically interacting qubits from Fig.\,\ref{PositionQubit}).}
\label{fig:spectra}
\end{figure}

The example of function dependence of eigenenergy spectra of 2 electrostatically interacting qubits on distance is given by Fig.\ref{fig:spectra}.

An important observation is that any element of matrix $\hat{H}(t')$ for $t' \in (t_0,t)$ denoted as $H_{k,l}(t')$ is transferred to element $\hat{U}_{k,l}(t,t_0)=e^{\frac{1}{\hbar i}\int_{t_0}^{t}dt'(H_{k,l}(t'))}$ of matrix $\hat{U}(t,t_0)$. 
We can easily generalize the presented reasoning for the system of N electrostatically coupled electrons confined by some local potentials. However we need to know the position dependent Hamiltonian eigenstate at the initial time $t_0$. In case $N>2$ finding such eigenstate is the numerical problem since
analytical solutions for roots of polynomials of one variable for higher order than 4 does not exist. Using numerical eigenstate at time instance $t_0$ we can compute the system quantum dynamics in analytical way.
This give us a strong and relatively simple mathematical tool giving full determination of quantum dynamical state at the any instance of time.
The act of measurement on position based qubit is represented by the operator $P_{Left}=\ket{1,0}_{E_1,E_2}\bra{1,0}_{E_1,E_2}$ and $P_{Right}=\ket{0,1}_{E_1,E_2}\bra{0,1}_{E_1,E_2}$.
\subsection{Simplified picture of symmetric Q-Swap gate}
\begin{figure}
\centering
\includegraphics[scale=0.3]{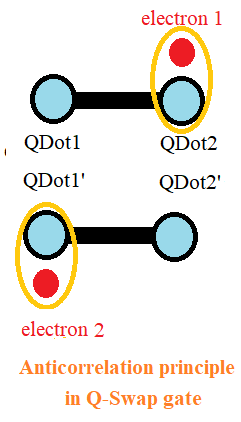}\includegraphics[scale=0.3]{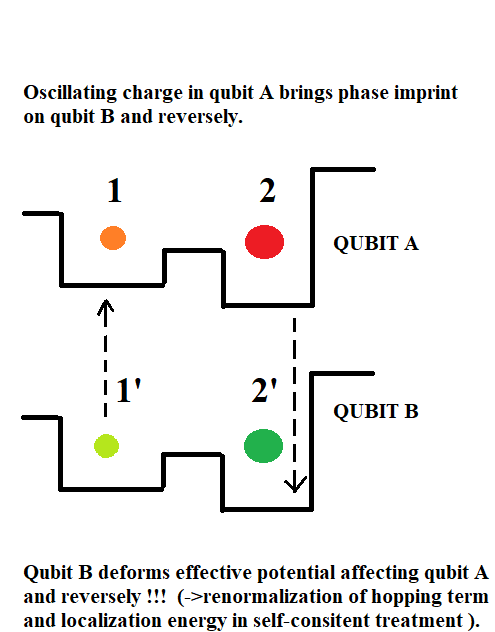}\includegraphics[scale=0.3]{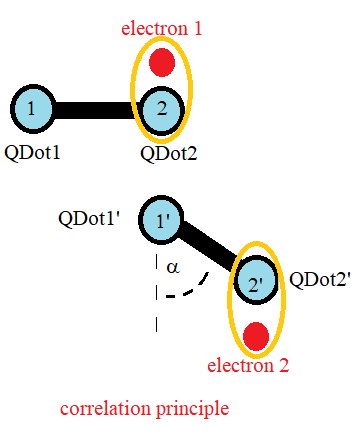}\includegraphics[scale=0.3]{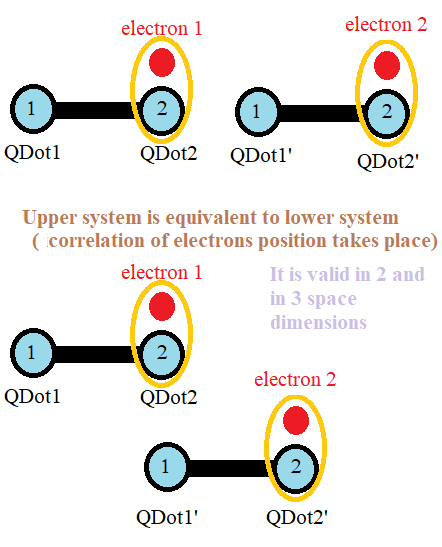}\includegraphics[scale=0.3]{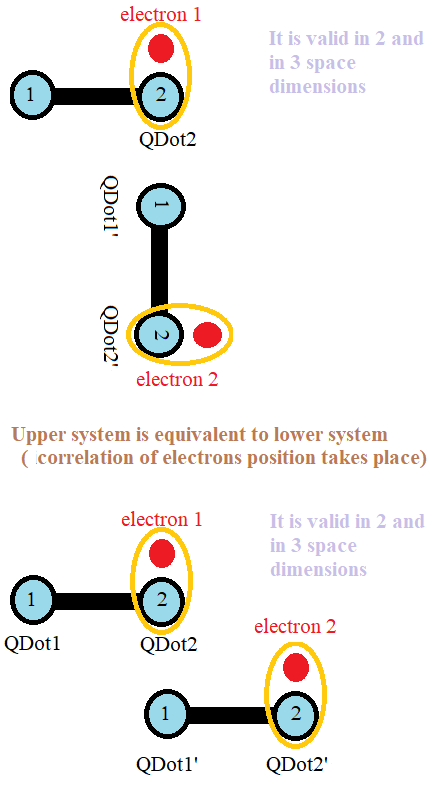}
\caption{(Very Left):Anticorrelation principle in Q-Swap gate, (Left): Scheme of renormalization procedure in the system of coupled qubits accounted by procedure given in "Analytic view on N body interaction in electrostatic
quantum gates and decoherence effects in
tight-binding model" (Arxiv:1912.01205), (Right):Illustration of anticorrelation principle in classical or quantum swap gate, (Right): Correlation principle in chain of coupled dots on one line (parallel lines), (Very Right):Correlation principle for the system of qubits in perpendicular alignment. This system is also equivalent to generalized electrostatic quantum swap gate from Fig.2.}
\label{renormalization}
\end{figure}

Now we need to find a system 4 eigenvalues and eigenstates \newline
 (4 orthogonal 4-dimensional vectors)  so we are dealing with a matrix eigenvalue problem) what is the subject of classical algebra. Let us assume that 2 double quantum dot systems are symmetric and biased by the same voltages generating potential bottoms $V_s$ so we have $E_{p1}=E_{p2}=E_{p1'}=E_{p2'}=E_p=V_s$ and that $t_{s12}=t_{s1'2'}=t_s$. Denoting $E_c(1,1')=E_c(2,2')=E_{c1}$ and $E_c(1,2')=E_c(2,1')=E_{c2}$  we are obtaining 4 orthogonal Hamiltonian eigenvectors
\begin{eqnarray}
\ket{E_1}=
\begin{pmatrix}
-1 \\
0 \\
0 \\
+1
\end{pmatrix}= \nonumber \\
=-\ket{1,0}_U\ket{1,0}_L+\ket{0,1}_U\ket{0,1}_L \nonumber \\
\neq (a_1\ket{1,0}_U+a_2\ket{0,1}_U)(a_3\ket{1,0}_U+a_4\ket{0,1}_U),
\end{eqnarray}
\begin{eqnarray}
\ket{E_2}=
\begin{pmatrix}
1 \\
0 \\
0 \\
-1
\end{pmatrix}= \nonumber \\
=\ket{1,0}_U\ket{0,1}_L-\ket{0,1}_U\ket{1,0}_L \neq (a_1\ket{1,0}_U+a_2\ket{0,1}_U)(a_3\ket{1,0}_U+a_4\ket{0,1}_U)\nonumber \\.
\end{eqnarray}
We observe that two first energetic states are degenerated so the same quantum state corresponds to 2 different eigenenergies $E_1$ and $E_2$.
 This degeneracy is non-present if we come back to Schroedinger picture and  observe that localized energy and hopping terms for one particle are depending on another particle presence that will bring renormalization of wavevectors.
Situation is depicted in Fig.\ref{renormalization}. Degeneracy of eigenstates is lifted if we set $E_{p1}(|\psi(1')|^2,|\psi(2')|^2), E_{p2}(|\psi(1')|^2,|\psi(1')|^2)$, $E_{p1'}(|\psi(1)|^2,|\psi(2)|^2), E_{p2'}(|\psi(1)|^2,|\psi(1)|^2)$ and $t_{1 \rightarrow 2}(|\psi(1')|^2,|\psi(2')|^2)$, \newline
 $t_{1 \rightarrow 2}(|\psi(1)|^2,|\psi(2)|^2)$.

The same argument is for another
wavevectors as given below.
\begin{eqnarray}
\ket{E_{3(4)}}=
\begin{pmatrix}
1 \\
\mp\frac{4 t_s}{\pm(-E_{c1} + E_{c2}) + \sqrt{(E_{c1} - E_{c2})^2 + 16 t_s^2}} \\
\mp\frac{4 t_s}{\pm(-E_{c1} + E_{c2}) + \sqrt{(E_{c1} - E_{c2})^2 + 16 t_s^2}} \\
1
\end{pmatrix}= \nonumber \\
=\ket{1,0}_U\ket{1,0}_L+\ket{0,1}_U\ket{0,1}_L+c (\ket{1,0}_U\ket{0,1}_L +\ket{0,1}_U\ket{1,0}_L )= \nonumber \\
=(\ket{1,0}_U+\ket{0,1}_U)(\ket{1,0}_L+\ket{0,1}_L)+(c-1)(\ket{1,0}_U\ket{0,1}_L +\ket{0,1}_U\ket{1,0}_L) \nonumber \\
\neq (a_1\ket{1,0}_U+a_2\ket{0,1}_U)(a_3\ket{1,0}_U+a_4\ket{0,1}_U), \nonumber \\
\end{eqnarray}
where c=$\mp\frac{4 t_s}{\pm(-E_{c1} + E_{c2}) + \sqrt{(E_{c1} - E_{c2})^2 + 16 t_s^2}}$.
First two $\ket{E_1}$ and $\ket{E_2}$ energy eigenstates are always entangled, while  $\ket{E_3}$ and $\ket{E_4}$ eigenenergies are only partially entangled if $\mp\frac{4 t_s}{\pm(-E_{c1} + E_{c2}) + \sqrt{(E_{c1} - E_{c2})^2 + 16 t_s^2}} \neq 1$. If
$c=1=\mp\frac{4 t_s}{\pm(-E_{c1} + E_{c2}) + \sqrt{(E_{c1} - E_{c2})^2 + 16 t_s^2}}$ last two energy eigenstates are not entangled. The situation of c=1 takes place when $E_{c1}=E_{c2}$ so when two qubits are infinitely far away so when they are electrostatically decoupled. Situation of c=0 is interesting because it means that $\ket{E_3}$ and $\ket{E_4}$ are maximally entangled and it occurs when $t_s=0$ so when two electrons are maximally localized in each of the qubit so there is no hopping between left and right well.

The obtained eigenenergy states correspond to 4 eigenenergies
\begin{eqnarray}
E_1=E_{c1} + 2 V_s, E_2=E_{c2} + 2 V_s, E_1 > E_2 \nonumber \\
E_3= \frac{1}{2} ( (E_{c1} + E_{c2})  - \sqrt{(E_{c1} -E_{c2})^2 + 16 t_s^2} + 4 V_s)=  \nonumber \\
=\frac{1}{2} ( (q^2(\frac{1}{d_1}+\frac{1}{\sqrt{d_1^2+(a+b)^2}} ))  - \sqrt{(q^2(\frac{1}{d_1}-\frac{1}{\sqrt{d_1^2+(a+b)^2}} ))^2 + 16 t_s^2} + 4 V_s), \nonumber \\
E_4 =\frac{1}{2} ( (E_{c1} + E_{c2})  + \sqrt{(E_{c1} -E_{c2})^2 + 16 t_s^2} + 4 V_s)=  \nonumber \\
 \frac{1}{2} ( (q^2(\frac{1}{d_1}+\frac{1}{\sqrt{d_1^2+(a+b)^2}})+ \sqrt{(q^2(\frac{1}{d_1}-\frac{1}{\sqrt{d_1^2+(a+b)^2}}))^2 + 16 t_s^2} + 4 V_s), E_4 > E_3 \nonumber \\ .
\end{eqnarray}

We also notice that the eigenenergy states $\ket{E_1}$, $\ket{E_2}$ ,$\ket{E_3}$, $\ket{E_4}$ do not have its classical counterpart since upper electron exists at both positions 1 and 2 and lower electron exists at both positions at the same time. We observe that when distance between two systems of double quantum dots goes into infinity the energy difference between quantum state corresponding to $\ket{E_3}$ and $\ket{E_4}$ goes to zero. This makes those two entangled states degenerated.




Normalized 4 eigenvectors of 2 interacting qubits in SWAP Q-Gate configuration are of the following form
\begin{eqnarray}
 \ket{E_1}_n=
\frac{1}{\sqrt{\left(8\left(\frac{t_{sr1}-t_{sr2}}{\sqrt{(E_{c1}-E_{c2})^2+4
   (t_{sr1}-t_{sr2})^2}-E_{c1}+E_{c2}}\right)\right)^2+2
   }}
\begin{pmatrix}
-1,\\
-\frac{2(t_{sr1}-t_{sr2})}{\sqrt{(E_{c1}-E_{c2})^2+4 (t_{sr1}-t_{sr2})^2}-E_{c1}+E_{c2}}, \\
\frac{2(t_{sr1}-t_{sr2})}{\sqrt{(E_{c1}-E_{c2})^2+4(t_{sr1}-t_{sr2})^2}-E_{c1}+E_{c2}} \\
1
\end{pmatrix}=\nonumber \\
\frac{1}{\sqrt{\left(8\left(\frac{t_{sr1}-t_{sr2}}{\sqrt{(E_{c1}-E_{c2})^2+4
   (t_{sr1}-t_{sr2})^2}-E_{c1}+E_{c2}}\right)\right)^2+2
   }} \ket{E_1}
\end{eqnarray}

\begin{eqnarray}
 \ket{E_2}_{n}= -\frac{1}{\sqrt{\left(8\left(\frac{t_{sr1}-t_{sr2}}{\sqrt{(E_{c1}-E_{c2})^2+4
   (t_{sr1}-t_{sr2})^2}+E_{c1}-E_{c2}}\right)\right)^2+2
   }}
\begin{pmatrix}
-1 \\
\frac{2
   (t_{sr1}-t_{sr2})}{\sqrt{(E_{c1}-E_{c2})^2+4
   (t_{sr1}-t_{sr2})^2}+E_{c1}-E_{c2}} \\
-\frac{2(t_{sr1}-t_{sr2})}{\sqrt{(E_{c1}-E_{c2})^2+4
   (t_{sr1}-t_{sr2})^2}+E_{c1}-E_{c2}} \\,1
\end{pmatrix}=\nonumber \\ =-
\frac{1}{\sqrt{\left(8\left(\frac{t_{sr1}-t_{sr2}}{\sqrt{(E_{c1}-E_{c2})^2+4
   (t_{sr1}-t_{sr2})^2}+E_{c1}-E_{c2}}\right)\right)^2+2
   }} \ket{E_2}
\end{eqnarray}

\begin{eqnarray}
\ket{E_3}_{n}= \frac{1}{\sqrt{\left(8\left(\frac{\text{tsr1}+\text{tsr2}}{\sqrt{(\text{Ec1}-\text{Ec2})^2+4
   (\text{tsr1}+\text{tsr2})^2}-\text{Ec1}+\text{Ec2}}\right)\right)^2+2
   }}
\begin{pmatrix}
1, \\
-\frac{2(t_{sr1}+t_{sr2})}{\sqrt{(E_{c1}-E_{c2})^2+4
   (t_{sr1}+t_{sr2})^2}-E_{c1}+E_{c2}}, \\
-\frac{2(t_{sr1}+t_{sr2})}{\sqrt{(E_{c1}-E_{c2})^2+4
   (t_{sr1}+t_{sr2})^2}-E_{c1}+E_{c2}}, \\
1
\end{pmatrix}= \nonumber \\ =
\frac{1}{\sqrt{\left(8\left(\frac{\text{tsr1}+\text{tsr2}}{\sqrt{(\text{Ec1}-\text{Ec2})^2+4
   (\text{tsr1}+\text{tsr2})^2}-\text{Ec1}+\text{Ec2}}\right)\right)^2+2
   }} \ket{E_3}
\end{eqnarray}

\begin{eqnarray}
\ket{E_4}_{n}=  \frac{1}{\sqrt{\left(8\left(\frac{t_{sr1}+t_{sr2}}{\sqrt{(E_{c1}-E_{c2})^2+4
   (t_{sr1}+t_{sr2})^2}-E_{c2}+E_{c1}}\right)\right)^2+2
   }}
\begin{pmatrix}
1, \\
\frac{2(t_{sr1}+t_{sr2})}{\sqrt{(E_{c1}-E_{c2})^2+4(t_{sr1}+t_{sr2})^2}+E_{c1}-E_{c2}}, \\
\frac{2(t_{sr1}+t_{sr2})}{\sqrt{(E_{c1}-E_{c2})^2+4(t_{sr1}+t_{sr2})^2}+E_{c1}-E_{c2}},  \\
1
\end{pmatrix}= \nonumber \\
\frac{1}{\sqrt{\left(8\left(\frac{t_{sr1}+t_{sr2}}{\sqrt{(E_{c1}-E_{c2})^2+4
   (t_{sr1}+t_{sr2})^2}-E_{c2}+E_{c1}}\right)\right)^2+2
   }} \ket{E_4}.
   \end{eqnarray}

We are obtaining simplifications after assuming $t_{sr1}(t)=t_{sr2}(t)$ so we obtain
\begin{equation}
\ket{E_1}_{n}=\frac{1}{\sqrt{2}}
\begin{pmatrix}
-1 \\
0 \\
0 \\
1
\end{pmatrix},
\ket{E_2}_{n}=\frac{1}{\sqrt{2}}
\begin{pmatrix}
1 \\
0 \\
0 \\
-1
\end{pmatrix},
\end{equation}
\begin{eqnarray}
\ket{E_3}_{n}=\sqrt{\frac{4t_s}{(E_{c2}-E_{c1}) + 8 t_s - \sqrt{ (E_{c1} - E_{c2})^2 + 16 t_s^2}}}
\begin{pmatrix}
1 \\
-\frac{4 t_s}{(-E_{c1} + E_{c2}) + \sqrt{(E_{c1} - E_{c2})^2 + 16 t_s^2}} \\
-\frac{4 t_s}{(-E_{c1} + E_{c2}) + \sqrt{(E_{c1} - E_{c2})^2 + 16 t_s^2}} \\
1
\end{pmatrix}, \nonumber \\
\end{eqnarray}

\begin{eqnarray}
\ket{E_4}_{n}=\sqrt{\frac{4t_s}{(E_{c1}-E_{c2}) + 8 t_s - \sqrt{ (E_{c1} - E_{c2})^2 + 16 t_s^2}}}
\begin{pmatrix}
1 \\
\frac{4 t_s}{(E_{c1} - E_{c2}) + \sqrt{(E_{c1} - E_{c2})^2 + 16 t_s^2}} \\
\frac{4 t_s}{(E_{c1} - E_{c2}) + \sqrt{(E_{c1} - E_{c2})^2 + 16 t_s^2}} \\
1
\end{pmatrix}, \nonumber \\.
\end{eqnarray}

It is worth mentioning that if we want to bring two electrostatic qubits to the entangled state we need to cool down (or heat-up)  the system of interacting qubits to the energy $E_1$ (or to energy $E_2$).
Otherwise we might also wish to disentangle two electrostatically interacting qubits. In such way one of the scenario is to bring the quantum system either to energy $E_3$ or $E_4$ so only partial entanglement will be achieved.
Other scenario would be by bringing the occupancy of different energetic levels so net entanglement is reduced. One can use the entanglement witness in quantifying the existence of entanglement. One of the simplest q-state entanglement measurement is von Neumann entanglement entropy as it is expressed by formula \ref{entropyS} that requires the knowledge of q-system density matrix with time. Such matrices can be obtained analytically for the case of 2 electrostatically interacting qubits.

It is interesting to spot the dependence of eigenergies on distance between interacting qubits in the general case as it is depicted in Fig.6. Now we are moving towards description the procedure of cooling down or heating up in Q-Swap gate.
The procedure was discussed previously in the case of single qubit. Now it is exercised in the case of 2-qubit electrostatic interaction. For the sake of simplicity we will change the occupancy of the energy level $E_1$ and energy level level $E_2$ and keep
the occupancy of other energy levels unchanged.  We can write the $\ket{E_2}\bra{E_1} $ as
\begin{eqnarray}
\ket{E_2}_n\bra{E_1}_n=\frac{1}{2}
\begin{pmatrix}
1 \\ 0 \\ 0 \\ -1
\end{pmatrix}
\begin{pmatrix}
-1 & 0 & 0 & 1
\end{pmatrix}=
\begin{pmatrix}
-1 & 0 & 0 & +1 \\
0 & 0 & 0 & 0 \\
0 & 0 & 0 & 0 \\
+1 & 0 & 0 & -1 \\
\end{pmatrix}, \nonumber \\
\ket{E_1}_n\bra{E_2}_n=\frac{1}{2}
\begin{pmatrix}
-1 \\ 0 \\ 0 \\ 1
\end{pmatrix}
\begin{pmatrix}
1 & 0 & 0 & -1
\end{pmatrix}=
\begin{pmatrix}
-1 & 0 & 0 & +1 \\
0 & 0 & 0 &  0 \\
0 & 0 & 0 & 0 \\
+1 & 0 & 0 & -1 \\
\end{pmatrix}. \nonumber \\
\end{eqnarray}
We are introducing $f_1$ and $f_2$ real valued functions of small magnitude $f(t)=f_1(t)=f_2(t), (|f_1|,|f_2|<<(E_1,E_2))$  and we are considering the following Hamiltonian having $H_0$ that is time-independent and other part dependent part as
\begin{eqnarray}
\hat{H}=\hat{H}_0+f_1(t)\ket{E_2}_n\bra{E_1}_n+f_2(t)\ket{E_1}_n\bra{E_2}_n= \nonumber \\
=E_1\ket{E_1}\bra{E_1}+E_2\ket{E_2}\bra{E_2}+f_1(t)\ket{E_2}_n\bra{E_1}_n+f_2(t)\ket{E_1}_n\bra{E_2}_n= \nonumber
\end{eqnarray}
\begin{eqnarray}
=
\begin{pmatrix}
2E_{p}+ \frac{q^2}{d_1} & t_{s} & t_{s} & 0 \\
t_{s}^{*} & 2E_{p}+\frac{q^2}{\sqrt{(d1)^2+(b+a)^2}} & 0 & t_{s} \\
t_{s}^{*} & 0 & 2E_{p}+ \frac{q^2}{\sqrt{(d1)^2+(b+a)^2}} & t_{s} \\
0 & t_{s}^{*} & t_{s}^{*} & 2 E_{p}+ \frac{q^2}{d1} \\
\end{pmatrix}   \nonumber \\
+\frac{1}{2}\left(f_1
\begin{pmatrix}
-1 & 0 & 0 & 1 \\
0 & 0 & 0 & 0 \\
0 & 0 & 0 & 0 \\
1 & 0 & 0 & -1 \\
\end{pmatrix}
+f_2
\begin{pmatrix}
-1 & 0 & 0 & 1 \\
0 & 0 & 0 &  0 \\
0 & 0 & 0 & 0 \\
1 & 0 & 0 & -1 \\
\end{pmatrix}\right)=
\nonumber \\
=
\begin{pmatrix}
2E_{p}+ \frac{q^2}{d_1}-f(t) & t_{s} & t_{s} & f(t) \\
t_{s}^{*} & 2E_{p}+\frac{q^2}{\sqrt{(d1)^2+(b+a)^2}} & 0 & t_{s} \\
t_{s}^{*} & 0 & 2E_{p}+ \frac{q^2}{\sqrt{(d1)^2+(b+a)^2}} & t_{s} \\
f(t) & t_{s}^{*} & t_{s}^{*} & 2 E_{p}+ \frac{q^2}{d1}-f(t) \\
\end{pmatrix} = \nonumber \\
=\hat{H}(t)_{E_1<->E_2,Q-Swap}.\nonumber \\
\end{eqnarray}

\normalsize
Initially we have established the following parameters of tight-binding model as $t_{s12}=t_{s1'2'}$. Changing $t_{s12}$ into $t_{s12}-\frac{f(t)}{2}$ and $t_{s1'2'}$ into $t_{s1'2'}+\frac{f(t)}{2}$ while keeping other parameters of tight-binding model unchanged will result
in the heating up (cooling down) of q-state of SWAP gate so population of energy level $E_1$ and $E_2$ are time-depenent, while populations of energy levels $E_3$ and $E_4$ are unchanged. Practically our results mean that we need to keep all our confiment potential bottoms constant, while changing barrier height between neighbouring q-dots in each of position based qubits.
In such way we have established the procedure of perturbative cooling (heating up) of q-state. Non-perturbative approach is absolutely possible but it requires full knowledge of time dependent eigenstates and eigenenergies (solutions of eigenenergies of 4th order polynomial are very lengthy in general case) and therefore corresponding expression are very lengthy. In similar fashion we can heat up or cool down two coupled Single Electron Lines \cite{SEL} as in Fig.1 or any other q-system having N interacting q-bodies that can be represented by the system of N-interacting position based qubits.
\subsection{Case of density matrix in case of 2 interacting particles in symmetric case}
We consider the simplifying matrix and highly symmetric matrix of the form
\tiny
\begin{eqnarray}
\hat{H}(t)= \nonumber \\
\begin{pmatrix}
2E_{p}(t)+ \frac{q^2}{d_1}=q_{11}+q_{22}  & t_{sr2}(t) & t_{sr1}(t) & 0 \\
t_{sr2}(t)  & 2E_{p}(t)+\frac{q^2}{\sqrt{(d_1)^2+(b+a)^2}}=q_{11}-q_{22} & 0 & t_{sr1}(t) \\
t_{sr1}(t) & 0 & 2E_{p}(t)+ \frac{q^2}{\sqrt{(d_1)^2+(b+a)^2}}=q_{11}-q_{22} & t_{sr2}(t) \\
0 & t_{sr1}(t) & t_{sr2}(t) & 2 E_{p}(t)+ \frac{q^2}{d1}=q_{11}+q_{22} \\
\end{pmatrix} =\nonumber \\
=\hat{\sigma}_0 \times \hat{\sigma}_0 q_{11} + \hat{\sigma}_{3} \times \hat{\sigma}_{3} q_{22} +t_{sr2}(t)  \hat{\sigma}_0 \times \hat{\sigma}_3 + t_{sr1}(t) \hat{\sigma}_3  \times \hat{\sigma}_0 \nonumber \\
\end{eqnarray}
\normalsize
that has only real value components $H_{k,l}$ with $q_{11}=E_p(t)+\frac{E_{c1}+E_{c2}}{2}=E_p(t)+\frac{1}{2}(\frac{q^2}{d_1}+\frac{q^2}{\sqrt{(d_1)^2+(b+a)^2}})$, $q_{22}=\frac{E_{c1}-E_{c2}}{2}=\frac{1}{2}(\frac{q^2}{d_1}-\frac{q^2}{\sqrt{(d_1)^2+(b+a)^2}})$ and $Q_{11}(t)=\int_{t_0}^{t}dt'q_{11}(t')$, $Q_{22}(t)=\int_{t_0}^{t}dt'q_{22}(t')$, $TR1(t)=\int_{t_0}^{t}dt't_{sr1}(t')$, $TR2(t)=\int_{t_0}^{t}dt't_{sr2}(t')$.  We obtain the density matrix

\begin{eqnarray}
\hat{U}(t)=
\begin{pmatrix}
U_{1,1}(t) & U_{1,2}(t) & U_{1,3}(t) & U_{1,4}(t) \\
U_{2,1}(t) & U_{2,2}(t) & U_{2,3}(t) & U_{2,4}(t) \\
U_{3,1}(t) & U_{3,2}(t) & U_{3,3}(t) & U_{3,4}(t) \\
U_{4,1}(t) & U_{4,2}(t) & U_{4,3}(t) & U_{4,4}(t) \\
\end{pmatrix},
\hat{\rho}(t)=\hat{U}(t,t_0)
\begin{pmatrix}
\rho_{1,1}(t_0) & \rho_{1,2}(t_0) & \rho_{1,3}(t_0) & \rho_{1,4}(t_0) \\
\rho_{2,1}(t_0) & \rho_{2,2}(t_0) & \rho_{2,3}(t_0) & \rho_{2,4}(t_0) \\
\rho_{3,1}(t_0) & \rho_{3,2}(t_0) & \rho_{3,3}(t_0) & \rho_{3,4}(t_0) \\
\rho_{4,1}(t_0) & \rho_{4,2}(t_0) & \rho_{4,3}(t_0) & \rho_{4,4}(t_0) \\
\end{pmatrix}\hat{U}^{-1}(t,t_0)\nonumber \\
\end{eqnarray}
with the following components of unitary matrix
\begin{eqnarray}
U_{1,1}(t)=\frac{e^{-i \hbar Q_{11}(t)}}{2}\Bigg[
-iQ_{22}(t)\times \nonumber \\ \times \left(\frac{\sin \left(\hbar
   \sqrt{|Q_{22}(t)|^2+(TR1(t)-TR2(t))^2}\right)}{\sqrt{|Q_{22}(t)|^2+(TR1(t)-TR2(t))^2}}+\frac{\sin \left(\hbar
   \sqrt{|Q_{22}(t)|^2+(TR1(t)+TR2(t))^2}\right)}{\sqrt{|Q_{22}(t)|^2+(TR1(t)+TR2(t))^2}}\right)+ \nonumber \\ +\cos \left(\hbar
   \sqrt{|Q_{22}(t)|^2+(TR1(t)-TR2(t))^2}\right)+\cos \left(\hbar \sqrt{|Q_{22}(t)|^2+(TR1(t)+TR2(t))^2}\right) \Bigg].
\end{eqnarray}
\begin{eqnarray}
U_{1,2}(t)=\frac{i e^{-i \hbar Q_{11}(t)} \Bigg(  (TR1(t)-TR2(t))  \sin \left(\hbar
   \sqrt{  |Q_{22}(t)|^2  +(TR1(t)-TR2(t))^2}\right) }{2 \sqrt{|Q_{22}(t)|^2+(TR1(t)-TR2(t))^2}  }
\nonumber \\
-\frac{(TR1(t)+TR2(t))  \sin \left(\hbar
   \sqrt{|Q_{22}(t)|^2+(TR1(t)+TR2(t))^2}\right) \Bigg)  }{2  \sqrt{|Q_{22}(t)|^2+(TR1(t)+TR2(t))^2}},
\end{eqnarray}
\begin{eqnarray}
U_{1,3}(t)=-i e^{-i \hbar Q_{11}(t)}\frac{ \Bigg[ (TR1(t)-TR2(t))  \sin \left(\hbar
   \sqrt{|Q_{22}(t)|^2+(TR1(t)-TR2(t))^2}\right)}{2 \sqrt{|Q_{22}(t)|^2+(TR1(t)-TR2(t))^2} } \nonumber \\
+  \frac{(TR1(t)+TR2(t))  \sin \left(\hbar
   \sqrt{|Q_{22}(t)|^2+|TR1(t)+TR2(t)|^2}\right)\Bigg]  }{2  \sqrt{|Q_{22}(t)|^2+(TR1(t)+TR2(t))^2}}. 
\end{eqnarray}

\begin{eqnarray}
U_{1,4}(t)=
\frac{1}{2} e^{-i \hbar Q_{11}(t)} \Bigg[i Q_{22}(t) \Bigg[\frac{\sin \left(\hbar
   \sqrt{ |Q_{22}(t)|^2+(TR1(t)-TR2(t))^2}\right)}{\sqrt{|Q_{22}(t)|^2+(TR1(t)-TR2(t))^2}}  \nonumber \\
   -\frac{\sin \left(\hbar
   \sqrt{|Q_{22}(t)|^2+(TR1(t)+TR2(t))^2}\right)}{\sqrt{|Q_{22}(t)|^2+(TR1(t)+TR2(t))^2}}\Bigg] \nonumber \\ -\cos \left(\hbar
   \sqrt{|Q_{22}(t)|^2+(TR1(t)-TR2(t))^2}\right)+\cos \left(\hbar \sqrt{|Q_{22}(t)|^2+(TR1(t)+TR2(t))^2}\right)\Bigg]
\end{eqnarray}

\begin{eqnarray}
U_{2,1}(t)=-\frac{i}{2}e^{-i \hbar Q_{11}(t)}
\frac{\Bigg[ (TR1(t)-TR2(t)) \sin \left(\hbar
   \sqrt{|Q_{22}(t)|^2+(TR1(t)-TR2(t))^2}\right)}{\sqrt{|Q_{22}(t)|^2+(TR1(t)-TR2(t))^2}} \nonumber \\
-\frac{ (TR1(t)+TR2(t)) \sin \left(\hbar
   \sqrt{|Q_{22}(t)|^2+(TR1(t)+TR2(t))^2}\right)\Bigg] }{\sqrt{|Q_{22}(t)|^2+(TR1(t)+TR2(t))^2}}
\end{eqnarray}

\begin{eqnarray}
U_{2,2}(t)=
\frac{1}{2} e^{-i \hbar Q_{11}(t)} \Bigg[ i Q_{22}(t) \Bigg[\frac{\sin \left(\hbar
   \sqrt{|Q_{22}(t)|^2+(TR1(t) -TR2(t))^2}\right)}{\sqrt{|Q_{22}(t)|^2+(TR1(t)-TR2(t))^2}}+\nonumber \\ \frac{\sin \left(\hbar
   \sqrt{|Q_{22}(t)|^2+(TR1(t)+TR2(t))^2}\right)}{\sqrt{|Q_{22}(t)|^2+(TR1(t)+TR2(t))^2}}\Bigg]+  \nonumber \\ +\cos \left(\hbar
   \sqrt{|Q_{22}(t)|^2+(TR1(t)-TR2(t))^2}\right)+\cos \left(\hbar \sqrt{|Q_{22}(t)|^2+(TR1(t)+TR2(t))^2}\right)\Bigg]
\end{eqnarray}

\begin{eqnarray}
U_{2,3}(t)=e^{-i \hbar Q_{11}(t)} 
\Bigg[\frac{  
 -Q_{22}(t)  \sin \left(\hbar
   \sqrt{|Q_{22}(t)|^2+(TR1(t)-TR2(t))^2}\right) 
}{2 \sqrt{|Q_{22}(t)|^2+(TR1(t)-TR2(t))^2} } + \nonumber \\ +
\frac{Q_{22}(t)  \sin \left(\hbar
   \sqrt{|Q_{22}(t)|^2+(TR1(t)+TR2(t))^2}\right)}{2  \sqrt{|Q_{22}(t)|^2+(TR1(t)+TR2(t))^2}}+\nonumber \\ +\frac{ i   \cos \left(\hbar \sqrt{|Q_{22}(t)|^2+(TR1(t)-TR2(t))^2}
   \right)-i   \cos \left(\hbar
   \sqrt{|Q_{22}(t)|^2+(TR1(t)+TR2(t))^2}\right)}{2  }\Bigg]
\end{eqnarray}

\begin{eqnarray}
U_{2,4}(t)=-\frac{i e^{-i \hbar Q_{11}(t) } \Bigg[(TR1(t)-TR2(t)) \sin \left(\hbar
   \sqrt{|Q_{22}(t)|^2+(TR1(t)-TR2(t))^2}\right)}{2 \sqrt{|Q_{22}(t)|^2+(TR1(t)-TR2(t))^2} }+\nonumber \\+\frac{(TR1(t)+TR2(t))  \sin \left(\hbar
   \sqrt{|Q_{22}(t)|^2+(TR1(t)-TR2(t))^2}\right)\Bigg]}{2  \sqrt{|Q_{22}(t)|^2+(TR1(t)+TR2(t))^2}}
\end{eqnarray}

\begin{eqnarray}
U_{3,1}(t)=-i e^{-i \hbar Q_{11}(t) }\frac{ \Bigg[(TR1(t)-TR2(t))  \sin \left(\hbar
   \sqrt{|Q_{22}(t)|^2+(TR1(t)-TR2(t))^2}\right)}{2 \sqrt{|Q_{22}(t)|^2+(TR1(t)-TR2(t))^2} }+\nonumber\\+\frac{(TR1(t)+TR2(t))  \sin \left(\hbar
   \sqrt{|Q_{22}(t)|^2+(TR1(t)+TR2(t))^2}\right)\Bigg]}{2  \sqrt{|Q_{22}(t)|^2+(TR1(t)+TR2(t))^2}}
\end{eqnarray}

\begin{eqnarray}
U_{3,2}(t)= e^{-i \hbar Q_{11}(t) } 
\Bigg[ \frac{ -Q_{22}(t)  \sin \left(\hbar
   \sqrt{|Q_{22}(t)|^2+(TR1(t)-TR2(t))^2}\right) }{2 \sqrt{|Q_{22}(t)|^2+(TR1(t)-TR2(t))^2}  }+ \nonumber \\ \frac{Q_{22}(t)  \sin \left(\hbar
   \sqrt{|Q_{22}(t)|^2+(TR1(t)+TR2(t))^2}\right)}{2  \sqrt{|Q_{22}(t)|^2+(TR1(t)+TR2(t))^2}}+\nonumber \\ +\frac{i   \cos \left(\hbar
   \sqrt{|Q_{22}(t)|^2+(TR1(t)-TR2(t))^2}\right)-i   \cos \left(\hbar
   \sqrt{|Q_{22}(t)|^2+(TR1(t)+TR2(t))^2}\right)\Bigg]}{2  }
\end{eqnarray}

\begin{eqnarray}
U_{3,3}(t)=\frac{1}{2} e^{-i \hbar Q_{11}(t)} \Bigg[ i Q_{22}(t) \Bigg[\frac{\sin \left(\hbar
   \sqrt{|Q_{22}(t)|^2+(TR1(t)-TR2(t))^2}\right)}{\sqrt{|Q_{22}(t)|^2+(TR1(t)-TR2(t))^2}}\nonumber \\ +\frac{\sin \left(\hbar
   \sqrt{|Q_{22}(t)|^2+(TR1(t)+TR2(t))^2}\right)}{\sqrt{|Q_{22}(t)|^2+(TR1(t)+TR2(t))^2}}\Bigg]+\nonumber \\ +\cos \left(\hbar
   \sqrt{|Q_{22}(t)|^2+(TR1(t)-TR2(t))^2}\right)+\cos \left(\hbar \sqrt{|Q_{22}(t)|^2+(TR1(t)+TR2(t))^2}\right)\Bigg]
\end{eqnarray}

\begin{eqnarray}
U_{3,4}(t)=(\sin (\hbar Q_{11}(t))+i \cos (\hbar Q_{11}(t) ))\frac{ \Bigg[(TR1(t)-TR2(t))  \sin \left(\hbar
   \sqrt{|Q_{22}(t)|^2+(TR1(t)-TR2(t))^2}\right)}{2 \sqrt{|Q_{22}(t)|^2 +(TR1(t)-TR2(t))^2} }+\nonumber \\   -\frac{(TR1(t)+TR2(t))  \sin \left(\hbar
   \sqrt{|Q_{22}(t)|^2+(TR1(t)+TR2(t))^2}\right)\Bigg]}{2  \sqrt{|Q_{22}(t)|^2+(TR1(t)+TR2(t))^2}}
\end{eqnarray}

\begin{eqnarray}
U_{4,1}(t)=\frac{1}{2} e^{-i \hbar Q_{11}(t)} \Bigg[i Q_{22}(t) \Bigg[\frac{\sin \left(\hbar
   \sqrt{|Q_{22}(t)|^2+(TR1(t)-TR2(t))^2}\right)}{\sqrt{|Q_{22}(t)|^2+(TR1(t)-TR2(t))^2}}\nonumber \\ -\frac{\sin \left(\hbar
   \sqrt{|Q_{22}(t)|^2+(TR1(t)+TR2(t))^2}\right)}{\sqrt{|Q_{22}(t)|^2+(TR1(t)+TR2(t))^2}}\Bigg]+\nonumber \\  -\cos \left(\hbar
   \sqrt{|Q_{22}(t)|^2+(TR1(t)-TR2(t))^2}\right)+\cos \left(\hbar \sqrt{|Q_{22}(t)|^2+(TR1(t)+TR2(t))^2}\right)\Bigg]
\end{eqnarray}

\begin{eqnarray}
U_{4,2}(t)=-\frac{i e^{-i \hbar Q_{11}(t) } \Bigg[(TR1(t)-TR2(t))  \sin \left(\hbar
   \sqrt{|Q_{22}(t)|^2+(TR1(t)-TR2(t))^2}\right)}{2 \sqrt{|Q_{22}(t)|^2+(TR1(t)-TR2(t))^2} }+\nonumber \\+\frac{(TR1(t)+TR2(t))  \sin \left(\hbar
   \sqrt{|Q_{22}(t)|^2+(TR1(t)+TR2(t))^2}\right)\Bigg]}{2  \sqrt{|Q_{22}(t)|^2+(TR1(t)+TR2(t))^2}}
\end{eqnarray}

\begin{eqnarray}
U_{4,3}(t)=i e^{-i \hbar Q_{11}(t)}\Bigg[ \frac{  (TR1(t)-TR2(t))  \sin \left(\hbar
   \sqrt{|Q_{22}(t)|^2+(TR1(t)-TR2(t))^2}\right)}{2 \sqrt{|Q_{22}(t)|^2+(TR1(t)-TR2(t))^2} }+\nonumber \\ -\frac{(TR1(t)+TR2(t))  \sin \left(\hbar
   \sqrt{|Q_{22}(t)|^2+(TR1(t)+TR2(t))^2}\right)}{2  \sqrt{|Q_{22}(t)|^2+(TR1(t)+TR2(t))^2}}\Bigg]
\end{eqnarray}

\begin{eqnarray}
U_{4,4}(t)=\frac{1}{2} e^{-i \hbar Q_{11}(t)} \Bigg[-i Q_{22}(t) \Bigg[\frac{\sin \left(\hbar
   \sqrt{|Q_{22}(t)|^2+(TR1(t)-TR2(t))^2}\right)}{\sqrt{|Q_{22}(t)|^2+(TR1(t)-TR2(t))^2}}+ \nonumber \\
   \frac{\sin \left(\hbar
   \sqrt{|Q_{22}(t)|^2+(TR1(t)+TR2(t))^2}\right)}{\sqrt{|Q_{22}(t)|^2+(TR1(t)+TR2(t))^2}}\Bigg]+ \nonumber \\ + \cos \left(\hbar
   \sqrt{|Q_{22}(t)|^2+(TR1(t)-TR2(t))^2}\right)+\cos \left(\hbar \sqrt{|Q_{22}(t)|^2+(TR1(t)+TR2(t))^2}\right)\Bigg] \nonumber \\
\end{eqnarray}
We set the quantum state to be $\ket{\psi,t_0}=\ket{E_1}$ at time $t_0$ so it is maximally entangled and its density matrix is $\rho(t_0)=\ket{\psi,t_0}\bra{\psi,t_0}=\ket{E_1}\bra{E_1}=\frac{1}{2}
\begin{pmatrix}
+1 & 0 & 0 & -1 \\
0 & 0 & 0 & 0  \\
0 & 0 & 0 & 0  \\
-1 & 0 & 0 & 1 \\
\end{pmatrix}
$.
 Finally we obtain the following density matrix
\begin{eqnarray}
\rho_{1,1}(t)=\frac{(TR1(t)-TR2(t))^2 \cos \left(2 \hbar \sqrt{|Q_{22}(t)|^2+(TR1(t)-TR2(t))^2}\right)+2 |Q_{22}(t)|^2+(TR1(t)-TR2(t))^2}{4
   \left(|Q_{22}(t)|^2+(TR1(t)-TR2(t))^2\right)}\nonumber \\
\end{eqnarray}

\begin{eqnarray}
\rho_{1,2}(t)=\frac{(TR1(t)-TR2(t)) \Bigg[-i \sqrt{|Q_{22}(t)|^2+(TR1(t)-TR2(t))^2} \sin \left(2 \hbar \sqrt{|Q_{22}(t)|^2+(TR1(t)-TR2(t))^2}\right)}{4 \left(|Q_{22}(t)|^2+(TR1(t)-TR2(t))^2\right)}+ \nonumber \\ +\frac{Q_{22}(t) \cos
   \left(2 \hbar \sqrt{|Q_{22}(t)|^2+(TR1(t)-TR2(t))^2}\right)-Q_{22}(t)\Bigg]}{4 \left(|Q_{22}(t)|^2+(TR1(t)-TR2(t))^2\right)}\nonumber \\
\end{eqnarray}

\begin{eqnarray}
\rho_{1,3}(t)=-(TR1(t)-TR2(t))\frac{ \Bigg[-i \sqrt{|Q_{22}(t)|^2+(TR1(t)-TR2(t))^2} \sin \left(2 \hbar \sqrt{|Q_{22}(t)|^2+(TR1(t)-TR2(t))^2}\right)}{4 \left(|Q_{22}(t)|^2+(TR1(t)-TR2(t))^2\right)}+ \nonumber \\ +\frac{Q_{22}(t) \cos
   \left(2 \hbar \sqrt{|Q_{22}(t)|^2+(TR1(t)-TR2(t))^2}\right)-Q_{22}(t)\Bigg]}{4 \left(|Q_{22}(t)|^2+(TR1(t)-TR2(t))^2\right)}\nonumber \\
\end{eqnarray}

\begin{eqnarray}
\rho_{1,4}(t)=-\frac{(TR1(t)-TR2(t))^2 \cos \left(2 \hbar \sqrt{|Q_{22}(t)|^2+(TR1(t)-TR2(t))^2}\right)+2 |Q_{22}(t)|^2+(TR1(t)-TR2(t))^2}{4
   \left(|Q_{22}(t)|^2+(TR1(t)-TR2(t))^2\right)}\nonumber \\
\end{eqnarray}

\begin{eqnarray}
\rho_{2,1}(t)=
\frac{(TR1(t)-TR2(t)) \Bigg(i \sqrt{|Q_{22}(t)|^2+(TR1(t)-TR2(t))^2} \sin \left(2 \hbar \sqrt{|Q_{22}(t)|^2+(TR1(t)-TR2(t))^2}\right)}{4 \left(|Q_{22}(t)|^2+(TR1(t)-TR2(t))^2\right)}+\nonumber \\+\frac{Q_{22}(t) \cos
   \left(2 \hbar \sqrt{|Q_{22}(t)|^2+(TR1(t)-TR2(t))^2}\right)-Q_{22}(t)\Bigg]}{4 \left(|Q_{22}(t)|^2+(TR1(t)-TR2(t))^2\right)}\nonumber \\
\end{eqnarray}

\begin{eqnarray}
\rho_{2,2}(t)=\frac{(TR1(t)-TR2(t))^2 \sin ^2\left(\hbar \sqrt{|Q_{22}(t)|^2+(TR1(t)-TR2(t))^2}\right)}{2 \left(|Q_{22}(t)|^2+(TR1(t)-TR2(t))^2\right)}
\end{eqnarray}

\begin{eqnarray}
\rho_{2,3}(t)=-\frac{(TR1(t)-TR2(t))^2 \sin ^2\left(\hbar \sqrt{|Q_{22}(t)|^2+(TR1(t)-TR2(t))^2}\right)}{2 \left(|Q_{22}(t)|^2+(TR1(t)-TR2(t))^2\right)}
\end{eqnarray}

\begin{eqnarray}
\rho_{2,4}(t)=-\frac{(TR1(t)-TR2(t)) \Bigg[ i \sqrt{|Q_{22}(t)|^2+(TR1(t)-TR2(t))^2} \sin \left(2 \hbar \sqrt{|Q_{22}(t)|^2+(TR1(t)-TR2(t))^2}\right)}{4 \left(|Q_{22}(t)|^2+(TR1(t)-TR2(t))^2\right)}+\nonumber \\+\frac{Q_{22}(t)\cos
   \left(2 \hbar \sqrt{|Q_{22}(t)|^2+(TR1(t)-TR2(t))^2}\right)-Q_{22}(t)\Bigg]}{4 \left(|Q_{22}(t)|^2+(TR1(t)-TR2(t))^2\right)}\nonumber \\
\end{eqnarray}

\begin{eqnarray}
\rho_{3,1}(t)=-(TR1(t)-TR2(t))\frac{ \Bigg[i \sqrt{|Q_{22}(t)|^2+(TR1(t)-TR2(t))^2} \sin \left(2 \hbar \sqrt{|Q_{22}(t)|^2+(TR1(t)-TR2(t))^2}\right)}{4 \left(|Q_{22}(t)|^2+(TR1(t)-TR2(t))^2\right)}+ \nonumber \\ +\frac{Q_{22}(t) \cos
   \left(2 \hbar \sqrt{|Q_{22}(t)|^2+(TR1(t)-TR2(t))^2}\right)-Q_{22}(t)\Bigg]}{4 \left(|Q_{22}(t)|^2+(TR1(t)-TR2(t))^2\right)}\nonumber \\
\end{eqnarray}

\begin{eqnarray}
\rho_{3,2}(t)=-\frac{(TR1(t)-TR2(t))^2 \sin ^2\left(\hbar \sqrt{|Q_{22}(t)|^2+(TR1(t)-TR2(t))^2}\right)}{2 \left(|Q_{22}(t)|^2+(TR1(t)-TR2(t))^2\right)}
\end{eqnarray}

\begin{eqnarray}
\rho_{3,3}(t)=\frac{(TR1(t)-TR2(t))^2 \sin ^2\left(\hbar \sqrt{|Q_{22}(t)|^2+(TR1(t)-TR2(t))^2}\right)}{2 \left(|Q_{22}(t)|^2+(TR1(t)-TR2(t))^2\right)}
\end{eqnarray}

\begin{eqnarray}
\rho_{3,4}(t)=\frac{(TR1(t)-TR2(t)) \Bigg(i \sqrt{|Q_{22}(t)|^2+(TR1(t)-TR2(t))^2} \sin \left(2 \hbar \sqrt{|Q_{22}(t)|^2+(TR1(t)-TR2(t))^2}\right)}{4 \left(|Q_{22}(t)|^2+(TR1(t)-TR2(t))^2\right)}+ \nonumber \\ \frac{Q_{22}(t) \cos
   \left(2 \hbar \sqrt{|Q_{22}(t)|^2+(TR1(t)-TR2(t))^2}\right)-Q_{22}(t)\Bigg]}{4 \left(|Q_{22}(t)|^2+(TR1(t)-TR2(t))^2\right)}\nonumber \\
\end{eqnarray}

\begin{eqnarray}
\rho_{4,1}(t)=-\frac{(TR1(t)-TR2(t))^2 \cos \left(2 \hbar \sqrt{|Q_{22}(t)|^2+(TR1(t)-TR2(t))^2}\right)+2|Q_{22}(t)|^2+(TR1(t)-TR2(t))^2}{4
   \left(|Q_{22}(t)|^2+(TR1(t)-TR2(t))^2\right)}\nonumber \\
\end{eqnarray}

\begin{eqnarray}
\rho_{4,2}(t)=-(TR1(t)-TR2(t))\frac{ \Bigg[ -i \sqrt{|Q_{22}(t)|^2+(TR1(t)-TR2(t))^2} \sin \left(2 \hbar \sqrt{|Q_{22}(t)|^2+(TR1(t)-TR2(t))^2}\right)}{4 \left(|Q_{22}(t)|^2+(TR1(t)-TR2(t))^2\right)}+ \nonumber \\ + \frac{Q_{22}(t) \cos
   \left(2 \hbar \sqrt{|Q_{22}(t)|^2+(TR1(t)-TR2(t))^2}\right)-Q_{22}(t)\Bigg]}{4 \left(|Q_{22}(t)|^2+(TR1(t)-TR2(t))^2\right)}\nonumber \\
\end{eqnarray}

\begin{eqnarray}
\rho_{4,3}(t)=(TR1(t)-TR2(t))\frac{ \Bigg[-i \sqrt{|Q_{22}(t)|^2+(TR1(t)-TR2(t))^2} \sin \left(2 \hbar \sqrt{|Q_{22}(t)|^2+(TR1(t)-TR2(t))^2}\right)}{4 \left(|Q_{22}(t)|^2+(TR1(t)-TR2(t))^2\right)}+\nonumber \\ \frac{Q_{22}(t) \cos
   \left(2 \hbar \sqrt{|Q_{22}(t)|^2+(TR1(t)-TR2(t))^2}\right)-Q_{22}(t)\Bigg]}{4 \left(|Q_{22}(t)|^2+(TR1(t)-TR2(t))^2\right)}\nonumber \\
\end{eqnarray}

\begin{eqnarray}
\rho_{4,4}(t)=\frac{(TR1(t)-TR2(t))^2 \cos \left(2 \hbar \sqrt{|Q_{22}(t)|^2+(TR1(t)-TR2(t))^2}\right)+2|Q_{22}(t)|^2+(TR1(t)-TR2(t))^2}{4
   \left(|Q_{22}(t)|^2+(TR1(t)-TR2(t))^2\right)}\nonumber \\
\end{eqnarray}

It turns out that $\rho^n(t)=\rho(t)$ so one deals with a pure quantum state.
Now we are obtaining reduced matrices describing the state of particle B  from 2 particle density matrix.

\begin{eqnarray}
\rho_{B}(t)=
\begin{pmatrix}
\rho_{11}(t)+\rho_{22}(t) & \rho_{13}(t)+\rho_{24}(t) \\
\rho_{31}(t)+\rho_{42}(t) & \rho_{33}(t)+\rho_{44}(t) \\
\end{pmatrix}= \nonumber \\
\begin{pmatrix}
\frac{1}{2} & \frac{Q_{22}(t) (TR1(t)-TR2(t)) \sin ^2\left(\hbar \sqrt{|Q_{22}(t)|^2+(TR1(t)-TR2(t))^2}\right)}{|Q_{22}(t)|^2+(TR1(t)-TR2(t))^2} \\
\frac{ Q_{22}(t) (TR1(t)-TR2(t)) \sin ^2\left(\hbar \sqrt{|Q_{22}(t)|^2+(TR1(t)-TR2(t))^2}\right)}{|Q_{22}(t)|^2+(TR1(t)-TR2(t))^2} & \frac{1}{2} \\
\end{pmatrix}. \nonumber \\
\end{eqnarray}
Consequently we can compute entanglement entropy.
At first we evaluate
\begin{eqnarray}
Log(\rho_{B}(t))=
\begin{pmatrix}
a & b \\
c & d \\
\end{pmatrix},
\end{eqnarray}
\small
\begin{eqnarray*}
a=\frac{1}{2} \Bigg[\log \Bigg[ Q_{22}(t) (TR1(t)-TR2(t)) \cos \left(2 \hbar \sqrt{|Q_{22}(t)|^2+(TR1(t)-TR2(t))^2}\right)+\nonumber \\ +|Q_{22}(t)|^2+Q_{22}(t)
   (TR2(t)-TR1(t))+(TR1(t)-TR2(t))^2 \Bigg] \nonumber \\
   -2\log \Bigg[
   |Q_{22}(t)|^2+(TR1(t)-TR2(t))^2
   \Bigg] \nonumber \\
   +\log \Bigg[\Bigg[Q_{22}(t) (TR2(t)-TR1(t)) \cos \left(2 \hbar
   \sqrt{|Q_{22}(t)|^2+(TR1(t)-TR2(t))^2}\right)+ \nonumber \\
   |Q_{22}(t)|^2+Q_{22}(t)
   (TR1(t)-TR2(t))+(TR1(t)-TR2(t))^2\Bigg] 
   -\log (4)\Bigg]
\end{eqnarray*}
\normalsize
$b=-\tanh ^{-1}\left(\frac{Q_{22}(t)(TR1(t)-TR2(t)) \left(\cos \left(2 \hbar
   \sqrt{|Q_{22}(t)|^2+(TR1(t)-TR2(t))^2}\right)-1\right)}{|Q_{22}(t)|^2+(TR1(t)-TR2(t))^2}\right)=c$

\small
\begin{eqnarray}
d=\frac{1}{2}  \Bigg[\log\Bigg[ Q_{22}(t) (TR1(t)-TR2(t)) \cos \left(2 \hbar
\sqrt{| Q_{22}(t)|^2+(TR1(t)-TR2(t))^2}\right)+\nonumber \\ | Q_{22}(t)|^2+Q_{22}(t)
   (TR2(t)-TR1(t))+(TR1(t)-TR2(t))^2\Bigg] \nonumber \\  -2\log\Bigg[|Q_{22}(t)|^2+(TR1(t)-TR2(t))^2\Bigg]\Bigg] \nonumber \\
+\log \Bigg[] Q_{22}(t)(TR2(t)-TR1(t)) \cos \left(2 \hbar
   \sqrt{|Q_{22}(t)|^2+(TR1(t)-TR2(t))^2}\right)+ \nonumber \\ |Q_{22}(t)|^2+Q_{22}(t)
   (TR1(t)-TR2(t))+(TR1(t)-TR2(t))^2\Bigg] 
-\log (4)\Bigg]
\end{eqnarray}
\normalsize
and we obtain the formula when we start from $TR1(t_0)=TR2(t_0)$ as
\small
\begin{eqnarray}
S_{B}(t)=Tr[\rho_{B}(t) Log[\rho_{B}(t)]]= \nonumber \\
=Tr \Bigg[
\begin{pmatrix}
\frac{1}{2} & \frac{Q_{22}(t) (TR1(t)-TR2(t)) \sin ^2\left(\hbar \sqrt{|Q_{22}(t)|^2+(TR1(t)-TR2(t))^2}\right)}{|Q_{22}(t)|^2+(TR1(t)-TR2(t))^2} \\
\frac{ Q_{22}(t) (TR1(t)-TR2(t)) \sin ^2\left(\hbar \sqrt{|Q_{22}(t)|^2+(TR1(t)-TR2(t))^2}\right)}{|Q_{22}(t)|^2+(TR1(t)-TR2(t))^2} & \frac{1}{2} \\
\end{pmatrix} \times \nonumber \\
Log \Bigg[
\begin{pmatrix}
\frac{1}{2} & \frac{Q_{22}(t) (TR1(t)-TR2(t)) \sin ^2\left(\hbar \sqrt{|Q_{22}(t)|^2+(TR1(t)-TR2(t))^2}\right)}{|Q_{22}(t)|^2+(TR1(t)-TR2(t))^2} \\
\frac{ Q_{22}(t) (TR1(t)-TR2(t)) \sin ^2\left(\hbar \sqrt{|Q_{22}(t)|^2+(TR1(t)-TR2(t))^2}\right)}{|Q_{22}(t)|^2+(TR1(t)-TR2(t))^2} & \frac{1}{2} \\
\end{pmatrix}
\Bigg] \Bigg]=\nonumber \\
=-\log (4)\frac{1}{2}+ 
\frac{1}{2} \Bigg[\log \Bigg[Q_{22}(t) (TR1(t)-TR2(t)) \cos \left(2 \hbar \sqrt{|Q_{22}(t)|^2+(TR1(t)-TR2(t))^2}\right)+\nonumber \\ +|Q_{22}(t)|^2+Q_{22}(t)
   (TR2(t)-TR1(t))+(TR1(t)-TR2(t))^2\Bigg]+
   \nonumber \\ +\log \Bigg[Q_{22}(t)(TR2(t)-TR1(t)) \cos \left(2 \hbar
   \sqrt{|Q_{22}(t)|^2+(TR1(t)-TR2(t))^2}\right) \nonumber \\ +|Q_{22}(t)|^2+Q_{22}(t)
   (TR1(t)-TR2(t))+(TR1(t)-TR2(t))^2\Bigg]\nonumber \\ -2\log\Bigg[
   |Q_{22}(t)|^2+(TR1(t)-TR2(t))^2
   \Bigg] \nonumber \\ +\frac{4 Q_{22}(t) (TR2(t)-TR1(t)) \sin ^2\left(\hbar
   \sqrt{|Q_{22}(t)|^2+(TR1(t)-TR2(t))^2}\right) }{|Q_{22}(t)|^2+(TR1(t)-TR2(t))^2} \times \nonumber \\
\times   \tanh ^{-1}\left(\frac{Q_{22}(t) (TR1(t)-TR2(t)) \left(\cos \left(2 \hbar
   \sqrt{|Q_{22}(t)|^2+(TR1(t)-TR2(t))^2}\right)-1\right)}{|Q_{22}(t)|^2+(TR1(t)-TR2(t))^2}\right)\Bigg]
   \nonumber \\
\end{eqnarray}
\normalsize
\subsection{Analytical treatment of 2 interacting particles in asymmetric case}
We consider the situation as depicted in Fig.2. We obtain the following Hamiltonian
\begin{eqnarray}
H=
\begin{pmatrix}
E_c(1,1')+E_{p}(1)+E_{p}(1') & t_{s1'2p'} & t_{s12} & 0 \\
t_{s1'2p'}^{*} & E_c(1,2')+E_{p}(1)+E_{p}(2') & 0 &  t_{s12} \\
t_{s12}^{*} & 0 & E_c(2,1')+E_{p}(2)+E_{p}(1') & t_{s1'2p'} \\
0 & t_{s12}^{*} & t_{s1'2p'}{*} & E_c(2,2')+E_{p}(2)+E_{p}(2')
\end{pmatrix}= \nonumber \\
\begin{pmatrix}
\frac{q^2}{\sqrt{d^2+(a+b)^2}} & t_{s1'2p'} & t_{s12} & 0 \\
t_{s1'2p'}^{*} & \frac{q^2}{\sqrt{(d+Cos(\alpha)(a+b))^2+(1+Sin(\alpha))^2(a+b)^2}} & 0 &  t_{s12} \\
t_{s12}^{*} & 0 & \frac{q^2}{d} & t_{s1'2p'} \\
0 & t_{s12}^{*} & t_{s1'2p'}{*} & \frac{q^2}{\sqrt{(d+Cos(\alpha)(a+b))^2+(Sin(\alpha))^2(a+b)^2}}
\end{pmatrix}+\nonumber \\
\begin{pmatrix}
E_{p}(1)+E_{p}(1') & 0 & 0 & 0 \\
0 & E_{p}(1)+E_{p}(2') & 0 &  0 \\
0 & 0 & E_{p}(2)+E_{p}(1') & 0 \\
0 & 0 & 0 &E_{p}(2)+E_{p}(2')
\end{pmatrix}=\nonumber \\
\begin{pmatrix}
E_f(1,1') & t_{s1'2p'} & t_{s12} & 0 \\
t_{s1'2p'}^{*} & E_f(1,2') & 0 &  t_{s12} \\
t_{s12}^{*} & 0 & E_f(2,1') & t_{s1'2p'} \\
0 & t_{s12}^{*} & t_{s1'2p'}{*} & E_f(2,2')
\end{pmatrix}. \nonumber \\
\end{eqnarray}
In general situation all eigenvalues and eigenstates of this matrix can be determined in analytical way since roots of polynomials of 4th are given by explicit formulas. However those formulas are very complicated and thus no so practical. We will make radical simplification leading to simple formulas for quantum eigenstates and eigenenergies. First main assumption is setting $t_{s1'2p'}=t_{s12}=1$. We obtain simplified Hamiltonian of the form
\begin{eqnarray}H=
\begin{pmatrix}
E_f(1,1') & 1 & 1 & 0 \\
1 & E_f(1,2') & 0 & 1 \\
1 & 0 & E_f(2,1') & 1 \\
0 & 1 & 1 & E_f(2,2')
\end{pmatrix}. \nonumber \\
\end{eqnarray}
Now we will consider certain cases.
\subsection{Qubit-Qubit interaction with 2 symmetric conditions}
\subsubsection{Case I: $E_f(1,1')=E_f(1,2')=U$, $E_f(2,1')=E_f(2,2')=U_1$ }
We postulate
\begin{eqnarray}H=
\begin{pmatrix}
U & 1 & 1 & 0 \\
1 & U & 0 & 1 \\
1 & 0 & U_1 & 1 \\
0 & 1 & 1 & U_1
\end{pmatrix}. \nonumber \\
\end{eqnarray}
We obtain
\begin{eqnarray}
U=E_{p1}+E_{p1'}+\frac{q^2}{\sqrt{d^2+(a+b)^2}}, \nonumber \\
U=E_{p1}+E_{p2'}+\frac{q^2}{\sqrt{(d+Cos(\alpha)(a+b))^2+(1+Sin(\alpha))^2(a+b)^2}},
\end{eqnarray}
and it implies
\begin{eqnarray}
E_{p2'}-E_{p1'}=+\frac{q^2}{\sqrt{d^2+(a+b)^2}}-\frac{q^2}{\sqrt{(d+Cos(\alpha)(a+b))^2+(1+Sin(\alpha))^2(a+b)^2}}, \nonumber \\
\end{eqnarray}
We also have
\begin{eqnarray}
U_1=E_{p2}+E_{p1'}+\frac{q^2}{d}, \nonumber \\
U_1=E_{p2}+E_{p2'}+\frac{q^2}{\sqrt{(d+Cos(\alpha)(a+b))^2+(Sin(\alpha))^2(a+b)^2}}.
\end{eqnarray}
and we obtain
\begin{eqnarray}
E_{p2'}-E_{p1'}=+\frac{q^2}{d}-\frac{q^2}{\sqrt{(d+Cos(\alpha)(a+b))^2+(Sin(\alpha))^2(a+b)^2}}.
\end{eqnarray}
that implies condition
\begin{eqnarray}
\label{extra}
+\frac{q^2}{d}-\frac{q^2}{\sqrt{(d+Cos(\alpha)(a+b))^2+(Sin(\alpha))^2(a+b)^2}}= \nonumber \\
\frac{q^2}{\sqrt{d^2+(a+b)^2}}-\frac{q^2}{\sqrt{(d+Cos(\alpha)(a+b))^2+(1+Sin(\alpha))^2(a+b)^2}}.
\end{eqnarray}
that is fulfilled for two angles $\alpha$ as in accordance to numerical solutions. Usually $a+b << d$ since in most cases qubit size is small in comparison to the distance between qubits. Let us solve the problem with certain approximation by using Taylor expansion

\begin{eqnarray}
+\frac{q^2}{d}-\frac{q^2}{d}+\frac{q^2}{d^2}(\sqrt{d^2+2 d Cos(\alpha)(a+b)+(Cos(\alpha))^2(a+b)^2+(Sin(\alpha))^2(a+b)^2}-d) =\nonumber \\ 
\frac{q^2}{\sqrt{d^2+(a+b)^2}}-\frac{q^2}{d}+ \nonumber \\
+\frac{q^2}{d^2}(\sqrt{(Cos(\alpha))^2(a+b)^2+2(a+b)d Cos(\alpha)+(Sin(\alpha))^2(a+b)^2+(a+b)^2+d^2+2Sin(\alpha)(a+b)^2}-d). 
\end{eqnarray}
It implies
\begin{eqnarray}
+\frac{1}{d}+\frac{1}{d^2}(\sqrt{d^2+2 d Cos(\alpha)(a+b)+(a+b)^2}) =\nonumber \\ 
\frac{1}{\sqrt{d^2+(a+b)^2}}
+\frac{1}{d^2}(\sqrt{2(a+b)d Cos(\alpha)+2(a+b)^2+d^2+2Sin(\alpha)(a+b)^2}). 
\end{eqnarray}
and hence we have
\begin{eqnarray}
+1+(\sqrt{1+2 Cos(\alpha)\frac{(a+b)}{d}+(\frac{a+b}{d})^2}) =\nonumber \\ 
\frac{d}{\sqrt{d^2+(a+b)^2}}
+(\sqrt{2\frac{(a+b)}{d}Cos(\alpha)+2(\frac{a+b}{d})^2+1+2Sin(\alpha)(\frac{a+b}{d})^2}). 
\end{eqnarray}
Using Taylor expansion for square root function we obtain relation
\begin{eqnarray}
+1+1+\frac{1}{2}(2 Cos(\alpha)\frac{(a+b)}{d}+(\frac{a+b}{d})^2) =\nonumber \\ 
\frac{d}{\sqrt{d^2+(a+b)^2}}+1
+\frac{1}{2}(2\frac{(a+b)}{d}Cos(\alpha)+2(\frac{a+b}{d})^2+2Sin(\alpha)(\frac{a+b}{d})^2). 
\end{eqnarray}
that can be simplified into simple relation for sinusoidal function of the form
\begin{eqnarray}
\frac{d^2}{(a+b)^2}[+1-\frac{1}{2}(\frac{a+b}{d})^2-\frac{d}{\sqrt{d^2+(a+b)^2}}] 
=Sin(\alpha). 
\end{eqnarray}
Two angles fulfills such relation.
\newline \newline

We can also solve equation \ref{extra} in rigorous way (without approximation) by equivalent to the following relation
\begin{eqnarray}
+\frac{1}{d}-\frac{1}{d_1}= 
\frac{1}{d_2}-\frac{1}{d_1+x} \text{ or } \frac{1}{d}-\frac{1}{d_2}=\frac{1}{d_1}-\frac{1}{d_1+x}=\frac{x}{d_1 (d_1+x)}.
\end{eqnarray}
and this implies
\begin{eqnarray}
d_1^2\frac{d_2-d}{d_2 d}+xd_1\frac{d_2-d}{d_2 d}=x \text{ or }+(d_1\frac{d_2-d}{d_2 d}-1)x=-d_1^2\frac{d_2-d}{d_2 d}.
\end{eqnarray}
and finally we have
\begin{eqnarray}
(\frac{d_1d_2-d d_1 - d_2 d}{d_2 d})x=-d_1^2\frac{d_2-d}{d_2 d}, x=-d_1^2\frac{d_2-d}{d_1d_2-d d_1 - d_2 d}=\frac{-d_1^2 d_2+d d_1^2}{d_1d_2-d d_1 - d_2 d},.
\end{eqnarray}
We obtain
\begin{equation}
\frac{x+d_1}{d_1}=\frac{-d_1^2 d_2+d d_1^2+ d_1(d_1d_2-d d_1 - d_2 d)}{ (d_1d_2-d d_1 - d_2 d) d_1}=\frac{-d d_1 d_2 }{(d_1d_2-d d_1 - d_2 d) d_1}=\frac{-d d_2 }{(d_1d_2-d d_1 - d_2 d)}=\frac{d d_2 }{d_1(-d_2+d) + d_2 d}
\end{equation}
This implies
\begin{eqnarray}
(\frac{x+d_1}{d_1})^2=\frac{ (d d_2)^2 }{ [d_1(-d_2+d) + d_2 d ]^2}=\frac{(d+Cos(\alpha)(a+b))^2+(1+Sin(\alpha))^2(a+b)^2}{(d+Cos(\alpha)(a+b))^2+(Sin(\alpha))^2(a+b)^2}= \nonumber \\
=\frac{ (d \sqrt{ d^2 + (a+b)^2})^2 }{ [\sqrt{(d+Cos(\alpha)(a+b))^2+(Sin(\alpha))^2(a+b)^2}(-\sqrt{ d^2 + (a+b)^2}+d) + (d \sqrt{ d^2 + (a+b)^2}) ]^2}.
\end{eqnarray}

We recognize that we have 3 free parameters $E_{p1}\in R, E_{p2}\in R, E_{p1'}\in R$ that predetermines $E_{p2'}$ given as
\begin{eqnarray}
E_{p2'}=E_{p1'}+\frac{q^2}{d}-\frac{q^2}{\sqrt{(d+Cos(\alpha)(a+b))^2+(Sin(\alpha))^2(a+b)^2}}.
\end{eqnarray}
Using 3 free parameters $E_{p1}\in R, E_{p2}\in R, E_{p1'}\in R$ we obtain
\begin{eqnarray}
U_1=2E_{p1'}+\frac{2q^2}{d}-\frac{q^2}{\sqrt{(d+Cos(\alpha)(a+b))^2+(Sin(\alpha))^2(a+b)^2}}, 
\end{eqnarray}
\begin{eqnarray}
U=E_{p1}+E_{p1'}+\frac{q^2}{\sqrt{d^2+(a+b)^2}}.
\end{eqnarray}
We have 4 energy eigenvalues
\begin{equation}
E_1=\frac{1}{2} \left(-\sqrt{(U-U_1)^2+4}+U+U_1)-2 \right),
\end{equation}
\begin{equation}
E_2=\frac{1}{2} \left(-\sqrt{(U-U_1)^2+4}+U+U_1)+2 \right),
\end{equation}
\begin{equation}
E_3=\frac{1}{2} \left(+\sqrt{(U-U_1)^2+4}+U+U_1)-2 \right),
\end{equation}
\begin{equation}
E_4=\frac{1}{2} \left(+\sqrt{(U-U_1)^2+4}+U+U_1)+2 \right).
\end{equation}
and we obtain the following eigenstates
\begin{equation}
\ket{E_1}=
\begin{pmatrix}
\frac{1}{2} \left(\sqrt{U^2-2 U U_1+U_1^2+4}-U+U_1\right) \\
\frac{1}{2} \left(-\sqrt{U^2-2 U U_1+U_1^2+4}+U-U_1\right) \\
-1 \\
+1
\end{pmatrix},
\end{equation}
\begin{equation}
\ket{E_2}=
\begin{pmatrix}
\frac{1}{2} \left(-\sqrt{U^2-2 U U_1+U_1^2+4}+(U-U_1)\right) \\
\frac{1}{2} \left(-\sqrt{U^2-2 U U_1+U_1^2+4}+(U-U_1)\right) \\
1 \\
1 \\
\end{pmatrix},
\end{equation}
\begin{equation}
\ket{E_3}=
\begin{pmatrix}
-\frac{1}{2} \left(\sqrt{U^2-2 U U_1+U_1^2+4}+(U-U_1)\right) \\
+\frac{1}{2} \left(\sqrt{U^2-2 U U_1+U_1^2+4}+(U-U_1)\right) \\
-1 \\
+1 \\
\end{pmatrix},
\end{equation}
\begin{equation}
\ket{E_4}=
\begin{pmatrix}
\frac{1}{2} \left(\sqrt{U^2-2 U U_1+U_1^2+4}+U-U_1\right) \\
\frac{1}{2} \left(\sqrt{U^2-2 U U_1+U_1^2+4}+U-U_1\right) \\
+1 \\
+1 \\
\end{pmatrix}.
\end{equation}
\subsubsection{Case II: $E_f(1,1')=E_f(2,2')=U$, $E_f(1,2')=E_f(2,1')=U_1$ }
We postulate
\begin{eqnarray}H=
\begin{pmatrix}
U & 1 & 1 & 0 \\
1 & U_1 & 0 & 1 \\
1 & 0 & U_1 & 1 \\
0 & 1 & 1 & U
\end{pmatrix}, \nonumber \\
\end{eqnarray}
that implies
\begin{eqnarray}
U=E_{p1}+E_{p1'}+\frac{q^2}{\sqrt{d^2+(a+b)^2}}, \nonumber \\
U=E_{p2}+E_{p2'}+\frac{q^2}{\sqrt{(d+Cos(\alpha)(a+b))^2+(Sin(\alpha))^2(a+b)^2}}.
\end{eqnarray}
and consequently we obtain
\begin{eqnarray}
(E_{p1}-E_{p2})+(E_{p1'}-E_{p2'})=\frac{q^2}{\sqrt{(d+Cos(\alpha)(a+b))^2+(Sin(\alpha))^2(a+b)^2}} 
-\frac{q^2}{\sqrt{d^2+(a+b)^2}}.
\end{eqnarray}

We also have
\begin{eqnarray}
U_1=E_{p1}+E_{p2'}+\frac{q^2}{\sqrt{(d+Cos(\alpha)(a+b))^2+(1+Sin(\alpha))^2(a+b)^2}}, \nonumber \\
U_1=E_{p2}+E_{p1'}+\frac{q^2}{d},
\end{eqnarray}
and it implies
\begin{eqnarray}
-(E_{p1}-E_{p2})+(E_{p1'}-E_{p2'})=\frac{q^2}{\sqrt{(d+Cos(\alpha)(a+b))^2+(1+Sin(\alpha))^2(a+b)^2}}-\frac{q^2}{d}.
\end{eqnarray}
This results in
\begin{eqnarray}
(E_{p1'}-E_{p2'})=\frac{1}{2}[\frac{q^2}{\sqrt{(d+Cos(\alpha)(a+b))^2+(1+Sin(\alpha))^2(a+b)^2}}-\frac{q^2}{d} \nonumber \\
+\frac{q^2}{\sqrt{(d+Cos(\alpha)(a+b))^2+(Sin(\alpha))^2(a+b)^2}}-\frac{q^2}{\sqrt{d^2+(a+b)^2}}].
\end{eqnarray}
and
\begin{eqnarray}
(E_{p1}-E_{p2})=\frac{1}{2}[\frac{q^2}{\sqrt{(d+Cos(\alpha)(a+b))^2+(1+Sin(\alpha))^2(a+b)^2}}-\frac{q^2}{d} \nonumber \\
-\frac{q^2}{\sqrt{(d+Cos(\alpha)(a+b))^2+(Sin(\alpha))^2(a+b)^2}}+\frac{q^2}{\sqrt{d^2+(a+b)^2}}].
\end{eqnarray}
Therefore we have 3 open parameters $E_{p1} \in R$, $E_{p1'} \in R$ and for any $\alpha \in (0, 2\pi)$ we have predetermined conditions for $E_{p2}$ and $E_{p2'}$ given as
\begin{eqnarray}
E_{p2}=E_{p1}-\frac{1}{2}[\frac{q^2}{\sqrt{(d+Cos(\alpha)(a+b))^2+(1+Sin(\alpha))^2(a+b)^2}}-\frac{q^2}{d} \nonumber \\
-\frac{q^2}{\sqrt{(d+Cos(\alpha)(a+b))^2+(Sin(\alpha))^2(a+b)^2}}+\frac{q^2}{\sqrt{d^2+(a+b)^2}}].
\end{eqnarray}
and
\begin{eqnarray}
E_{p2'}=E_{p1'}-\frac{1}{2}[\frac{q^2}{\sqrt{(d+Cos(\alpha)(a+b))^2+(1+Sin(\alpha))^2(a+b)^2}}-\frac{q^2}{d} \nonumber \\
+\frac{q^2}{\sqrt{(d+Cos(\alpha)(a+b))^2+(Sin(\alpha))^2(a+b)^2}}-\frac{q^2}{\sqrt{d^2+(a+b)^2}}].
\end{eqnarray}
and implies $U \in R$
\begin{eqnarray}
U=E_{p1}+E_{p1'}+\frac{q^2}{\sqrt{d^2+(a+b)^2}},
\end{eqnarray}
as well as
\begin{eqnarray}
U_1=E_{p2}+E_{p1'}+\frac{q^2}{d}= \nonumber
2E_{p1'}+\frac{q^2}{d} 
-\frac{1}{2}[\frac{q^2}{\sqrt{(d+Cos(\alpha)(a+b))^2+(1+Sin(\alpha))^2(a+b)^2}}-\frac{q^2}{d} \nonumber \\
+\frac{q^2}{\sqrt{(d+Cos(\alpha)(a+b))^2+(Sin(\alpha))^2(a+b)^2}}-\frac{q^2}{\sqrt{d^2+(a+b)^2}}].
\end{eqnarray}

We have the following eigenvalues
\begin{eqnarray}
E_1=U, E_2=U_1, E_3=\frac{(U+U_1-\sqrt{(4)^2+(U-U_1)^2})}{2}, E_4=\frac{(U+U_1+\sqrt{(4)^2+(U-U_1)^2})}{2}.
\end{eqnarray}
and eigenstates
\begin{eqnarray} \ket{E_1}=
\begin{pmatrix}
-1 \\ 0 \\ 0 \\ 1 \\
\end{pmatrix}, 
\end{eqnarray}
\begin{eqnarray} \ket{E_2}=
\begin{pmatrix}
0 \\ -1 \\ +1 \\ 0 \\
\end{pmatrix}, 
\end{eqnarray}
\begin{eqnarray} \ket{E_3}=
\begin{pmatrix}
1 \\ \frac{4}{-U+U_1+\sqrt{(4)^2+(U-U_1)^2}} \\ \frac{4}{-U+U_1+\sqrt{(4)^2+(U-U_1)^2}} \\ 1 \\
\end{pmatrix}, 
\end{eqnarray}
\begin{eqnarray} \ket{E_4}=
\begin{pmatrix}
1 \\ \frac{4}{U-U_1+\sqrt{(4)^2+(U-U_1)^2}} \\ \frac{4}{U-U_1+\sqrt{(4)^2+(U-U_1)^2}} \\ 1 \\
\end{pmatrix}, 
\end{eqnarray}
\subsubsection{Case III: $E_f(1,1')=E_f(2,1')=U$, $E_f(1,2')=E_f(2,2')=U_1$ }
We postulate
\begin{eqnarray}H=
\begin{pmatrix}
U & 1 & 1 & 0 \\
1 & U_1 & 0 & 1 \\
1 & 0 & U & 1 \\
0 & 1 & 1 & U_1
\end{pmatrix}. \nonumber \\
\end{eqnarray}
that implies $U=\frac{q^2}{\sqrt{d^2+(a+b)^2}}+E_{p}(1)+E_{p}(1')$, $U=\frac{q^2}{d}+E_{p2}+E_{p}(1')$
and we obtain
\begin{eqnarray}
E_{p2}=E_{p1}+\frac{q^2}{\sqrt{d^2+(a+b)^2}}-\frac{q^2}{d},
\end{eqnarray}
.
We also have
\begin{eqnarray}
U_1=\frac{q^2}{\sqrt{(d+Cos(\alpha)(a+b))^2+(1+Sin(\alpha))^2(a+b)^2}}+E_{p1}+E_{p2'}, \nonumber \\
U_1=\frac{q^2}{\sqrt{(d+Cos(\alpha)(a+b))^2+(Sin(\alpha))^2(a+b)^2}}+E_{p2}+E_{p2'},
\end{eqnarray}
what implies
\begin{eqnarray}
E_{p2}=E_{p1}+\frac{q^2}{\sqrt{(d+Cos(\alpha)(a+b))^2+(1+Sin(\alpha))^2(a+b)^2}}-\frac{q^2}{\sqrt{(d+Cos(\alpha)(a+b))^2+(Sin(\alpha))^2(a+b)^2}}.
\end{eqnarray}
and consequently we obtain the condition
\begin{eqnarray}
\frac{q^2}{\sqrt{(d+Cos(\alpha)(a+b))^2+(1+Sin(\alpha))^2(a+b)^2}}-\frac{q^2}{\sqrt{(d+Cos(\alpha)(a+b))^2+(Sin(\alpha))^2(a+b)^2}}= \nonumber \\
\frac{q^2}{\sqrt{d^2+(a+b)^2}}-\frac{q^2}{d}
\end{eqnarray}
that is fulfilled for one angle $\alpha$.
It gives us 3 controlling parameters $E_{p1}$, $E_{p1'}$, $E_{p2'}$ under given $(a+b),d$ that determine
\begin{eqnarray}
U_1=\frac{q^2}{\sqrt{(d+Cos(\alpha)(a+b))^2+(1+Sin(\alpha))^2(a+b)^2}}+E_{p1}+E_{p2'}, \nonumber
U=\frac{q^2}{d}+E_{p2}+E_{p}(1').
\end{eqnarray}
We obtain the following Hamiltonian eigenvalues and eigenstates
\begin{eqnarray}
E_1=\frac{-2+U+U_1-\sqrt{(4)^2+(U-U_1)^2}}{2},  \nonumber \\
E_2=\frac{+2+U+U_1-\sqrt{(4)^2+(U-U_1)^2}}{2},  \nonumber \\
E_3=\frac{-2+U+U_1+\sqrt{(4)^2+(U-U_1)^2}}{2},   \nonumber \\
E_4=\frac{+2+U+U_1+\sqrt{(4)^2+(U-U_1)^2}}{2}.
\end{eqnarray}
and
\begin{eqnarray}
\ket{E_1}=
\begin{pmatrix}
\frac{1}{2}(-U + \sqrt{4 + (U - U_1)^2} + U_1) \\
-1 \\
\frac{1}{2}(U - \sqrt{4 + (U - U_1)^2} - U_1) \\
1 \\
\end{pmatrix},
\end{eqnarray}
\begin{eqnarray}
\ket{E_2}=
\begin{pmatrix}
\frac{1}{2}(U - \sqrt{4 + (U - U_1)^2} - U_1)\\
1 \\
\frac{1}{2}(U - \sqrt{4 + (U - U_1)^2} - U_1)\\
1
\end{pmatrix},
\end{eqnarray}
\begin{eqnarray}
\ket{E_3}=
\begin{pmatrix}
\frac{1}{2}(-U - \sqrt{4 + (U - U_1)^2} + U_1) \\
-1 \\
\frac{1}{2}(U + \sqrt{4 + (U - U_1)^2} - U_1) \\
1 \\
\end{pmatrix},
\end{eqnarray}
\begin{eqnarray}
\ket{E_4}=
\begin{pmatrix}
\frac{1}{2}(-U - \sqrt{4 + (U - U_1)^2} + U_1) \\
1 \\
\frac{1}{2}(U + \sqrt{4 + (U - U_1)^2} - U_1) \\
1 \\
\end{pmatrix},
\end{eqnarray}
\section{Analytical method for prediction of correlation-anticorrelation in generalized Swap Gate}
We can predict correlation or anticorrelation by monitoring the sign of the correlation function $f(C)$ with correlation function operator C
\begin{eqnarray}
  C=\frac{N_{1,1'}+N_{2,2'}-N_{1,2'}-N_{2,1'}}{N_{1,1'}+N_{2,2'}+N_{1,2'}+N_{2,1'}}= 
  \begin{pmatrix}
  1 & 0 & 0 & 0 \\
  0 & -1 & 0 & 0 \\
  0 & 0 & -1 & 0 \\
  0 & 0 & 0 & 1
  \end{pmatrix}, 1=N_{1,1'}+N_{2,2'}+N_{1,2'}+N_{2,1'},
\end{eqnarray}
and under the circumstances of time-independent Hamiltonian we have
\begin{eqnarray}
f(C)=\bra{\psi(t)}C\ket{\psi(t)}=(\bra{E_1}c_{E1}^{*}e^{-i\phi_{E1}}e^{-\frac{E_1 t}{i \hbar}}+\bra{E_2}c_{E2}^{*}e^{-i\phi_{E2}}e^{-\frac{E_2 t}{i\hbar}}+\bra{E_3}c_{E3}^{*}e^{-i\phi_{E3}}e^{-\frac{E_3 t}{i\hbar}}+\bra{E_4}c_{E4}^{*}e^{-i\phi_{E4}}e^{-\frac{E_4 t}{\hbar}}) \nonumber \\
C(\ket{E_1}c_{E1}e^{i\phi_{E1}}e^{\frac{E_1 t}{i\hbar}}+\ket{E_2}c_{E2}e^{i\phi_{E2}}e^{\frac{E_2 t}{i\hbar}}+\ket{E_3}c_{E3}e^{i\phi_{E3}}e^{\frac{E_3 t}{i\hbar}}+\ket{E_4}c_{E4}e^{i\phi_{E4}}e^{\frac{E_4 t}{i\hbar}}). 
\end{eqnarray}

When $f(C)$ has negative values we are dealing with Swap Gate, while ANTISWAP gate requires positive value.
\subsubsection{Correlation function for case I:$E_f(1,1')=E_f(1,2')=U$, $E_f(2,1')=E_f(2,2')=U_1$}
For qubit-qubit Hamiltonian of structure
\begin{eqnarray}H=
\begin{pmatrix}
U & 1 & 1 & 0 \\
1 & U & 0 & 1 \\
1 & 0 & U_1 & 1 \\
0 & 1 & 1 & U_1
\end{pmatrix}. \nonumber \\
\end{eqnarray}
we obtain the following correlation function
\begin{eqnarray}
C= \frac{1}{((4 + (U - U_1)^2)^{(\frac{3}{2})})}2 e^{-i(\phi_{E1} + \phi_{E2} + \phi_{E3} +
    \phi_{E4} + (E_{1} + E_{2} + E_{3} +
       E_{4}) t)} (c_{E1} c_{E2} (4 + (U - U_1)^2) (U - U_1) \times \nonumber \\ \times
         Cos(
     \phi_{E1} - \phi_{E2} - E_{1}t + E_{2}t) + \nonumber \\
   c_{E2} c_{E3} \sqrt{4 + (U - U_1)^2} \sqrt{
    4 + (U - U_1) (U + \sqrt{4 + (U - U_1)^2} - U_1)} \sqrt{
    4 - (U - U_1) (-U +\sqrt{4 + (U - U_1)^2} + U_1)} \times \nonumber \\ \times
     Cos[\phi_{E2} - \phi_{E3} - E_{2} t + E_{3} t] + \nonumber \\
   c_{E1} c_{E4} \sqrt{4 + (U - U_1)^2} \sqrt{
    4 + (U - U1) (U + \sqrt{4 + (U - U_1)^2} - U_1)} \sqrt{
    4 - (U - U1) (-U + Sqrt[4 + (U - U_1)^2] + U_1)} \times \nonumber \\ \times
     Cos(\phi_{E1} - \phi_{E_4} - E_1 t  + E_4 t)  \nonumber \\
   -c_{E3} c_{E4} (4 + (U - U_1)^2) (U - U_1) Cos(
     \phi_{E3} - \phi_{E4} - E_3t + E_4 t)) (Cos[
    \phi_{E1} + \phi_{E2} + \phi_{E3} +
     \phi_{E4} + (E_{1} + E_{2} + E_{3} + E_{4})t] + \nonumber \\
   i Sin(\phi_{E1} + \phi_{E2} + \phi_{E3} +
      \phi_{E4} + (E_1 + E_2 + E_3 + E_4) t)),\nonumber \\
\end{eqnarray}
where $|c_{E1}|^2$,$|c_{E2}|^2$, $|c_{E3}|^2$, $|c_{E4}|^2$ are probabilities for the system of coupled qubits to occupy energy level $E_1$, $E_2$, $E_3$, $E_4$ and normalization condition is fulfilled $1=|c_{E1}|^2+|c_{E2}|^2+|c_{E3}|^2+|c_{E4}|^2$. We encounter mixture of time-dependent cos or sin oscillations of different frequencies as $(E_{1} + E_{2} + E_{3} + E_{4})$, $- E_{1} + E_{2}$,  $- E_{2} + E_{3}$, $- E_1 + E_4$, $- E_3+ E_4$, where quantum state is given as
\begin{eqnarray}
\ket{\psi}_t=c_{E1}e^{\frac{E_1 t}{i}} e^{i \phi_{E1}}\ket{E_1}+c_{E2}e^{\frac{E_2 t}{i}} e^{i \phi_{E2}}\ket{E_2}+c_{E3}e^{\frac{E_3 t}{i}} e^{i \phi_{E3}}\ket{E_3}+c_{E4}e^{\frac{E_4 t}{i}} e^{i \phi_{E4}}\ket{E_4}.
 \end{eqnarray}
 Here we have tunning coefficients $U, U_1$, $\phi_{E1}$,$\phi_{E2}$, $\phi_{E3}$, $\phi_{E4}$.
The probability of occupancy of $1,1'$ is of the following form
\small
\begin{eqnarray}
\rho_{1,1}=p(1,1')=\frac{1}{ [4(4 + (U - U_1)^2)]} \times \nonumber \\
\times \Bigg[ ((c_{E3}^2 + c_{E4}^2) (4 + (U - U_1) (U + \sqrt{4 + (U - U_1)^2} - U_1)) +
    c_{E1}^2 (4 - (U - U_1) (-U + \sqrt{4 + (U - U_1)^2} + U_1)) + \nonumber \\
    +c_{E2}^2 (4 - (U - U_1) (-U + \sqrt{4 + (U - U_1)^2} +
          U_1)) \Bigg] \nonumber \\
           - \frac{(c_{E1} c_{E2} (-U + \sqrt{
      4 + (U - U_1)^2} + U_1) Cos[
     \phi_{E1} - \phi_{E2} + (-E_1 + E_2) t])}{(2 \sqrt{
    4 + (U -
       U_1)^2})} \nonumber \\
        - \frac{(2 c_{E_1} (c_{E_3} Cos[
        \phi_{E1} - \phi_{E3} - E_{1} t + E_{3} t] -
      c_{E4} Cos[\phi_{E1} - \phi_{E4} - E_1 t + E_4 t]))}{(\sqrt{
    4 + (U - U1) (U + \sqrt{4 + (U - U_1)^2} - U_1)} \sqrt{
    4 - (U - U1) (-U + \sqrt{4 + (U - U_1)^2} + U_1)})} + \nonumber \\
    +\frac{(2 c_{E2} (c_{E3} Cos[
        \phi_{E2} - \phi_{E3} - E_2 t + E_3 t] -
      c_{E4} Cos[\phi_{E2} - \phi_{E4} - E_2 t + E_4 t]))}{(\sqrt{
    4 + (U - U_1) (U + \sqrt{4 + (U - U_1)^2} - U_1)} \sqrt{
    2 + 1/2 (-U + \sqrt{4 + (U - U_1)^2} + U_1)^2})} \nonumber \\
    - \frac{(c_{E3} c_{E4} (U + \sqrt{
      4 + (U - U_1)^2} - U_1) Cos[
     \phi_{E3} - \phi_{E4} + (-E_3 + E_4) t])}{(2 \sqrt{4 + (U - U_1)^2})},
\end{eqnarray}

\begin{eqnarray}
\rho_{2,2}=p(1,2')=\frac{1}{(4 (4 + (U - U_1)^2))} ((c_{E3}^2 +
       c_{E4}^2) (4 + (U - U_1) (U + \sqrt{4 + (U - U_1)^2} - U_1)) + \nonumber \\
+  c_{E1}^2 (4 - (U - U_1) (-U + \sqrt{4 + (U - U_1)^2} + U_1)) +
    c_{E2}^2 (4 - (U - U_1) (-U + \sqrt{4 + (U - U_1)^2} +
          U_1)))+   \nonumber \\
 + \frac{(c_{E1} c_{E2} (-U + \sqrt{4 + (U - U_1)^2} + U_1) Cos[
     \phi_{E1} - \phi_{E2} + (-E_1 + E_2) t])}{(2 \sqrt{[
    4 + (U -
       U_1)^2]})}    \nonumber \\
 - \frac{(2 c_{E1} (c_{E3} Cos[
        \phi_{E1} - \phi_{E3} - E_1 t + E_3 t] +
      c_{E4} Cos[\phi_{E1} - \phi_{E4} - E_1 t + E_4 t]))}{(\sqrt{[
    4 + (U - U_1) (U + \sqrt{4 + (U - U_1)^2} - U_1)]} \sqrt{[
    4 - (U - U_1) (-U + \sqrt{4 + (U - U_1)^2} + U_1)]})}   \nonumber \\
 - \frac{(2 c_{E2} (c_{E3} Cos[
        \phi_{E2} - \phi_{E3} - E_2 t + E_3 t] +
      c_{E4} Cos[\phi_{E2} - \phi_{E4} - E_2 t + E_4 t]))}{(\sqrt{
    4 + (U - U_1) (U + \sqrt{4 + (U - U_1)^2} - U_1)} \sqrt{
    2 + \frac{1}{2} (-U + \sqrt{[4 + (U - U_1)^2]} + U_1)^2})} \nonumber \\
 + \frac{(c_{E3} c_{E4} (U + \sqrt{
      4 + (U - U_1)^2} - U_1) Cos[
     \phi_{E3} -\phi_ {E4} + (-E_3 + E_4) t])}{(2 \sqrt{4 + (U - U_1)^2})},
\end{eqnarray}

and 
\begin{eqnarray}
\rho_{3,3}=p(2,1')=\frac{(c_{E3}^2 + c_{E4}^2)}{(4 + (U - U_1) (U + \sqrt{[4 + (U - U_1)^2]} - U_1))} + \nonumber \\
 + \frac{(
 c_{E1}^2 + c_{E2}^2)}{(4 - (U - U_1) (-U + \sqrt{[4 + (U - U_1)^2]} + U_1))} + \nonumber \\
 -\frac{ (
 2 c_{E1} c_{E2} Cos[ \phi_{E1} -\phi_{E2} + (-E_1 + E_2) t])}{(
 4 - (U - U_1) (-U + \sqrt{[4 + (U - U_1)^2]} +
     U_1))}+  \nonumber \\
 + \frac{(2 c_{E1} (c_{E3} Cos[ \phi_{E1} -\phi_{E3} - E_1 t + E_3 t] -
      c_{E4} Cos[\phi_{E1} - \phi_{E4} - E_1 t + E_4 t]))}{(\sqrt{[
    4 + (U - U_1) (U + \sqrt{[4 + (U - U_1)^2]} - U_1)]} \sqrt{[
    4 - (U - U_1) (-U + \sqrt{[4 + (U - U_1)^2]} + U_1)]})} \nonumber \\
-\frac{ (2 c_{E2} (c_{E3} Cos[
        \phi_{E2} - \phi_{E3} - E_2 t + E_3 t]
 -  c_{E4} Cos[ \phi_{E2} - \phi_{E4} - E_2 t + E_4 t]))}{(\sqrt{[
    4 + (U - U_1) (U + \sqrt{[4 + (U - U_1)^2]} - U_1)]} \sqrt{[
    2 + \frac{1}{2} (-U + \sqrt[4 + (U - U_1)^2] + U_1)^2]})} \nonumber \\
 - \frac{(
 2 c_{E3} c_{E4} Cos[\phi_{E3} - \phi_{E4} + (-E_3 + E_4) t])}{(
 4 + (U - U1) (U + \sqrt{4 + (U - U_1)^2} - U1))}
\end{eqnarray}

\begin{eqnarray}
\rho_{4,4}(t)=p(2,2')=\frac{(c_{E3}^2 + c_{E4}^2)}{(4 + (U - U_1) (U + \sqrt{4 + (U - U_1)^2} - U_1))} + \frac{(
 c_{E1}^2 + c_{E2}^2)}{(4 - (U - U_1) (-U + \sqrt{4 + (U - U_1)^2} + U_1))} \nonumber \\
 - \frac{(
 2 c_{E1} c_{E2} Cos[
   \phi_{E1} -
    \phi_{E2} + (-E_1 + E_2) t])}{(-4 + (U - U_1) (-U + \sqrt{
     4 + (U - U_1)^2} +
     U_1))} + \nonumber \\
+\frac{(2 c_{E1} (c_{E3} Cos[\phi_{E1} - \phi_{E3} - E_1 t + E_3 t] +
      c_{E4} Cos[\phi_{E1} - \phi_{E4} - E_1 t + E_4 t]))}{(\sqrt{
    4 + (U - U_1) (U + \sqrt{[4 + (U - U_1)^2]} - U_1)} \sqrt{[
    4 - (U - U_1) (-U + \sqrt{[4 + (U - U_1)^2]} + U_1)]})}+ \nonumber \\ + \frac{(2 c_{E2} (c_{E3} Cos[
        \phi_{E2} - \phi_{E3} - E_2 t + E_3 t] +
      c_{E4} Cos[\phi_{E2} -\phi_ {E4} - E_2 t + E_4 t]))}{(\sqrt{
    4 + (U - U_1) (U + \sqrt{4 + (U - U_1)^2} - U_1)} \sqrt{
    2 + \frac{1}{2} (-U + \sqrt{4 + (U - U_1)^2} + U_1)^2})} + \nonumber \\ +\frac{(
 2 c_{E3} c_{E4} Cos[\phi_{E3} - \phi_{E4} + (-E_3 + E_4) t])}{(
 4 + (U - U_1) (U + \sqrt{4 + (U - U_1)^2} - U_1))}
\end{eqnarray}

Denisty matix of 2 interacting parciles allow us to obtain density matrices of particles A and B. The elements of density matrix A are given as
\begin{eqnarray}
\rho_{A: (1,1)}=\frac{((c_{E3}^2 + c_{E4}^2) (U + \sqrt{4 + (U - U_1)^2} - U_1) +
    c_{E1}^2 (-U + \sqrt{[4 + (U - U_1)^2]} + U_1) +
    c_{E2}^2 (-U + \sqrt{[4 + (U - U_1)^2]} + U_1))}{(2 \sqrt{[
    4 + (U -
       U_1)^2]})} \nonumber \\  - \frac{(4 (c_{E1} c_{E3} Cos[
        \phi_{E1} - \phi_{E3} - E_1 t + E_3 t] +
      c_{E2} c_{E4} Cos[\phi_{E2} - \phi_{E4} - E_2 t + E_4 t]))}{(\sqrt{[
    4 + (U - U_1) (U + \sqrt{[4 + (U - U_1)^2]} - U_1)]} \sqrt{[
    4 - (U - U_1) (-U + \sqrt{[4 + (U - U_1)^2]} + U_1)]})}=p(1,t),
\end{eqnarray}

\begin{eqnarray}
\rho_{A: (2,2)}= (   
\frac{  ((c_{E1}^2+c_{E2}^2) (U + \sqrt{[4 + (U - U_1)^2]} - U_1) +
 (c_{E3}^2 + c_{E4}^2) (-U +
         \sqrt{[4 + (U - U_1)^2]} +
         U_1))}{2 \sqrt{[
     4 + (U - U_1)^2]}} + \nonumber \\
+\frac{(4 (c_{E1} c_{E3} Cos[\phi_{E1} -\phi_{E3} - E_1 t + E_3 t] +
         c_{E2} c_{E4} Cos[\phi_{E2} - \phi_{E4} - E_2 t + E_4 t]))}{(\sqrt{[
      4 + (U - U_1) (U + \sqrt{[4 + (U - U_1)^2]} - U_1)]} \sqrt{[
      4 - (U - U_1) (-U + \sqrt{[4 + (U - U_1)^2]} + U_1)]})})=p(2,t).
\end{eqnarray}

\begin{eqnarray}
\rho_{A: (1,2)}=\frac{1}{(\sqrt{[
   4 + (U - U_1)^2]} \sqrt{[4 + (U - U_1) (U + \sqrt{[4 + (U - U_1)^2]} - U_1)]}
    \sqrt{[2 + \frac{1}{2} (-U + \sqrt{[4 + (U - U_1)^2]} + U_1)^2]})} \times \nonumber \\
\times \Bigg[(-c_{E1}^2 - c_{E2}^2 + c_{E3}^2 + c_{E4}^2) \sqrt{[
    4 + (U - U_1) (U + \sqrt{[4 + (U - U_1)^2]} - U_1)]} \sqrt{[
    4 - (U - U_1) (-U + \sqrt{[4 + (U - U_1)^2]} + U_1)]} + \nonumber \\
   +2 (c_{E1} c_{E3} \sqrt{[
       4 + (U - U_1)^2]} (U - U_1) Cos[
        \phi_{E1} - \phi_{E3} - E_1 t + E_3 t] + \nonumber \\
      +c_{E2} c_{E4} \sqrt{[
       4 + (U - U_1)^2]} (U - U_1) Cos[
        \phi_{E2} - \phi_{E4} - E_2 t + E_4 t] + \nonumber \\
     + i (4 + (U - U_1)^2) (c_{E1} c_{E3} Sin[
           \phi_{E1} - \phi_{E3} - E_1 t + E_3 t] +
         c_{E2} c_{E4} Sin[\phi_{E2} - \phi_{E4} - E_2 t + E_4 t]))\Bigg]
\end{eqnarray}


\begin{eqnarray}
\rho_{A: (2,1)}= \frac{1}{(\sqrt{[
   4 + (U - U_1)^2]} \sqrt{[4 + (U - U_1) (U + \sqrt{[4 + (U - U_1)^2]} - U_1)]}
    \sqrt{[2 + 1/2 (-U + \sqrt{[4 + (U - U_1)^2]} + U_1)^2]})}
\times \nonumber \\ \times
\Bigg[(-c_{E1}^2 - c_{E2}^2 + c_{E3}^2 + c_{E4}^2) \sqrt{[
    4 + (U - U_1) (U + \sqrt{[4 + (U - U_1)^2]} - U_1)]} \sqrt{[
    4 - (U - U_1) (-U + \sqrt{[4 + (U - U_1)^2]} + U_1)]} + \nonumber \\
   +2 (c_{E1} c_{E3} \sqrt{[
       4 + (U - U_1)^2]} (U - U_1) Cos[
        \phi_{E1} - \phi_{E3} - E_1 t + E_3 t] + \nonumber \\
     + c_{E2} c_{E4} \sqrt{[
       4 + (U - U_1)^2]} (U - U_1) Cos[
        \phi_{E2} - \phi_{E4} - E_2 t + E_4 t] \nonumber \\
-  i (4 + (U - U_1)^2) (c_{E1} c_{E3} Sin[
           \phi_{E1} - \phi_{E3} - E_1 t + E_3 t] +
         c_{E2} c_{E4} Sin[\phi_{E2} - \phi_{E4} - E_2 t + E_4 t]))\Bigg]
\end{eqnarray}
We have also obtained the density matrix of particle B (electron at nodes 1' and 2') in the form as
\begin{eqnarray}
\rho_{B: (1,1)}=\frac{1}{2}(1 - 2 c_{E1} c_{E2} Cos[\phi_{E1} -\phi_{E2} - E_1 t + E_2 t]  \nonumber \\ -
   2 c_{E3} \sqrt{[1 - c_{E1}^2 - c_{E2}^2 - c_{E3}^2]}
     Cos[\phi_{E3} - \phi_{E4} - E_3 t + E_4 t])
\end{eqnarray}

\begin{eqnarray}
\rho_{B: (2,2)}=\frac{1}{2}(1 + 2 c_{E1} c_{E2} Cos[\phi_{E1} - \phi_{E2} - E_1 t + E_2 t]  \nonumber \\ +
   2 c_{E3} \sqrt{[1 - c_{E1}^2 - c_{E2}^2 - c_{E3}^2]}
     Cos[\phi_{E3} - \phi_{E4} - E_3 t + E_4 t])
\end{eqnarray}

\begin{eqnarray}
\rho_{B: (1,2)}=\frac{1}{2} (1 - 2 c_{E1}^2 - 2 c_{E3}^2 +
   2 i c_{E1} c_{E2} Sin[\phi_{E1} - \phi_{E2} - E_1 t + E_2 t] + \nonumber \\
   2 i c_{E3} \sqrt{[1 - c_{E1}^2 - c_{E2}^2 - c_{E3}^2]}
     Sin[\phi_{E3} - \phi_{E4} - E_3 t + E_4 t])
\end{eqnarray}

\begin{eqnarray}
\rho_{B: (2,1)}=\frac{1}{2}(1 - 2 c_{E1}^2 - 2 c_{E3}^2 -
   2 i c_{E1} c_{E2} Sin[\phi_{E1} - \phi_{E2} - E_1 t + E_2 t] \nonumber \\
 -    2 i c_{E3} \sqrt{[1 - c_{E1}^2 - c_{E2}^2 - c_{E3}^2]}
     Sin[\phi_{E3} - \phi_{E4} - E_3 t + E_4 t])
\end{eqnarray}
It  remarkable to notice that hopping constant that was constant for position dependent qubit was modified and it has two parts that depends on the frequency $E_2-E_1=\frac{1}{2} \left(-\sqrt{(U-U_1)^2+4}+U+U_1)+2 \right)-\frac{1}{2} \left(-\sqrt{(U-U_1)^2+4}+U+U_1)-2 \right)$ and $E_4-E_3=\frac{1}{2} \left(+\sqrt{(U-U_1)^2+4}+U+U_1)+2 \right)-\frac{1}{2} \left(+\sqrt{(U-U_1)^2+4}+U+U_1)-2 \right)$
For given density matrix
\begin{equation}
\rho=
\begin{pmatrix}
\rho_{1,1}=1-\rho_{2,2} & \rho_{1,2} \\
\rho_{2,1}=\rho_{1,2}^{*} & \rho_{2,2} \\
\end{pmatrix},
\end{equation}
where $\rho_{1,1}$ , $\rho_{2,2}$, $\rho(1,2)_r$, $\rho(1,2)_i$ in R with $\rho_{1,2}=\rho(1,2)_r + i\rho(1,2)_i $
we identify von Neumann entanglement $S$ entropy expressed as
\begin{eqnarray}
\label{entropyS}
-S(\rho_{2,2},|\rho_{1,2}|)= \frac{1}{(2 \sqrt{
  4 (|\rho(1,2)_r|^2+|\rho(1,2)_i|^2) + (1 -
     2 \rho_{2,2})^2})} \times [((-1 - 4 (|\rho(1,2)_r|^2+|\rho(1,2)_i|^2)+ \nonumber \\ + \sqrt{4 (|\rho(1,2)_r|^2+|\rho(1,2)_i|^2) + (1 - 2 \rho_{2,2})^2} -
      4 (-1 + \rho_{2,2}) \rho_{2,2}) Log[
     \frac{1}{2} (1 - \sqrt{4 (|\rho(1,2)_r|^2+|\rho(1,2)_i|^2) + (1 - 2 \rho_{2,2})^2})] + \nonumber \\ +(4 (|\rho(1,2)_r|^2+|\rho(1,2)_i|^2) + \sqrt{
      4 (|\rho(1,2)_r|^2+|\rho(1,2)_i|^2) + (1 - 2 \rho_{2,2})^2} + \nonumber \\ + (1 - 2 \rho_{2,2})^2) Log[
     \frac{1}{2} (1 + \sqrt{4 (|\rho(1,2)_r|^2+|\rho(1,2)_i|^2) + (1 - 2 \rho_{2,2})^2})])] .
     \end{eqnarray}
\begin{equation}
E_1=\frac{1}{2} \left(-\sqrt{(U-U_1)^2+4}+U+U_1)-2 \right),
E_2=\frac{1}{2} \left(-\sqrt{(U-U_1)^2+4}+U+U_1)+2 \right),
\end{equation}
\begin{equation}
E_3=\frac{1}{2} \left(+\sqrt{(U-U_1)^2+4}+U+U_1)-2 \right),
E_4=\frac{1}{2} \left(+\sqrt{(U-U_1)^2+4}+U+U_1)+2 \right).
\end{equation}
\normalsize
\subsubsection{Correlation function for case II: $E_f(1,1')=E_f(2,2')=U$, $E_f(1,2')=E_f(2,1')=U_1$ }
For qubit-qubit Hamiltonian
\begin{eqnarray}H=
\begin{pmatrix}
U & 1 & 1 & 0 \\
1 & U_1 & 0 & 1 \\
1 & 0 & U_1 & 1 \\
0 & 1 & 1 & U
\end{pmatrix}, \nonumber \\
\end{eqnarray}
that gives
\begin{eqnarray}
E_1=U, E_2=U_1, E_3=\frac{(U+U_1-\sqrt{(4)^2+(U-U_1)^2})}{2}, E_4=\frac{(U+U_1+\sqrt{(4)^2+(U-U_1)^2})}{2},
\end{eqnarray}
with
\begin{eqnarray}
U=E_{p1}+E_{p1'}+\frac{q^2}{\sqrt{d^2+(a+b)^2}}
\end{eqnarray}
and
\begin{eqnarray}
U_1=E_{p2}+E_{p1'}+\frac{q^2}{d}= \nonumber
2E_{p1'}+\frac{1}{2}\frac{q^2}{d} 
-\frac{1}{2}[\frac{q^2}{\sqrt{(d+Cos(\alpha)(a+b))^2+(1+Sin(\alpha))^2(a+b)^2}} \nonumber \\
+\frac{q^2}{\sqrt{(d+Cos(\alpha)(a+b))^2+(Sin(\alpha))^2(a+b)^2}}-\frac{q^2}{\sqrt{d^2+(a+b)^2}}].
\end{eqnarray}
We obtain the following correlation function
\begin{eqnarray}
C(t,U,U_1,c_{E1},c_{E2},c_{E3},c_{E4})=
[\left((U-U_1)^2+16\right) (\left(c_{E2}^2-c_{E1}^2\right) \left(-\sqrt{(U-U_1)^2+16}\right)-(c_{E3}^2-c_{E4}^2)
   (U-U_1))+ \nonumber \\
   +2 c_{E3} c_{E4} \sqrt{(U-U_1) \left(\sqrt{(U-U_1)^2+16}+U-U_1\right)+16} \sqrt{16-(U-U_1)
   \left(\sqrt{(U-U_1)^2+16}-U+U_1\right)} \times \nonumber \\
  \times \sqrt{(U-U_1)^2+16} \cos ( (-E_{3}+E_4)t
   +\phi_{E3}-\phi_{E4})]\frac{1}{\left((U-U_1)^2+16\right)^{3/2}}.
\end{eqnarray}

We have given the probability

$p(1,1')=\rho_{AB:(1,1)}=(\frac{1}{(4 \sqrt{
  16 + (U - U_1)^2})}  ) [(2 c_{E1}^2 \sqrt{16 + (U - U_1)^2} +
   c_{E4}^2 (U + \sqrt{16 + (U - U_1)^2} - U_1) +
   c_{E3}^2 (-U + \sqrt{16 + (U - U_1)^2} + U_1) -
   2 \sqrt{2}
     c_{E1} (c_{E3} \sqrt{16 - (U - U_1) (-U + \sqrt{16 + (U - U_1)^2} + U_1)}
        Cos[\phi_{E1} - \phi_{E3} - E_1 t + E_3 t] +
      c_{E4} \sqrt{[16 + (U - U_1) (U + \sqrt{16 + (U - U_1)^2} - U_1)]}
        Cos[\phi_{E1} - \phi_{E4} - E_1 t + E_4 t]))] + \frac{(
 8 c_{E3} c_{E4} Cos[\phi_{E3} - \phi_{E4} + (-E_3 + E_4) t])}{(
 \sqrt{16 + (U - U_1) (U + \sqrt{16 + (U - U_1)^2} - U_1)} \sqrt{
  16 - (U - U_1) (-U + \sqrt{16 + (U - U_1)^2} + U_1)})}$,

\begin{eqnarray}
p(1,2')=\rho_{AB:(2,2)}=
\frac{c_{E2}^2}{2} + \frac{(4 c_{E4}^2)}{(
 16 + (U - U_1) (U + \sqrt{16 + (U - U_1)^2} - U_1))}+ \nonumber \\
 + \frac{8 c_{E3}^2}{(
 16 - U + \sqrt{[16 + (U - U_1)^2]} + U_1)} + \frac{(
 4 c_{E2} c_{E3} Cos[\phi{E2} - \phi_{E3} - E_2 t + E_3 t])}{\sqrt{
 16 - U + \sqrt{16 + (U - U_1)^2} + U_1}}+ \nonumber \\
 -\frac{ (
 2 \sqrt{2} c_{E4} (c_{E2} \sqrt{[16 - U + \sqrt{16 + (U - U_1)^2} + U_1]}
      Cos[\phi_{E2} - \phi_{E4} - E_2 t + E_4 t] +
    4 c_{E3} Cos[\phi_{E3} - \phi_{E4} - E_3 t + E_4 t]))}{(
 \sqrt{16 + (U - U_1) (U + \sqrt{16 + (U - U_1)^2} - U_1)} \sqrt{
  16 - U + \sqrt{16 + (U - U_1)^2} + U_1})}
\end{eqnarray}

\begin{eqnarray}
p(2,1')=\rho_{AB:(3,3)}=
\frac{c_{E2}^2}{2} + \frac{(4 c_{E4}^2)}{(
 16 + (U - U_1) (U + \sqrt{[16 + (U - U_1)^2]} - U_1))}+ \nonumber \\
 + \frac{(8 c_{E3}^2)}{(
 16 - U + \sqrt{[16 + (U - U_1)^2]} + U_1)}+ \nonumber \\
 - \frac{(
 4 c_{E2} c_{E3} Cos[\phi_{E2} - \phi_{E3} - E_2 t + E_3 t])}{\sqrt{[
 16 - U + \sqrt{16 + (U - U_1)^2} + U_1]}}+ \nonumber \\
 + \frac{(
 2 \sqrt{2} c_{E4} (c_{E2} \sqrt{16 - U + \sqrt{16 + (U - U_1)^2} + U_1}
      Cos[\phi_{E2} - \phi_{E4} - E_2 t + E_4 t] -
    4 c_{E3} Cos[\phi_{E3} - \phi_{E4} - E_3 t + E_4 t]))}{(
 \sqrt{[16 + (U - U_1) (U + \sqrt{16 + (U - U_1)^2} - U_1)]} \sqrt{[
  16 - U + \sqrt{16 + (U - U1)^2} + U_1]})}
\end{eqnarray}

\begin{eqnarray}
p(2,2')=\rho_{AB:(4,4)}=
\frac{(c_{E1} c_{E3} (-U + \sqrt{[16 + (U - U_1)^2]} + U_1) Cos[
   \phi_{E1} - \phi_{E3} - E_1 t + E_3 t])}{\sqrt{[
 16 - U + \sqrt{[16 + (U - U_1)^2]} + U_1]}} + \nonumber \\
 \frac{1}{4}[ (2 c_{E1}^2 + c_{E4}^2 + \frac{(c_{E4}^2 (U - U1))}{\sqrt{[16 + (U - U_1)^2]}} + \nonumber \\
    2 c_{E3}^2 (-8 - U + \sqrt{[16 + (U - U1)^2]} + \frac{(
       16 \sqrt{[16 + (U - U_1)^2]})}{(-15 + 2 U - 2 U_1)} + U_1 + \frac{136}{(
       15 - 2 U + 2 U_1)})+ \nonumber \\
 + \frac{(
    2 \sqrt{2}
      c_{E4} (c_{E1} (U + \sqrt{[16 + (U - U_1)^2]} - U_1) Cos[
         \phi_{E1} - \phi_{E4} - E_1 t + E_4 t] + \frac{(
       16 c_{E3} Cos[\phi_{E3} - \phi_{E4} - E_3 t + E_4 t])}{\sqrt{[
       16 - U + \sqrt{[16 + (U - U_1)^2]} + U_1]}}))}{\sqrt{[
    16 + (U - U_1) (U + \sqrt{[16 + (U - U_1)^2]} - U_1)]}})] \nonumber \\
\end{eqnarray}

and we obtain denisty matrix of A particle by relations $\rho_{A}(1,1)=\rho_{AB}(1,1)+\rho_{AB}(2,2)$, $\rho_{A}(2,2)=\rho_{AB}(3,3)+\rho_{AB}(4,4)$, $\rho_{A}(1,2)=\rho_{AB}(1,3)+\rho_{AB}(2,4)$, $\rho_{A}(2,1)=\rho_{AB}(3,1)+\rho_{AB}(4,2)$ so we have

\begin{eqnarray}
\rho_{A (1,1)}=\frac{c_{E2}^2}{2} + \frac{(4 c_{E4}^2)}{(
 16 + (U - U_1) (U + \sqrt{[16 + (U - U_1)^2]} - U_1))}+ \nonumber \\ + \frac{(8 c_{E3}^2)}{(
 16 - U + \sqrt{[16 + (U - U_1)^2]} + U_1)} + \frac{(
 4 c_{E2} c_{E3} Cos[\phi_{E2} - \phi_{E3} - E_2 t + E_3 t])}{\sqrt{[
 16 - U + \sqrt{[16 + (U - U_1)^2]} + U_1]}} + (\frac{1}{(
 4 \sqrt{[16 + (U - U_1)^2]})}) \nonumber \\
(2 c_{E1}^2 \sqrt{[16 + (U - U_1)^2]} +
   c_{E4}^2 (U + \sqrt{[16 + (U - U_1)^2]} - U_1) + \nonumber \\
   c_{E3}^2 (-U + \sqrt{[16 + (U - U_1)^2]} + U_1)+ \nonumber \\
 -  2 \sqrt{2}
     c_{E1} (c_{E3}\sqrt{[16 - (U - U_1) (-U + \sqrt{[16 + (U - U_1)^2]} + U_1)]}
        Cos[\phi_{E1} - \phi_{E3} - E_1 t + E_3 t] + \nonumber \\
      c_{E4} \sqrt{[16 + (U - U_1) (U + \sqrt{[16 + (U - U_1)^2]} - U_1)]}
        Cos[\phi_{E1} - \phi_{E4} - E_1 t + E_4 t]))+ \nonumber \\
 - \frac{(
 2 \sqrt{2} c_{E4} (c_{E2} \sqrt{[16 - U + \sqrt{16 + (U - U_1)^2} + U_1]}
      Cos[\phi_{E2} - \phi_{E4} - E_2 t + E_4 t] +
    4 c_{E3} Cos[\phi_{E3} - \phi_{E4} - E_3 t + E_4 t]))}{(
 \sqrt{[16 + (U - U_1) (U + \sqrt{[16 + (U - U_1)^2]} - U_1)]} \sqrt{[
  16 - U + \sqrt{[16 + (U - U_1)^2]} + U_1]})}+ \nonumber \\
 + \frac{(
 8 c_{E3} c_{E4} Cos[\phi_{E3} - \phi_{E4} + (-E_3 + E_4) t])}{(
 \sqrt{[16 + (U - U_1) (U + \sqrt{[16 + (U - U_1)^2]} - U_1)]} \sqrt{[
  16 - (U - U_1) (-U + \sqrt{[16 + (U - U_1)^2]} + U_1)]})} \nonumber \\
\end{eqnarray}

$\rho_{A (2,2)}=
\frac{c_{E2}^2}{2} + \frac{(4 c_{E4}^2)}{(
 16 + (U - U_1) (U + \sqrt{[16 + (U - U_1)^2]} - U_1))} + \frac{(8 c_{E3}^2)}{(
 16 - U + \sqrt{[16 + (U - U_1)^2]} + U_1)} + \frac{(
 c_{E1} c_{E3} (-U + \sqrt{[16 + (U - U_1)^2]} + U_1) Cos[
  \phi{E1} - \phi_{E3} - E_1 t + E_3 t])}{\sqrt{[
 16 - U + \sqrt{[16 + (U - U_1)^2]} + U_1]}} - \frac{(
 4 c_{E2} c_{E3} Cos[\phi_{E2} - \phi_{E3} - E_2 t + E_3 t])}{\sqrt{[
 16 - U + \sqrt{[16 + (U - U_1)^2]} + U_1]}} + \frac{(
 2 \sqrt{2} c_{E4} (c_{E2} \sqrt{[16 - U + \sqrt{[16 + (U - U1)^2]} + U_1]}
      Cos[\phi_{E2} -\phi_{E4} - E_2 t + E_4 t] -
    4 c_{E3} Cos[\phi_{E3} - \phi_{E4} - E_3 t+ E_4 t]))}{(
 \sqrt{[16 + (U - U1) (U + \sqrt{[16 + (U - U_1)^2]} - U_1)]}\sqrt{[
  16 - U + \sqrt{16 + (U - U_1)^2} + U_1]})} +
 \frac{1}{4} (2 c_{E1}^2 + c_{E4}^2 + \frac{(c_{E4}^2 (U - U_1))}{\sqrt{[16 + (U - U1)^2]}} +
    2 c_{E3}^2 (-8 - U + \sqrt{[16 + (U - U_1)^2]} + \frac{(
       16 \sqrt{[16 + (U - U_1)^2]})}{(-15 + 2 U - 2 U_1)} + U_1 + 136/(
       15 - 2 U + 2 U_1)) + \frac{(
    2 \sqrt{2}
      c_{E4} (c_{E1} (U + \sqrt{[16 + (U - U_1)^2]} - U_1) Cos[
         \phi_{E1} - \phi_{E4} - E_1t+E_4t] + \frac{(
       16 c_{E3} Cos[\phi_{E3} - \phi_{E4} - E_3t+E_4t])}/\sqrt{[
       16 - U + \sqrt{[16 + (U - U_1)^2]} + U_1]}))}{\sqrt{[
    16 + (U - U1) (U + \sqrt{[16 + (U - U_1)^2]} - U_1)]}})
$

$\rho_{A (1,2)}=
- \frac{1}{2} c_{E1} c_{E2} e^{(i (-\phi_{E1} + \phi_{E2} + (E_1 - E_2) t))} -
 \frac{1}{2} c_{E1} c_{E2} e^{(i (\phi_{E1} - \phi_{E2} + (-E_1 + E_2) t))} + \frac{(
 2 c_{E4}^2)}{\sqrt{[16 + (U - U_1)^2]}} + \frac{(
 2 \sqrt{2} c_{E1} c_{E4} e^{(i (-\phi_{E1} + \phi_{E4} + (E_1 - E_4) t) )} )}{\sqrt{
 32 + 2 (U + \sqrt{16 + (U - U_1)^2} - U_1)^2}} - \frac{(
 2 \sqrt{2} c_{E1} c_{E4} e^(i (\phi_{E1} - \phi_{E4} + (-E_1 + E_4) t)))}{\sqrt{[
 32 + 2 (U + \sqrt{[16 + (U - U_1)^2]}- U_1)^2]}} + \frac{(
 c_{E2} c_{E4} e^{(
  i(-\phi_{E2} + \phi_{E4} + (E_2 - E4_) t))} (U + \sqrt{[16 + (U - U_1)^2]} -
    U1))}{(\sqrt{2} \sqrt{[32 + 2 (U + \sqrt{16 + (U - U1)^2} - U1)^2]})} - \frac{(
 c_{E2} c_{E4} e^{(
  i (\phi_{E2} - \phi_{E4} + (-E_2 + E_4) t))} (U + \sqrt{[16 + (U - U_1)^2]} -
    U_1))}{(\sqrt{2} \sqrt{[32 + 2 (U + \sqrt{[16 + (U - U_1)^2]} - U_1)^2]})} + \frac{(
 2 c_{E3}^2 (-1 - U + \sqrt{[16 + (U - U_1)^2]} + U_1))}{(-15 + 2 U - 2 U_1)} - \frac{(
 2 c_{E1} c_{E3} e^{(i (-\phi_{E1} + \phi_{E3} + (E_1-E_3) t))} )}{\sqrt{[
 16 - U + \sqrt{[16 + (U - U_1)^2]} + U_1]}} + \frac{(
 2 c_{E1} c_{E3} e^{(i (\phi_{E1} - \phi_{E3} + (-E_1 + E_3) t))} )}{\sqrt{[
 16 - U + \sqrt{[16 + (U - U_1)^2]} + U_1]}} - \frac{(
 c_{E2} c_{E3} e^{(
  i (\phi_{E2} - \phi_{E3} + (-E_2 + E_3) t))} (-U + \sqrt{[16 + (U - U_1)^2]} +
     U_1))}{(2 \sqrt{[16 - U + \sqrt{[16 + (U - U_1)^2]} + U_1]})} - \frac{(
 4 c_{E3} c_{E4} e^{(
  i (- \phi_{E3} + \phi_{E4} + (E_3 - E_4) t) )} (U + \sqrt{[16 + (U - U_1)^2]} -
    U_1))}{(\sqrt{[32 + 2 (U + \sqrt{[16 + (U - U_1)^2]} - U_1)^2]}
   \sqrt{[-2 U + 2 (16 + \sqrt{[16 + (U - U_1)^2]} + U_1)]} )} + \frac{(
 c_{E2} c_{E3} e^{(
  i (-\phi_{E2} + \phi_{E3} + (E_2 - E_3) t) )}  (-U + \sqrt{[16 + (U - U_1)^2]} +
     U_1))}{( \sqrt{2} \sqrt{[
  32 + 2 (-U + \sqrt{[16 + (U - U_1)^2]} + U_1)^2]} )} + \frac{(
 4 c_{E3} c_{E4} e^{(
  i (\phi_{E3} - \phi_{E4} + (-E_3 + E_4)t) )}  (-U + \sqrt{[16 + (U - U_1)^2]} +
     U_1))}{(\sqrt{[32 + 2 (U + \sqrt{16 + (U - U_1)^2} - U_1)^2]} \sqrt{[
  32 + 2 (-U + \sqrt{[16 + (U - U_1)^2]} + U_1)^2]})}+ \nonumber \\
 -\frac{(
 4 c_{E3}^2 (-U + \sqrt{[16 + (U - U_1)^2]} + U_1))}{(
 \sqrt{[-2 U + 2 (16 + \sqrt{[16 + (U - U_1)^2]} + U_1)]} \sqrt{[
  32 + 2 (-U + \sqrt{[16 + (U - U_1)^2]} + U_1)^2]}  )}+ \nonumber \\
 - \frac{(
 2 \sqrt{2} c_{E3} c_{E4} ((U - U_1) Cos[ \phi_{E3} - \phi_{E4} - E_3 t + E_4 t] +
    i \sqrt{[16 + (U - U_1)^2]} Sin[ \phi_{E3} - \phi_{E4} - E_3 t + E_4 t]))}{(
 \sqrt{[16 + (U - U_1) (U + \sqrt{[16 + (U - U_1)^2]} - U_1)]} \sqrt{[
  16 - U + \sqrt{[16 + (U - U_1)^2]} + U_1]} )}
$

$\rho_{A (2,1)}=
-(1/2) c_{E1} c_{E2} e^{(i (-\phi_{E1} + \phi_{E2} + (E_1 - E_2) t))} -
 \frac{1}{2}c_{E1} c_{E2} e^{(i (\phi_{E1} - \phi_{E2} + (-E_1 + E_2) t))} + \frac{(
 2 c_{E4}^2)}{\sqrt{16 + (U - U_1)^2}} - \frac{(
 2 \sqrt{2} c_{E1} c_{E4} e^(i (-\phi_{E1} + \phi_{E4} + (E_1 - E_4) t)))}{\sqrt{[
 32 + 2 (U + \sqrt{[16 + (U - U_1)^2]} - U_1)^2]}} + \frac{(
 2 \sqrt{2} c_{E1} c_{E4} e^{(i (\phi_{E1} - \phi_{E4} + (-E_1 + E_4) t) )} )}{\sqrt{[
 32 + 2 (U + \sqrt{[16 + (U - U_1)^2]} - U_1)^2]}} - \frac{(
 c_{E2} c_{E4} e^{(
  i (-\phi_{E2} + \phi_{E4} + (E2 - E4) t))} (U + \sqrt{[16 + (U - U_1)^2]} -
    U_1))}{(\sqrt{[2]} \sqrt{[32 + 2 (U + \sqrt{[16 + (U - U_1)^2]} - U_1)^2]})} + \frac{(
 c_{E2} c_{E4} e^(
  i (\phi_{E2} - \phi_{E4} + (-E_2 + E_4) t)) (U + \sqrt{[16 + (U - U_1)^2]} -
    U_1))}{(\sqrt{2} \sqrt{[32 + 2 (U + \sqrt{[16 + (U - U_1)^2]} - U_1)^2]})} + \frac{(
 2 c_{E3}^2 (-1 - U + \sqrt{[16 + (U - U_1)^2]} + U_1))}{(-15 + 2 U - 2 U_1)} + \frac{(
 2 c_{E1} c_{E3} e^{(i (-\phi_{E1} + \phi_{E3} + (E_1 - E_3) t ) )} )}{\sqrt{[
 16 - U + \sqrt{[16 + (U - U_1)^2]} + U_1]}} - \frac{(
 2 c_{E1} c_{E3} e^{(i ( \phi_{E1} - \phi_{E3} + (-E_1 + E_3) t) )} )}{\sqrt{[
 16 - U +\sqrt{16 + (U - U_1)^2} + U_1]}} - \frac{(
 c_{E2} c_{E3} e^{(
  i (- \phi_{E2} + \phi_{E3} + (E_2 - E_3) t) )}  (-U + \sqrt{[16 + (U - U_1)^2]} +
     U_1))}{(2 \sqrt{[16 - U + \sqrt{[16 + (U - U_1)^2]} + U_1]} )}  - \frac{(
 4 c_{E3} c_{E4} e^{(
  i ( \phi_{E3} -  \phi_{E4} + (-E_3 + E_4) t))} (U + \sqrt{[16 + (U - U_1)^2]} -
    U_1))}{( \sqrt{[32 + 2 (U + \sqrt{[16 + (U - U_1)^2]} - U_1)^2]}
   \sqrt{[-2 U + 2 (16 + \sqrt{[16 + (U - U_1)^2]} + U_1)]}   )} + \frac{(
 c_{E2} c_{E3} e^{(
  i (\phi_{E2} - \phi_{E3} + (-E_2 + E_3) t))} (-U + \sqrt{[16 + (U - U_1)^2]} +
     U_1))}{(\sqrt{2} \sqrt{[
  32 + 2 (-U + \sqrt{[16 + (U - U_1)^2]} + U_1)^2]})} + \nonumber \\
 +\frac{(
 4 c_{E3} c_{E4} e^{(
  i (-\phi_{E3} + \phi_{E4} + (E_3 - E_4) t))} (-U + \sqrt{[16 + (U - U_1)^2]} +
     U_1))}{(\sqrt{[32 + 2 (U + \sqrt{[16 + (U - U_1)^2]} - U_1)^2]} \sqrt{[
  32 + 2 (-U +\sqrt{[16 + (U - U_1)^2]} + U_1)^2]})}+  \nonumber \\
 - \frac{(
 4 c_{E3}^2 (-U + \sqrt{[16 + (U - U_1)^2]} + U_1))}{(
 \sqrt{[-2 U + 2 (16 + \sqrt{[16 + (U - U_1)^2]} + U_1)]} \sqrt{[
  32 + 2 (-U + \sqrt{[16 + (U - U_1)^2]} + U_1)^2]})}+ \nonumber \\
 + \frac{(
 2 \sqrt{2} c_{E3} c_{E4} ((-U + U_1) Cos[\phi_{E3} - \phi_{E4} - E_3 t + E_4 t] +
     i \sqrt{[16 + (U - U_1)^2]}
      Sin[\phi_{E3} - \phi_{E4} - E_3 t + E_4 t]))}{(
 \sqrt{16 + (U - U_1) (U + \sqrt{[16 + (U - U_1)^2]} - U_1)]} \sqrt{[
  16 - U + \sqrt{[16 + (U - U_1)^2]} + U_1]})}
$

Since $\rho_{B}(1,1)=\rho_{AB}(1,1)+\rho_{AB}(3,3)$, $\rho_{B}(2,2)=\rho_{AB}(2,2)+\rho_{AB}(4,4)$, $\rho_{B}(1,2)=\rho_{AB}(1,2)+\rho_{AB}(3,4)$, $\rho_{B}(2,1)=\rho_{AB}{2,1}+\rho_{AB}(4,3)$ we
obtain density matrix B for particle between 1' and 2' nodes in the form as given below. In particular the probability of finding electron B at node 1' is given as

$\rho_{B:(1,1)}=p(1')=\frac{c_{E2}^2}{2} + \frac{(4 c_{E4}^2)}{(
 16 + (U - U_1) (U + \sqrt{[16 + (U - U_1)^2]} - U_1))} + \frac{(8 c_{E3}^2)}{(
 16 - U + \sqrt{[16 + (U - U_1)^2]} + U_1)}+ $ \newline
  $ - \frac{(
 4 c_{E2} c_{E3} Cos[ \phi{E2} - \phi_{E3} - E_2 t + E_3 t])}{\sqrt{[
 16 - U + \sqrt{[16 + (U - U_1)^2]} + U_1]}}+$ \newline $ + [(
 2 c_{E1}^2 \sqrt{[16 + (U - U_1)^2]} +
  c_{E4}^2 (U + \sqrt{[16 + (U - U_1)^2]} - U_1) +
  c_{E3}^2 (-U + \sqrt{[16 + (U - U_1)^2]} + U_1) -
  2 \sqrt{2} c_{E1} (c_{E3} \sqrt{[
      16 - (U - U_1) (-U + \sqrt{[16 + (U - U_1)^2]} + U_1)]}
       Cos[ \phi_{E1} - \phi_{E3} - E_1 t + E_3 t] +
     c_{E4} \sqrt{[16 + (U - U_1) (U + \sqrt{[16 + (U - U_1)^2]} - U_1)]}
       Cos[\phi_{E1} - \phi_{E4} - E_1 t + E_4 t]))]\frac{1}{(
 4 \sqrt{[16 + (U - U_1)^2]})}+ $ \newline $ + \frac{(
 2 \sqrt{2} c_{E4} (c_{E2} \sqrt{[16 - U + \sqrt{[16 + (U - U_1)^2]} + U_1]}
      Cos[\phi_{E2} - \phi_{E4} - E_2 t + E_4 t] -
    4 c_{E3} Cos[\phi_{E3} - \phi_{E4} - E_3 t + E_4 t]))}{(
 \sqrt{[16 + (U - U_1) (U + \sqrt{[16 + (U - U_1)^2]} - U_1)]} \sqrt{[
  16 - U + \sqrt{[16 + (U - U_1)^2]} + U_1]})} +$ $ +\frac{(
 8 c_{E3} c_{E4} Cos[\phi_{E3} - \phi_{E4} + (-E_3 + E_4) t])}{(
 \sqrt{[16 + (U - U_1) (U + \sqrt{[16 + (U - U_1)^2]} - U_1)]} \sqrt{[
  16 - (U - U_1) (-U + \sqrt{[16 + (U - U_1)^2]} + U_1)]})}$ \newline
and probability of finding electron B at point 2' is given as \newline
$\rho_{B:(2,2)}=p(2')=c_{E2}^2/2 + \frac{(4 c_{E4}^2)}{(
 16 + (U - U_1) (U + \sqrt{[16 + (U - U_1)^2]} - U_1))} + $ \newline
  $ +\frac{(8 c_{E3}^2)}{(
 16 - U + \sqrt{[16 + (U - U_1)^2]} + U_1)} +$
 \newline $ +(
 c_{E1} c_{E3} (-U + \sqrt{[16 + (U - U_1)^2]} + U_1) Cos[
   \phi_{E1} - \phi_{E3} - E_1 t + E_3 t])/\sqrt{[
 16 - U + \sqrt{[16 + (U - U_1)^2]} + U_1]} + (
 4 c_{E2} c_{E3} Cos[\phi_{E2} - \phi_{E3} - E_2 t + E_3 t])/\sqrt{[
 16 - U + \sqrt{[16 + (U - U_1)^2]} + U_1]} - \frac{(
 2 \sqrt{2} c_{E4} (c_{E2} \sqrt{[16 - U + \sqrt{[16 + (U - U_1)^2]} + U_1]}
      Cos[ \phi_{E2} - \phi_{E4} - E_2 t + E_4 t] +
    4 c_{E3} Cos[ \phi_{E3} - \phi_{E4} - E_3 t + E_4 t]))}{(
 \sqrt{[16 + (U - U_1) (U + \sqrt{16 + (U - U_1)^2} - U_1)]} \sqrt{[
  16 - U + \sqrt{[16 + (U - U_1)^2]} + U_1]})} +
 \frac{1}{4}(2 c_{E1}^2 + c_{E4}^2 + (c_{E4}^2 (U - U_1))/\sqrt{[16 + (U - U_1)^2]} +
    2 c_{E3}^2 (-8 - U + \sqrt{[16 + (U - U_1)^2]}+ (
       16 \sqrt{[16 + (U - U_1)^2]})/(-15 + 2 U - 2 U_1) + U_1 + 136/(
       15 - 2 U + 2 U_1)) + \frac{(
    2 \sqrt{2}
      c_{E4} (c_{E1} (U + \sqrt{[16 + (U - U_1)^2]} - U_1) Cos[
         \phi_{E1} - \phi_{E4} - E_1 t + E_4 t] + (
       16 c_{E3} Cos[ \phi_{E3} - \phi_{E4} - E_3 t + E_4 t])/\sqrt{[
       16 - U + \sqrt{[16 + (U - U_1)^2]} + U_1]}))}{\sqrt{[
    16 + (U - U_1) (U + \sqrt{[16 + (U - U_1)^2]} - U_1)]}})$
The non-diagonal terms of density matrix B are given as \newline
$\rho_{B:(2,1)}=\frac{1}{2} c_{E1} c_{E2}  e^{(i (- \phi_{E1} + \phi_{E2} + (E_1 - E_2) t) )} + \frac{c_{E4}^2}{\sqrt{[
 16 + (U - U_1)^2]}}+ $ \newline $ - \frac{(
 2 \sqrt{2} c_{E1} c_{E4} e^{(i (- \phi_{E1} + \phi_{E4} + (E1 - E4) t) )} )}{\sqrt{[
 32 + 2 (U + \sqrt{[16 + (U - U_1)^2]} - U_1)^2]}} - \frac{(
 c_{E2} c_{E4} e^{(
  i ( \phi_{E2} - \phi_{E4} + (-E_2 + E_4) t))} (U + \sqrt{[16 + (U - U_1)^2]} -
    U_1))}{(\sqrt{[2]} \sqrt{[32 + 2 (U + \sqrt{[16 + (U - U_1)^2]} - U_1)^2]})}+$ \newline $ + (
 2 c_{E1} c_{E3} e^{(i (- \phi_{E1} + \phi_{E3} + (E_1 - E_3) t ) )}  )/\sqrt{[
 16 - U + \sqrt{[16 + (U - U_1)^2]} + U_1]}+ $ \newline
  $- \frac{(
 4 c_{E3} c_{E4} e^(
  i (\phi_{E3} - \phi_{E4} + (-E_3 + E_4) t)) (U + \sqrt{[16 + (U - U_1)^2]} -
    U_1))}{(\sqrt{[32 + 2 (U + \sqrt{[16 + (U - U_1)^2]} - U_1)^2]}
   \sqrt{[-2 U + 2 (16 + \sqrt{[16 + (U - U_1)^2]} + U_1)]})} - \frac{(
 c_{E2} c_{E3} e^{(
  i (\phi_{E2} - \phi_{E3} + (-E_2 + E_3) t))} (-U + \sqrt{[16 + (U - U1)^2]} +
     U_1))}{(\sqrt{2} \sqrt{[
  32 + 2 (-U + \sqrt{[16 + (U - U_1)^2]} + U_1)^2]})} + \frac{(
 4 c_{E3} c_{E4} e^{(
  i (-\phi_{E3} + \phi_{E4} + (E_3 - E_4) t))} (-U + \sqrt{[16 + (U - U_1)^2]} +
     U_1))}{(\sqrt{[32 + 2 (U + \sqrt{[16 + (U - U_1)^2]} - U_1)^2]} \sqrt{[
  32 + 2 (-U + \sqrt{[16 + (U - U_1)^2]} + U_1)^2}))}+$ \nonumber \\  $- \frac{(
 4 c_{E3}^2 (-U + \sqrt{[16 + (U - U_1)^2]} + U_1))}{(
 \sqrt{[-2 U + 2 (16 + \sqrt{[16 + (U - U_1)^2]} + U_1)]} \sqrt{[
  32 + 2 (-U + \sqrt{[16 + (U - U_1)^2]} + U_1)^2]})} +$ \newline
 $+1/4 ( \frac{(c_{E4} e^(-i (\phi_{E2} + E_4 t)) (4 c_{E4} e^{(i (\phi_{E2} + E4 t))} +
       Sqrt[2] c_{E2} e^{(i (\phi_{E4} + E_2 t))} \sqrt{[
        16 + (U - U_1) (U + \sqrt{[16 + (U - U_1)^2]} - U_1)]}))}{\sqrt{[
    16 + (U - U_1)^2]}} + \frac{(
    8 c_{E3}^2 (-1 - U + \sqrt{[16 + (U - U_1)^2]} + U_1))}{(-15 + 2 U -
     2 U_1)} + $ $ +\frac{(
    2 c_{E2} c_{E3} e^{(
     i (-\phi_{E2} + \phi_{E3} + E_2 t - E_3 t))} (-U + \sqrt{[
       16 + (U - U_1)^2]} + U_1))}{\sqrt{[
    16 - U + \sqrt{[16 + (U - U_1)^2]} + U_1]}} +
    2 c_{E1} e^{(
     i (\phi_{E1} - E_1 t))} (c_{E2} e^{(i (- \phi_{E2} + E2 t))} + \frac{(
       2 \sqrt{2} c_{E4} e^{(i (-\phi_{E4} + E_4 t) )}
)}{\sqrt{[
       16 + (U - U_1) (U + \sqrt{[16 + (U - U_1)^2]} - U_1)]}} -
       \frac{(
       4 c_{E3} e^{(i (-\phi_{E3} + E_3 t))} )}{\sqrt{[
       16 - U + \sqrt{[16 + (U - U_1)^2]} + U_1]}})+$ \newline $+ \frac{(
 2 \sqrt{2} c_{E3} c_{E4} ((-U + U_1) Cos[\phi_{E3} - \phi_{E4} - E_3 t + E_4 t ] +
     i \sqrt{[16 + (U - U_1)^2]}
      Sin[ \phi_{E3} - \phi_{E4} - E_3 t + E_4 t]))}{(
 \sqrt{[16 + (U - U_1) (U + \sqrt{[16 + (U - U_1)^2]} - U_1)]} \sqrt{[
  16 - U + \sqrt{16 + (U - U_1)^2} + U_1]})}$ \newline
and another non-diagonal element of single particle density matrix is of the form \newline
$\rho_{B:(1,2)}=
\frac{1}{2} c_{E1} c_{E2} e^{(i (-\phi_{E1} + \phi_{E2} + (E_1 - E_2) t))} +
 \frac{1}{2} c_{E1} c_{E2} e^{(i ( \phi_{E1} - \phi_{E2} + (-E_1 + E_2) t))} + \frac{(
 2 c_{E4}^2)}{\sqrt{[16 + (U - U_1)^2]}} + \frac{(
 2 \sqrt{2} c_{E}1 c_{E4} e^{(i (- \phi_{E1} + \phi_{E4} + (E_1 - E_4) t) )}
)}{\sqrt{[
 32 + 2 (U + \sqrt{[16 + (U - U_1)^2]} - U_1)^2]}} - \frac{(
 2 \sqrt{2} c_{E1} c_{E4} e^{(i (\phi_{E1} - \phi_{E4} + (-E_1 + E_4) t) )} )}{\sqrt{[
 32 + 2 (U + \sqrt{[16 + (U - U_1)^2]} - U_1)^2]}} - \frac{(
 c_{E2} c_{E4} e^{(
  i (- \phi_{E2} + \phi_{E4} + (E_2 - E_4) t))} (U + \sqrt{[16 + (U - U_1)^2]} -
    U_1))}{(\sqrt{2} \sqrt{[32 + 2 (U + \sqrt{[16 + (U - U_1)^2]} - U_1)^2]} )} + \frac{(
 c_{E2} c_{E4} e^{(
  i (\phi_{E2} - \phi_{E4} + (-E_2 + E_4) t))} (U + \sqrt{[16 + (U - U_1)^2]} -
    U_1))}{(\sqrt{2} \sqrt{[32 + 2 (U + \sqrt{[16 + (U - U_1)^2]} - U_1)^2]} )} + \frac{(
 2 c_{E3}^2 (-1 - U + \sqrt{[16 + (U - U_1)^2]} + U_1))}{(-15 + 2 U - 2 U_1)} - \frac{(
 2 c_{E1} c_{E3} e^{(i (- \phi_{E1} + \phi_{E3} + (E_1 - E_3) t) )} )}{\sqrt{[
 16 - U + \sqrt{[16 + (U - U_1)^2]} + U_1]}} + \frac{(
 2 c_{E1} c_{E3} e^{(i ( \phi_{E1} - \phi_{E3} + (-E_1 + E_3) t) )} )}{\sqrt{[
 16 - U + \sqrt{[16 + (U - U_1)^2]} + U_1}} + \frac{(
 c_{E2} c_{E3} e^{(
  i ( \phi_{E2} - \phi_{E3} + (-E_2 + E_3) t))} (-U + \sqrt{[16 + (U - U_1)^2]} +
     U_1))}{(2 \sqrt{[16 - U + \sqrt{[16 + (U - U_1)^2]} + U_1]})} - \frac{(
 4 c_{E3} c_{E4} e^{(
  i (-\phi_{E3} + \phi_{E4} + (E_3 - E_4) t ))} (U + \sqrt{[16 + (U - U_1)^2]} -
    U_1))}{( \sqrt{[32 + 2 (U + \sqrt{[16 + (U - U_1)^2]} - U_1)^2]}
   \sqrt{[-2 U + 2 (16 + \sqrt{[16 + (U - U_1)^2]} + U_1)]})} - \frac{(
 c_{E2} c_{E3} e^{(
  i (-\phi_{E2} + \phi_{E3} + (E_2 - E_3) t))} (-U + \sqrt{[16 + (U - U_1)^2]} +
     U_1))}{(  \sqrt{2} \sqrt{[
  32 + 2 (-U + \sqrt{[16 + (U - U_1)^2]} + U_1)^2]})} + \frac{(
 4 c_{E3} c_{E4} e^{(
  i (\phi_{E3} - \phi_{E4} + (-E_3 + E_4) t))} (-U + \sqrt{[16 + (U - U_1)^2]} +
     U_1))}{(\sqrt{[32 + 2 (U + \sqrt{[16 + (U - U_1)^2]} - U_1)^2]} \sqrt{[
  32 + 2 (-U + \sqrt{[16 + (U - U_1)^2]} + U_1)^2]})}+ $ \newline $- \frac{(
 4 c_{E3}^2 (-U + \sqrt{[16 + (U - U_1)^2]} + U_1))}{(
 \sqrt{[-2 U + 2 (16 + \sqrt{[16 + (U - U_1)^2]} + U_1)]} \sqrt{[
  32 + 2 (-U + \sqrt{[16 + (U - U_1)^2]} + U_1)^2]})}+ $ \newline $- \frac{(
 2 \sqrt{2} c_{E3} c_{E4} ((U - U_1) Cos[\phi_{E3} - \phi_{E4} - E_3 t + E_4 t] +
    i \sqrt{[16 + (U - U_1)^2]} Sin[\phi_{E3} - \phi_{E4} - E_3 t + E_4 t]))}{(
 \sqrt{[16 + (U - U_1) (U + \sqrt{[16 + (U - U_1)^2]} - U_1)]} \sqrt{[
  16 - U + \sqrt{[16 + (U - U_1)^2]} + U_1]})}
$

\subsubsection{Correlation function for case III: $E_f(1,1')=E_f(2,1')=U$, $E_f(1,2')=E_f(2,2')=U_1$ }
For the case 3 we have
\begin{eqnarray}
C(t,\phi_{E_1},\phi_{E_2},\phi_{E_3},\phi_{E_4},U,U_1,E_1(U,U_1),E_2(U,U_1),E_3(U,U_1),E_4(U,U_1))= \nonumber \\
(2 (c_{E1}c_{E2}\sqrt{(U-U_1)^2+4}(U-U_1) \cos (-E_{1n}t+E_{2n}t+\phi_{E1}-\phi_{E2})+ \nonumber \\
+c_{E4} (c_{E1} \sqrt{(U-U_1) \left(\sqrt{(U-U_1)^2+4}+U-U_1\right)+4} \sqrt{4-(U-U_1) \left(\sqrt{(U-U_1)^2+4}-U+U_1\right)}\times \nonumber \\ \times  \cos
   ((- E_{1}+E_4)t+\phi_{E1}-\phi_{E4})+  \nonumber \\
   +c_{E3} \sqrt{(U-U_1)^2+4} (U_1-U) \cos ( (- E_3+E_{4})t
   +\phi_{E3}-\phi_{E4}))+ \nonumber \\
   +c_{E2} c_{E3} \sqrt{(U-U_1) \left(\sqrt{(U-U_1)^2+4}+U-U_1\right)+4} \sqrt{4-(U-U_1)
   \left(\sqrt{(U-U_1)^2+4}-U+U_1\right)} \times \nonumber \\
 \times \cos ((-E_{2}+E_{3})t+\phi_{E2}-\phi_{E3}))) \frac{1}{(U-U_1)^2+4},
\end{eqnarray}
where
\begin{eqnarray}H=
\begin{pmatrix}
U & 1 & 1 & 0 \\
1 & U_1 & 0 & 1 \\
1 & 0 & U & 1 \\
0 & 1 & 1 & U_1
\end{pmatrix}. \nonumber \\
\end{eqnarray}
and
\begin{eqnarray}
U_1=\frac{q^2}{\sqrt{(d+Cos(\alpha)(a+b))^2+(1+Sin(\alpha))^2(a+b)^2}}+E_{p1}+E_{p2'}, \nonumber
U=\frac{q^2}{d}+E_{p2}+E_{p}(1'),
\end{eqnarray}
\begin{eqnarray}
E_1=\frac{-2+U+U_1-\sqrt{(4)^2+(U-U_1)^2}}{2},  \nonumber \\
E_2=\frac{+2+U+U_1-\sqrt{(4)^2+(U-U_1)^2}}{2},  \nonumber \\
E_3=\frac{-2+U+U_1+\sqrt{(4)^2+(U-U_1)^2}}{2},   \nonumber \\
E_4=\frac{+2+U+U_1+\sqrt{(4)^2+(U-U_1)^2}}{2}.
\end{eqnarray}

Now we generalize the Hamiltonian matrix of the form
\begin{eqnarray}H=
\begin{pmatrix}
U & t_{s2} & t_{s1} & 0 \\
t_{s2} & U_1 & 0 & t_{s1} \\
t_{s1} & 0 & U & t_{s2} \\
0 & t_{s1} & t_{s2} & U_1
\end{pmatrix}. \nonumber \\
\end{eqnarray}
and energy eigenvalues depending on $t_{s1}$, $t_{s2}$, $U$ and $U_1$ are given as
\begin{eqnarray}
E_1=\frac{1}{2}(-2t_{s1}+U+U_1-\sqrt{(U-U_1)^2+(2t_{s2})^2}), \nonumber \\
E_2=\frac{1}{2}(+2t_{s1}+U+U_1-\sqrt{(U-U_1)^2+(2t_{s2})^2}), \nonumber \\
E_3=\frac{1}{2}(-2t_{s1}+U+U_1+\sqrt{(U-U_1)^2+(2t_{s2})^2}), \nonumber \\
E_4=\frac{1}{2}(+2t_{s1}+U+U_1+\sqrt{(U-U_1)^2+(2t_{s2})^2}).
\end{eqnarray}
It is surprising to discover that the corresponding energy eigenstates are depending only on $t_{s2}$ hopping constant and are not depending on $t_{s1}$ hopping constant and are given as
\begin{eqnarray}
\ket{E_1}=
\begin{pmatrix}
\frac{\sqrt{[8 t_{s2}^2 + 2 (-U + \sqrt{[4 t_{s2}^2 + (U - U_1)^2]} + U_1)^2]}}{(
  4 \sqrt{[4 t_{s2}^2 + (U - U_1)^2]})}, \\
  -(\frac{(2 t_{s2})}{\sqrt{[
   8 t_{s2}^2 + 2 (-U + \sqrt{[4 t_{s2}^2 + (U - U_1)^2]} + U_1)^2]}}), \\
   -(\frac{\sqrt{[
   8 t_{s2}^2 + 2 (-U + \sqrt{[4 t_{s2}^2 + (U - U_1)^2]} + U_1)^2]}}{(
   4 \sqrt{[4 t_{s2}^2 + (U - U_1)^2]})} ), \\
   \frac{(2 t_{s2})}{\sqrt{[
  8 t_{s2}^2 + 2 (-U + \sqrt{[4 t_{s2}^2 + (U - U_1)^2]} + U_1)^2]}},
  \end{pmatrix}, \nonumber \\
\ket{E_2}=
\begin{pmatrix}
-\frac{\sqrt{[
   8 t_{s2}^2 + 2 (-U + \sqrt{[4 t_{s2}^2 + (U - U_1)^2]} + U_1)^2]}}{(
   4 \sqrt{[4t_{s2}^2 + (U - U_1)^2]})}, \\
+\frac{2 t_{s2} }{\sqrt{
  8 t_{s2}^2 + 2 (-U + \sqrt{[4 t_{s2}^2 + (U - U_1)^2]} + U_1)^2}}, \\
  -\frac{\sqrt{
   8 t_{s2}^2 + 2 (-U + \sqrt{[4 t_{s2}^2 + (U - U_1)^2]} + U_1)^2}}{(
   4 \sqrt{[4 t_{s2}^2 + (U - U_1)^2]})}, \\
 +  \frac{2t_{s2}}{\sqrt{[
  8 t_{s2}^2 + 2 (-U + \sqrt{[4 t_{s2}^2 + (U - U_1)^2]} + U_1)^2]}},
  \end{pmatrix}, \nonumber \\
\ket{E_3}=
\begin{pmatrix}
-\frac{\sqrt{[8 t_{s2}^2 + 2 (U + \sqrt{[4 t_{s2}^2 + (U - U_1)^2]} - U_1)^2]}}{(
  4 \sqrt{[4 t_{s2}^2 + (U - U_1)^2]})}, \\
   -\frac{((2 t_{s2})}{\sqrt{[
  8 t_{s2}^2 + 2 (U + \sqrt{[4t_{s2}^2 + (U - U_1)^2]} - U_1)^2]}}),   \\
  \frac{\sqrt{[
 8 t_{s2}^2 + 2 (U + \sqrt{[4 t_{s2}^2 + (U - U_1)^2]} - U_1)^2]}}{(
 4 \sqrt{[4 t_{s2}^2 + (U - U_1)^2]})},  \\
 \frac{2 t_{s2}}{\sqrt{[
 8 t_{s2}^2 + 2 (U + \sqrt{[4 t_{s2}^2 + (U - U_1)^2]} - U_1)^2]}}
\end{pmatrix}
,  \\
\ket{E_4}=
\begin{pmatrix}
\frac{\sqrt{[8 t_{s2}^2 + 2 t_{s2}^2 (U + \sqrt{[4 t_{s2}^2 + (U - U_1)^2]} - U_1)^2]}}{(
 4 \sqrt{[4 t_{s2}^2 + (U - U_1)^2]})}, \\
 \frac{2 t_{s2}}{\sqrt{[
 8 t_{s2}^2 + 2 t_{s2}^2 (U + \sqrt{[4 t_{s2}^2 + (U - U_1)^2]} - U_1)^2]}}, \\
 \frac{\sqrt{[
 8 t_{s2}^2 + 2 t_{s2}^2 (U + \sqrt{[4 t_{s2}^2 + (U - U_1)^2]} - U_1)^2]}}{(
 4 \sqrt{[4 t_{s2}^2 + (U - U_1)^2]})}, \\
 \frac{2 t_{s2}}{\sqrt{[
 8 t_{s2}^2 + 2 t_{s2}^2 (U + \sqrt{[4 t_{s2}^2 + (U - U_1)^2]} - U_1)^2]}}
 \end{pmatrix}.
\end{eqnarray}

The minimalist density matrix of two electrostatically interacting qubits A and B in tight binding model in case of time-independent Hamiltonian is given as
\begin{eqnarray}
\rho_{AB}(t)= |c_{E1}|^2 \ket{E_1}\bra{E_1}+|c_{E2}|^2 \ket{E_2}\bra{E_2}+|c_{E3}|^2 \ket{E_3}\bra{E_3}+|c_{E4}|^2 \ket{E_4}\bra{E_4}+ \nonumber \\
+c_{E1}c_{E2}(\ket{E_1}\bra{E_2}e^{i(\phi_{E1(t_0)}-\phi_{E2(t_0)})}e^{\frac{1}{i\hbar}(E_1-E_2)(t-t_0)}+ \ket{E_2}\bra{E_1})e^{i(\phi_{E2(t_0)}-\phi_{E1(t_0)})}e^{\frac{1}{i\hbar}(E_2-E_1)(t-t_0)} + \nonumber \\ +c_{E1}c_{E3}(e^{i(\phi_{E1(t_0)}-\phi_{E3(t_0)})}e^{\frac{1}{i\hbar}(E_1-E_3)(t-t_0)}\ket{E_1}\bra{E_3} + e^{i(\phi_{E3(t_0)}-\phi_{E1(t_0)})}e^{\frac{1}{i\hbar}(E_3-E_1)(t-t_0)}\ket{E_3}\bra{E_1}) + \nonumber \\
+c_{E1}c_{E4}(e^{i(\phi_{E1(t_0)}-\phi_{E4(t_0)})}e^{\frac{1}{i\hbar}(E_1-E_4)(t-t_0)}\ket{E_1}\bra{E_4} + e^{i(\phi_{E4(t_0)}-\phi_{E1(t_0)})}e^{\frac{1}{i\hbar}(E_4-E_1)(t-t_0)}\ket{E_4}\bra{E_1})+ \nonumber \\
+c_{E2}c_{E3}(e^{i(\phi_{E2(t_0)}-\phi_{E3(t_0)})}e^{\frac{1}{i\hbar}(E_2-E_3)(t-t_0)}\ket{E_2}\bra{E_3}  
+e^{i(\phi_{E3(t_0)}-\phi_{E2(t_0)})}e^{\frac{1}{i\hbar}(E_3-E_2)(t-t_0)}\ket{E_3}\bra{E_2})+ \nonumber \\
+c_{E2}c_{E4}(e^{i(\phi_{E2(t_0)}-\phi_{E4(t_0)})}e^{\frac{1}{i\hbar}(E_2-E_4)(t-t_0)}\ket{E_2}\bra{E_4}  
+e^{i(\phi_{E4(t_0)}-\phi_{E2(t_0)})}e^{\frac{1}{i\hbar}(E_4-E_2)(t-t_0)}\ket{E_4}\bra{E_2})+ \nonumber \\
c_{E3}c_{E4}(e^{i(\phi_{E3(t_0)}-\phi_{E4(t_0)})}e^{\frac{1}{i\hbar}(E_3-E_4)(t-t_0)}\ket{E_3}\bra{E_4}  
+e^{i(\phi_{E4(t_0)}-\phi_{E3(t_0)})}e^{\frac{1}{i\hbar}(E_4-E_3)(t-t_0)}\ket{E_4}\bra{E_3}),
\end{eqnarray}
where $|c_{E1}|^2$, $|c_{E2}|^2$, $|c_{E3}|^2$ and $|c_{E4}|^2$ are probabilities of occupancy of eigenenergies $E_1$, $E_2$, $E_3$ and $E_4$ and
we obtain two particle  density matrix diagonal elements in the detailed form given below. We have probability of finding electron A at node 1 and electron B at node 1' given by formula
\small
\begin{eqnarray*}
\rho(1,1)=p(1,1',t)=
\frac{1}{4}[c_{E2}^2 + c_{E3}^2 + \frac{(c_{E4}^2 t_{s2}^2 (2 + 2 t_{s2}^2 + (U - U_1)^2))}{(
    4 t_{s2}^2 + (U - U_1)^2)} ]\nonumber \\
+\frac{(-(c_{E2}^2 - c_{E3}^2 - c_{E4}^2 t_{s2}^2) (U -
       U_1) + c_{E1}^2 (-U + \sqrt{[4 t_{s2}^2 + (U - U_1)^2]} + U_1) )}{(4 \sqrt{[
    4 t_{s2}^2 + (U - U_1)^2]})}\nonumber \\
%
 - \frac{(c_{E1} c_{E2} (-U + \sqrt{[
      4 t_{s2}^2 + (U - U_1)^2]} + U_1) Cos[
     \phi_{E1} -  \phi_{E2} + (-E_1 + E_2) t] )}{( 2 \sqrt{[
    4 t_{s2}^2 + (U - U_1)^2]})}\nonumber \\
 -\frac{(c_{E1} c_{E3} \sqrt{[
    8 t_{s2}^2 + 2 (U  + \sqrt{[4 t_{s2}^2 + (U - U_1)^2]} - U_1)^2]} \sqrt{[
    8 t_{s2}^2 + 2 (-U + \sqrt{[4 t_{s2}^2 + (U - U_1)^2]} + U_1)^2]}
     Cos[\phi_{E1} -
      \phi_{E3} + (-E_1 + E_3) t] )}{( 8 (4 t_{s2}^2 + (U -
        U_1)^2))} \nonumber \\
+ \frac{(c_{E2} c_{E3} \sqrt{[
    8 t_{s2}^2 + 2 (U + \sqrt{[4 t_{s2}^2 + (U - U_1)^2]} - U_1)^2]} \sqrt{[
    8 t_{s2}^2 + 2 (-U + \sqrt{[4 t_{s2}^2 + (U - U_1)^2]} + U_1)^2]}
     Cos[ \phi_{E2} -
      \phi_{E3} + (-E_2 + E_3) t])}{(8 (4 t_{s2}^2 + (U -
        U_1)^2) )}\nonumber \\
 + \frac{(c_{E1} c_{E4}  \sqrt{[
    t_{s2}^2 (4 + (U + \sqrt{[4 t_{s2}^2 + (U - U_1)^2]} - U_1)^2)]} \sqrt{[
    8 t_{s2}^2 + 2 (-U + \sqrt{[4 t_{s2}^2 + (U - U_1)^2]} + U_1)^2]}
     Cos[\phi_{E1} - \phi_{E4} + (-E_1 + E_4) t])}{(4 \sqrt{2} (4 t_{s2}^2 + (U - U_1)^2))} \nonumber \\
- \frac{(c_{E2} c_{E4} \sqrt{[
    t_{s2}^2 (4 + (U + \sqrt{[4 t_{s2}^2 + (U - U_1)^2]} - U_1)^2)]} \sqrt{[
    8 t_{s2}^2 + 2 (-U + \sqrt{[4 t_{s2}^2 + (U - U_1)^2]} + U_1)^2]}
     Cos[\phi_{E2} - \phi_{E4} + (-E_2 + E_4) t] )}{( 4 \sqrt{
    2} (4 t_{s2}^2 + (U - U_1)^2) )} \nonumber \\  
- \frac{(c_{E3} c_{E4} \sqrt{[
    t_{s2}^2 (4 + (U + \sqrt{[4 t_{s2}^2 + (U - U_1)^2]} - U_1)^2)]} \sqrt{[
    8 t_{s2}^2 + 2 (U + \sqrt{[4 t_{s2}^2 + (U - U_1)^2]} - U_1)^2 ]}
     Cos[ \phi_{E3} - \phi_{E4} + (-E_3 + E_4) t] )}{( 4 \sqrt{
    2} (4 t_{s2}^2 + (U - U_1)^2) )}
\end{eqnarray*}
and probability of finding electron A at node 1 and electron B at node 2' given by formula
\normalsize
\begin{eqnarray}
\rho(2,2)=p(1,2',t)=
\frac{(2 c_{E4}^2)}{(4 + (U + \sqrt{[4 t_{s2}^2 + (U - U_1)^2]} - U_1)^2)}+ \nonumber \\
+ \frac{(4 c_{E3}^2 t_{s2}^2)}{(
 8 t_{s2}^2 +
  2 (U + \sqrt{[4 t_{s2}^2 + (U - U_1)^2]} - U_1)^2)} + \frac{((c_{E1}^2 +
    c_{E2}^2) t_{s2}^2)}{(
 4 t_{s2}^2 - (U - U_1) (-U + \sqrt{[4 t_{s2}^2 + (U - U_1)^2]} + U_1))} \nonumber \\
- \frac{(
 2 c_{E1} c_{E2} t_{s2}^2 Cos[\phi_{E1} - \phi_{E2} + (-E_1 + E_2) t])}{(
 4 t_{s2}^2 - (U - U_1) (-U + \sqrt{[4 t_{s2}^2 + (U - U_1)^2]} +
     U_1))} + \nonumber \\
+\frac{(2 c_{E1} c_{E3} t_{s2}^2 Cos[
     \phi_{E1} - \phi_{E3} + (-E_1 + E_3) t])}{(\sqrt{[
    4 t_{s2}^2 + (U - U_1) (U + \sqrt{[4 t_{s2}^2 + (U - U_1)^2]} - U_1)]} \sqrt{[
    4 t_{s2}^2 - (U - U_1) (-U + \sqrt{[4 t_{s2}^2 + (U - U_1)^2]} +
        U_1)]})} \nonumber \\
 - \frac{(2 c_{E2} c_{E3} t_{s2}^2 Cos[
     \phi_{E2} - \phi_{E3} + (-E_2 + E_3) t])}{(\sqrt{[
    4 t_{s2}^2 + (U - U_1) (U + \sqrt{[4 t_{s2}^2 + (U - U_1)^2]} - U_1)]} \sqrt{[
    4 t_{s2}^2 - (U - U_1) (-U + \sqrt{[4 t_{s2}^2 + (U - U_1)^2]} +
        U_1)]})} + \nonumber \\
 - \frac{(2 c_{E1} c_{E4} t_{s2}^2 Cos[
     \phi_{E1} - \phi_{E4} + (-E_1 + E_4) t])}{(\sqrt{[
    t_{s2}^2 (2 +
       2 t_{s2}^2 + (U - U_1) (U + \sqrt{[4 t_{s2}^2 + (U - U_1)^2]} - U_1))]}
     \sqrt{[4 t_{s2}^2 - (U - U_1) (-U + \sqrt{[4 t_{s2}^2 + (U - U_1)^2]} +
        U_1)]})} + \nonumber \\
+ \frac{(2 c_{E2} c_{E4} t_{s2}^2 Cos[
     \phi_{E2} - \phi_{E4} + (-E_2 + E_4) t])}{(\sqrt{[
    t_{s2}^2 (2 +
       2 t_{s2}^2 + (U - U_1) (U + \sqrt{[4 t_{s2}^2 + (U - U_1)^2]} - U_1))]}
     \sqrt{[4 t_{s2}^2 - (U - U_1) (-U + \sqrt{[4 t_{s2}^2 + (U - U_1)^2]} +
        U_1)]})} + \nonumber \\
- \frac{(2 c_{E3} c_{E4} t_{s2}^2 Cos[
     \phi_{E3} - \phi_{E4} + (-E_3 + E_4) t])}{(\sqrt{[
    t_{s2}^2 (2 +
       2 t_{s2}^2 + (U - U_1) (U + \sqrt{[4 t_{s2}^2 + (U - U_1)^2]} - U_1))]}
     \sqrt{[4 t_{s2}^2 + (U - U_1) (U + \sqrt{[4 t_{s2}^2 + (U - U_1)^2]} - U_1)]})}
\end{eqnarray}
\small
and probability of finding electron A at node 2 and electron B at node 1' given as
\begin{eqnarray*}
\rho(3,3)=p(2,1',t)=
\frac{1}{4} (c_{E2}^2 + c_{E3}^2 + \frac{(c_{E4}^2 t_{s2}^2 (2 + 2 t_{s2}^2 + (U - U_1)^2))}{(
    4 t_{s2}^2 + (U - U_1)^2)}) + \nonumber \\
 + \frac{(-(c_{E2}^2 - c_{E3}^2 - c_{E4}^2 t_{s2}^2) (U -
       U_1) + c_{E1}^2 (-U + \sqrt{[4 ts2^2 + (U - U_1)^2]} + U_1))}{(4 \sqrt{[
    4 t_{s2}^2 + (U - U_1)^2]} )}+   \nonumber \\
 + \frac{(c_{E1}c_{E2} (-U + \sqrt{[
      4 t_{s2}^2 + (U - U_1)^2]} + U_1) Cos[
     \phi_{E1} - \phi_{E2} + (-E_1 + E_2) t])}{(2 \sqrt{[
    4 t_{s2}^2 + (U - U_1)^2]})}+ \nonumber \\
- \frac{(c_{E1} c_{E3} \sqrt{[
    8 t_{s2}^2 + 2 (U + \sqrt{[4 t_{s2}^2 + (U - U_1)^2]} - U_1)^2]} \sqrt{[
    8 t_{s2}^2 + 2 (-U + \sqrt{[4 t_{s2}^2 + (U - U1)^2]} + U_1)^2]}
     Cos[\phi_{E1} -
      \phi_{E3} + (-E_1 + E_3) t])}{(8 (4 t_{s2}^2 + (U -
        U_1)^2))} \nonumber \\
 - \frac{(c_{E2} c_{E3} \sqrt{[
    8 t_{s2}^2 + 2 (U + \sqrt{[4 t_{s2}^2 + (U - U_1)^2]} - U_1)^2]} \sqrt{[
    8 t_{s2}^2 + 2 (-U + \sqrt{[4 t_{s2}^2 + (U - U_1)^2]} + U_1)^2]}
     Cos[\phi_{E2} - \phi_{E3} + (-E_2 + E_3) t])}{(8 (4 t_{s2}^2 + (U -
        U_1)^2))} \nonumber \\
- \frac{( c_{E1} c_{E4} \sqrt{[
    t_{s2}^2 (4 + (U + \sqrt{[4 t_{s2}^2 + (U - U_1)^2]} - U_1)^2)]} \sqrt{[
    8 t_{s2}^2 + 2 (-U + \sqrt{[4 t_{s2}^2 + (U - U_1)^2]} + U_1)^2]}
     Cos[\phi_{E1} - \phi_{E4} + (-E_1 + E_4) t])} {(4 \sqrt{
    2} (4 t_{s2}^2 + (U - U_1)^2))} \nonumber \\
- \frac{(c_{E2} c_{E4} \sqrt{[
    t_{s2}^2 (4 + (U + \sqrt{[4 t_{s2}^2 + (U - U_1)^2]} - U_1)^2)]} \sqrt{[
    8 t_{s2}^2 + 2 (-U + \sqrt{[4 t_{s2}^2 + (U - U_1)^2]} + U_1)^2]}
     Cos[\phi_{E2} - \phi_{E4} + (-E_2 + E_4) t])}{(4 \sqrt{
    2} (4 t_{s2}^2 + (U - U_1)^2))} \nonumber \\
+ \frac{(c_{E3} c_{E4} \sqrt{[
    t_{s2}^2 (4 + (U + \sqrt{[4 t_{s2}^2 + (U - U_1)^2]} - U_1)^2)]} \sqrt{[
    8 t_{s2}^2 + 2 (U + \sqrt{[4 t_{s2}^2 + (U - U_1)^2]} - U_1)^2]}
     Cos[ \phi_{E3} - \phi_{E4} + (-E_3 + E_4) t])}{(4 \sqrt{
    2} (4 t_{s2}^2 + (U - U1)^2) )} \nonumber \\
\end{eqnarray*}
and probability of finding electron A at 2 node and electron B at node 2' as
\small
\begin{eqnarray}
\rho(4,4,t)=p(2,2',t)=\frac{(2 c_{E4}^2)}{(4 + (U + \sqrt{[4 t_{s2}^2 + (U - U1)^2]} - U_1)^2)} + \nonumber \\
+ \frac{(
 4 c_{E3}^2 t_{s2}^2)}{(
 8 t_{s2}^2 +
  2 (U + \sqrt{[4 t_{s2}^2 + (U - U_1)^2]} - U_1)^2)}+ \nonumber \\
+ \frac{((c_{E1}^2 +
    c_{E2}^2) t_{s2}^2)}{(
 4 t_{s2}^2 - (U - U_1) (-U + \sqrt{4 t_{s2}^2 + (U - U_1)^2} + U_1))}+ \nonumber \\
  + \frac{(
 2 c_{E1}c_{E2} t_{s2}^2 Cos[\phi_{E1} - \phi_{E2} + (-E_1 + E_2) t])}{(
 4 t_{s2}^2 - (U - U_1) (-U + \sqrt{[4 t_{s2}^2 + (U - U_1)^2]} +
     U_1))}+ \nonumber \\
     + \frac{(2 c_{E1}c_{E3} t_{s2}^2 Cos[\phi_{E1} - \phi_{E3} + (-E_1 + E_3) t])}{(\sqrt{[
    4 t_{s2}^2 + (U - U_1) (U + \sqrt{[4 t_{s2}^2 + (U - U_1)^2]} - U_1)]} \sqrt{[
    4 t_{s2}^2 - (U - U_1) (-U + \sqrt{[4 t_{s2}^2 + (U - U_1)^2]} +
        U_1)]})}+ \nonumber \\
          + \frac{(2 c_{E2} c_{E3} t_{s2}^2 Cos[
     \phi_{E2} - \phi_{E3} + (-E_2 + E_3) t])}{(\sqrt{[
    4 t_{s2}^2 + (U - U1) (U + \sqrt{[4 t_{s2}^2 + (U - U_1)^2]} - U_1)]} \sqrt{[
    4 t_{s2}^2 - (U - U_1) (-U + \sqrt{[4 t_{s2}^2 + (U - U_1)^2]} +
        U1)}])}+ \nonumber \\
         + \frac{(2 c_{E1}c_{E4} t_{s2}^2 Cos[
     \phi_{E1} - \phi_{E4} + (-E_1 + E_4) t])}{(\sqrt{[
    t_{s2}^2 (2 +
       2 t_{s2}^2 + (U - U_1) (U + \sqrt{[4 t_{s2}^2 + (U - U_1)^2]} - U_1))]}
     \sqrt{[4 t_{s2}^2 - (U - U_1) (-U + \sqrt{[4 ts2^2 + (U - U_1)^2]} +
        U1)]})} + \nonumber \\
        +\frac{(2 c_{E2} c_{E4} t_{s2}^2 Cos[
     \phi_{E2} - \phi_{E4} + (-E_2 + E_4) t])}{(\sqrt{[
    t_{s2}^2 (2 +
       2 t_{s2}^2 + (U - U_1) (U + \sqrt{[4 ts2^2 + (U - U1)^2]} - U_1))]}
     \sqrt{[4 t_{s2}^2 - (U - U_1) (-U + \sqrt{[4 t_{s2}^2 + (U - U_1)^2]} +
        U_1)]})} + \nonumber \\
        +\frac{(2 c_{E3} c_{E4} t_{s2}^2 Cos[
     \phi_{E3} - \phi_{E4} + (-E_3 + E_4) t])}{(\sqrt{[
    t_{s2}^2 (2 +
       2 t_{s2}^2 + (U - U_1) (U + \sqrt{[4t_{s2}^2 + (U - U_1)^2]} - U_1))]}
     \sqrt{[4 t_{s2}^2 + (U - U_1) (U + \sqrt{[4t_{s2}^2 + (U - U_1)^2]} - U_1)]})} \nonumber \\
     \end{eqnarray}.
\normalsize
\section{Quasiclassical approach towards tunable swap gate}
\subsection{Quasiclassical approach towards symmetric swap gate with the same localizing potentials}
We assume that one particle is in the field of another particle that is given by integro-differential equations as often used in quantum chemistry.
In addition we will assume that kinetic energy of particles A and B (electrons in qubit A and B) is very small so $t_{s12} \rightarrow 0$ and $t_{s1'2'} \rightarrow 0$ however still $t_{s12} \neq 0$ and $t_{s1'2'} \neq 0$ so after certain long time particles have chance to find proper configuration. We assume that $p_{A1}$ is the probability of finding particle A at node 1 and that $p_{B1'}$ is the probability of finding particle B at node 1'.
We have effective Hamiltonian omitting kinetic terms in the form
\begin{eqnarray}
\label{HphSymmetricSwap}
H(p_{A1},p_{B1'},E_{p1},E_{p2},E_{p1'},E_{p2'})=p_{A1} E_{p_1}+(1-p_{A1})E_{p_2}+p_{B1'} E_{p_{1'}}+(1-p_{B1'})E_{p_{2'}}+ \nonumber \\
+p_{A1}p_{B1'} \frac{q^2}{d}+(1-p_{A1})(1-p_{B1'}) \frac{q^2}{d} +p_{A1}(1-p_{B1'}) \frac{q^2}{\sqrt{d^2+(a+b)^2}} +p_{B1'}(1-p_{A1'})\frac{q^2}{\sqrt{d^2+(a+b)^2}}. 
\end{eqnarray}
Well designed classical Swap and Q-Swap gate (inverting state 0 to 1 and 1 to 0) will have the property that function $H(p_{A1},p_{B1'},E_{p1},E_{p2},E_{p1'},E_{p2'})$ will reach its minima at $(p_{A1}=0,p_{B1'}=1)$ and at $(p_{A1}=1,p_{B1'}=0)$ and that $H(1,0,E_{p1},E_{p2},E_{p1'},E_{p2'})=H(0,1,E_{p1},E_{p2},E_{p1'},E_{p2'})$. Consequently we assume $H(1,1,E_{p1},E_{p2},E_{p1'},E_{p2'})>H(0,1,E_{p1},E_{p2},E_{p1'},E_{p2'})$ and $H(0,0,E_{p1},E_{p2},E_{p1'},E_{p2'})>H(0,1,E_{p1},E_{p2},E_{p1'},E_{p2'})$ what implies that logical states that are not allowed have higher energy.  
Further simplification can be done by setting for example $E_{p1}$ and $E_{p1'}$ to constant value (as $E_{p1}=const_1$ and $E_{p1'}=const_{1'}$) for example 0 or 1 or any other fixed real number. It shall be underlined that we can chose the one among 4 possible combinations $(E_{p1},E_{p1'})$, $(E_{p2},E_{p2'})$, $(E_{p1},E_{p2'})$, $(E_{p2},E_{p1'})$ whose values needs to be fixed. We set $q=1$, $d=1$, $a+b=0.2$, $E_{p2}=E_{p2'}=1$.
We obtain
\begin{eqnarray}
H(p_{A1},p_{B1'},E_{p1},1,E_{p1'},1)=p_{A1} E_{p_1}+(1-p_{A1})+p_{B1'} E_{p_{1'}}+(1-p_{B1'})+ \nonumber \\
+p_{A1}p_{B1'}+(1-p_{A1})(1-p_{B1'})+p_{A1}(1-p_{B1'}) \frac{1}{\sqrt{1.01}} +p_{B1'}(1-p_{A1'})\frac{1}{\sqrt{1.01}}. 
\end{eqnarray}
Now we need to trace the numerical behaviour of 4 functions
\begin{eqnarray}
H(1,0,E_{p1},1,E_{p1'},1)= E_{p_1}+1 
+ \frac{1}{\sqrt{1.01}} . 
\end{eqnarray}
\begin{eqnarray}
H(0,1,E_{p1},1,E_{p1'},1)=+1+ E_{p_{1'}} 
 +\frac{1}{\sqrt{1.01}}. 
\end{eqnarray}
\begin{eqnarray}
H(1,1,E_{p1},1,E_{p1'},1)=E_{p_1}+ E_{p_{1'}}+1 . 
\end{eqnarray}
\begin{eqnarray}
H(0,0,E_{p1},1,E_{p1'},1)=+1+1+1=3 . 
\end{eqnarray}
Since we impose $H(1,0,E_{p1},1,E_{p1'},1)=H(0,1,E_{p1},1,E_{p1'},1)$ we obtain $E_{p1'}=E_{p1}$.
Imposing $H(0,0,E_{p1},1,E_{p1'},1)=H(1,1,E_{p1},1,E_{p1'},1)$ we finally obtain $E_{p1'}=E_{p1}=1$.
We observe that $H(1,0,E_{p1},1,E_{p1'},1)<H(0,0,E_{p1},1,E_{p1'},1)$
\subsection{Quasi-classical approach towards transition from anticorrelated symmetric swap gate to the correlated swap gate}
We have effective Hamiltonian omitting kinetic terms for any angle $\alpha$ in the form
\begin{eqnarray}
H(p_{A1},p_{B1'},E_{p1},E_{p2},E_{p1'},E_{p2'})=p_{A1} E_{p_1}+(1-p_{A1})E_{p_2}+p_{B1'} E_{p_{1'}}+(1-p_{B1'})E_{p_{2'}}+ \nonumber \\
+p_{A1}p_{B1'} \frac{q^2}{\sqrt{d^2+(a+b)^2}}+(1-p_{A1})(1-p_{B1'}) \frac{q^2}{\sqrt{(d+Cos(\alpha)(a+b))^2+(a+b)^2Sin(\alpha)^2}} + \nonumber \\
p_{A1}(1-p_{B1'}) \frac{q^2}{\sqrt{(d+(a+b)Cos(\alpha))^2+((1+Sin(\alpha))(a+b))^2}} +p_{B1'}(1-p_{A1'})\frac{q^2}{\sqrt{d^2+(a+b)^2}}. 
\end{eqnarray}
Previously considered case was for the angle $\alpha=-\Pi/2$ that corresponds to the symmetric anticorrelated swap gate as specified by formula \ref{HphSymmetricSwap}. Let us set $E_{p2}$ and $E_{p2'}$ to be 1. We obtain following $H(p_{A1},p_{B1'},E_{p1},1,E_{p1'},1)$ values as
\begin{eqnarray}
H(p_{A1}=0,p_{B1'}=0,E_{p1},E_{p2}=1,E_{p1'},E_{p2'}=1)=1+1+ \nonumber \\
+ \frac{q^2}{\sqrt{(d+Cos(\alpha)(a+b))^2+(a+b)^2Sin(\alpha)^2}}=V_1, 
\end{eqnarray}
\begin{eqnarray}
H(p_{A1}=1,p_{B1'}=1,E_{p1},E_{p2}=1,E_{p1'},E_{p2'}=1)= E_{p_1}+ E_{p_{1'}}+ \nonumber \\
+ \frac{q^2}{\sqrt{d^2+(a+b)^2}}=V_1, 
\end{eqnarray}
\begin{eqnarray}
H(p_{A1}=1,p_{B1'}=0,E_{p1},E_{p2}=1,E_{p1'},E_{p2'}=1)=E_{p_1}+1+ \nonumber \\
 \frac{q^2}{\sqrt{(d+(a+b)Cos(\alpha))^2+((1+Sin(\alpha))(a+b))^2}}=V_2, 
\end{eqnarray}
\begin{eqnarray}
H(p_{A1}=0,p_{B1'}=1,E_{p1},E_{p2}=1,E_{p1'},E_{p2'}=1)=1+ E_{p_{1'}} 
 +\frac{q^2}{\sqrt{d^2+(a+b)^2}}=V_2, 
\end{eqnarray}
with condition $V_2 < V_1$ that implies
\begin{eqnarray}
1+1+\frac{q^2}{\sqrt{(d+Cos(\alpha)(a+b))^2+(a+b)^2Sin(\alpha)^2}}-\frac{q^2}{\sqrt{d^2+(a+b)^2}}= 
E_{p_1}+ E_{p_{1'}},
\end{eqnarray}

\begin{eqnarray}
+\frac{q^2}{\sqrt{d^2+(a+b)^2}}-\frac{q^2}{\sqrt{(d+(a+b)Cos(\alpha))^2+((1+Sin(\alpha))(a+b))^2}}=E_{p_1}-E_{p_{1'}}
\end{eqnarray}
and we obtain
\begin{eqnarray}
1+\frac{1}{2}(\frac{q^2}{\sqrt{(d+Cos(\alpha)(a+b))^2+(a+b)^2Sin(\alpha)^2}}-\frac{q^2}{\sqrt{(d+(a+b)Cos(\alpha))^2+((1+Sin(\alpha))(a+b))^2}}=E_{p1},
\end{eqnarray}
\begin{eqnarray}
1+\frac{1}{2}(\frac{q^2}{\sqrt{(d+Cos(\alpha)(a+b))^2+(a+b)^2Sin(\alpha)^2}}+\frac{q^2}{\sqrt{(d+(a+b)Cos(\alpha))^2+((1+Sin(\alpha))(a+b))^2}} \nonumber \\
-\frac{q^2}{\sqrt{d^2+(a+b)^2}}=E_{p1'},
\end{eqnarray}
We can check the designing condition $V_1>V_2$ that implies
\begin{eqnarray}
H(p_{A1}=1,p_{B1'}=1,E_{p1},E_{p2}=1,E_{p1'},E_{p2'}=1)= E_{p_1}+ E_{p_{1'}} 
+ \frac{q^2}{\sqrt{d^2+(a+b)^2}}=V_1>, \nonumber \\
H(p_{A1}=1,p_{B1'}=0,E_{p1},E_{p2}=1,E_{p1'},E_{p2'}=1)=E_{p_1}+1+ 
 \frac{q^2}{\sqrt{(d+(a+b)Cos(\alpha)))^2+((1+Sin(\alpha))(a+b))^2}}=V_2, \nonumber \\ 
\end{eqnarray}
and consequently we obtain
\begin{eqnarray}
E_{p_{1'}} > 
+1+ \frac{q^2}{\sqrt{(d+(a+b)Cos(\alpha))^2+((1+Sin(\alpha))(a+b))^2}}-\frac{q^2}{\sqrt{d^2+(a+b)^2}}
\end{eqnarray}
what brings the condition
\begin{eqnarray}
\frac{q^2}{\sqrt{(d+Cos(\alpha)(a+b))^2+(a+b)^2Sin(\alpha)^2}}>\frac{q^2}{\sqrt{(d+(a+b)Cos(\alpha)))^2+((1+Sin(\alpha))(a+b))^2}}.
\end{eqnarray}
that is equivalent to the condition $Sin(\alpha)^2<(1+Sin(\alpha))^2=1+Sin^2(\alpha)+2Sin(\alpha)$ or $0<Sin(\alpha)$ that $\alpha \in (0, \Pi)$.
The conducted considerations are visualized by Fig.\ref{GeometricTransition} that describes the dependence of phenomenological electrostatic Hamiltonian on logical values $p_{A1}$ and $p_{B1}$ what gives the prescription for construction fuzzy logic swap gate or quantum swap gate implemented in single electron devices.
\section{Electrostatic quantum antiswap gate}
We can use the fact that two qubits can be in maximum correlated states (if $p_{A1}=1$ we have $p_{B1}=1-p_{A1}$).  Due to repulsive electrostatic interaction we can build the electrostatic quantum antiswap gate so it corresponds to the situation when one qubits is being copied by another qubit. It shall be underlined that the same considerations can be done for electron-hole system ($q^2=-1$) or hole-hole system ($q^2=1$).
\begin{figure}
\centering 
 \includegraphics[scale=0.6]{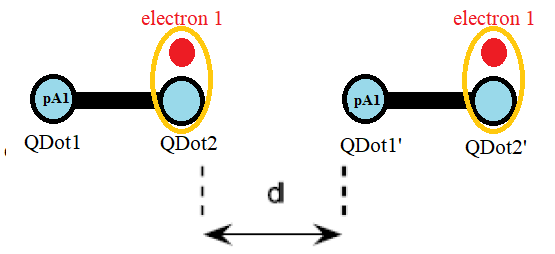}
 \caption{Scheme of electrostatic antiswap gate }
\end{figure}
We have
\begin{eqnarray}
H(p_{A1},p_{B1'},E_{p1},E_{p2},E_{p1'},E_{p2'})=p_{A1} E_{p_1}+(1-p_{A1})E_{p_2}+p_{B1'} E_{p_{1'}}+(1-p_{B1'})E_{p_{2'}}+ \nonumber \\
+p_{A1}p_{B1'} \frac{q^2}{d+a+b}+(1-p_{A1})(1-p_{B1'}) \frac{q^2}{d+a+b} +p_{A1}(1-p_{B1'}) \frac{q^2}{d+2(a+b)} +p_{B1'}(1-p_{A1'})\frac{q^2}{d}= \nonumber \\. 
\end{eqnarray}
We set $E_{p1}=1, E_{p2'}=1$ and we have
\begin{eqnarray}
H(p_{A1},p_{B1'},1,E_{p2},E_{p1'},1)=p_{A1}+(1-p_{A1})E_{p_2}+p_{B1'} E_{p_{1'}}+(1-p_{B1'})+ \nonumber \\
+p_{A1}p_{B1'} \frac{q^2}{d+a+b}+(1-p_{A1})(1-p_{B1'}) \frac{q^2}{d+a+b} +p_{A1}(1-p_{B1'}) \frac{q^2}{d+2(a+b)} +p_{B1'}(1-p_{A1'})\frac{q^2}{d}= \nonumber \\
=p_{A1}+(1-p_{A1})E_{p_2}+p_{B1'} E_{p_{1'}}+(1-p_{B1'})+ \nonumber \\
+(2p_{A1}p_{B1'}+1-(p_{A1}+p_{B1'}) )\frac{q^2}{d+a+b}+p_{A1}(1-p_{B1'}) \frac{q^2}{d+2(a+b)} +p_{B1'}(1-p_{A1'})\frac{q^2}{d}
\end{eqnarray}

In order to obtain quantum repeater we have $p_{A1}=p_{B1'}$ and such configuration is energetically favourable (minimizes Hamiltonian energy) so we have
\begin{eqnarray}
H(0,0,1,E_{p2},E_{p1'},1)=H(1,1,1,E_{p2},E_{p1'},1)=V_1, \nonumber \\
H(1,0,1,E_{p2},E_{p1'},1)=H(0,1,1,E_{p2},E_{p1'},1)=V_2, V_2<V_1.
\end{eqnarray}

We have the condition
\begin{eqnarray}
H(0,0,1,E_{p2},E_{p1'},1)=E_{p_2}+1+\frac{q^2}{d+a+b}=V_1,
\end{eqnarray}

\begin{eqnarray}
H(1,1,1,E_{p2},E_{p1'},1)
=1+ E_{p_{1'}} 
+(2+1-2 )\frac{q^2}{d+a+b}=1+ E_{p_{1'}}+\frac{q^2}{d+a+b}=V_1.
\end{eqnarray}
and it implies $E_{p_{1'}}=E_{p_2}=E_p$. We have

\begin{eqnarray}
H(1,0,1,E_{p2},E_{p1'},1)=1+1 
+ \frac{q^2}{d+2(a+b)} =V_2,
\end{eqnarray}

\begin{eqnarray}
H(0,1,1,E_{p2},E_{p1'},1)=+E_{p_2}+E_{p_{1'}}+\frac{q^2}{d}=V_2.
\end{eqnarray}
We have
\begin{eqnarray}
E_p=1+\frac{1}{2}(\frac{q^2}{d+2(a+b)}-\frac{q^2}{d})=E_{p_2}=E_{p_{1'}}.
\end{eqnarray}
and it implies
\begin{eqnarray}
V_2=2+(\frac{q^2}{d+2(a+b)}-\frac{q^2}{d})+\frac{q^2}{d}=2+\frac{q^2}{d+2(a+b)},
\end{eqnarray}
\begin{eqnarray}
V_1=2+\frac{q^2}{d+a+b}+\frac{1}{2}(\frac{q^2}{d+2(a+b)}-\frac{q^2}{d}).
\end{eqnarray}
what
\begin{eqnarray}
V_2-V_1=(\frac{q^2}{d+2(a+b)}-\frac{q^2}{d+a+b})+\frac{1}{2}(\frac{q^2}{d+2(a+b)}-\frac{q^2}{d})<0,
\end{eqnarray}
since $(\frac{q^2}{d+2(a+b)}-\frac{q^2}{d+a+b})<0$ and $\frac{1}{2}(\frac{q^2}{d+2(a+b)}-\frac{q^2}{d})<0$ so condition $V_2-V_1<0$ is fulfilled.

\section{Concept of programmable matter and conclusions}
\begin{figure}
\centering 
 \includegraphics[scale=0.6]{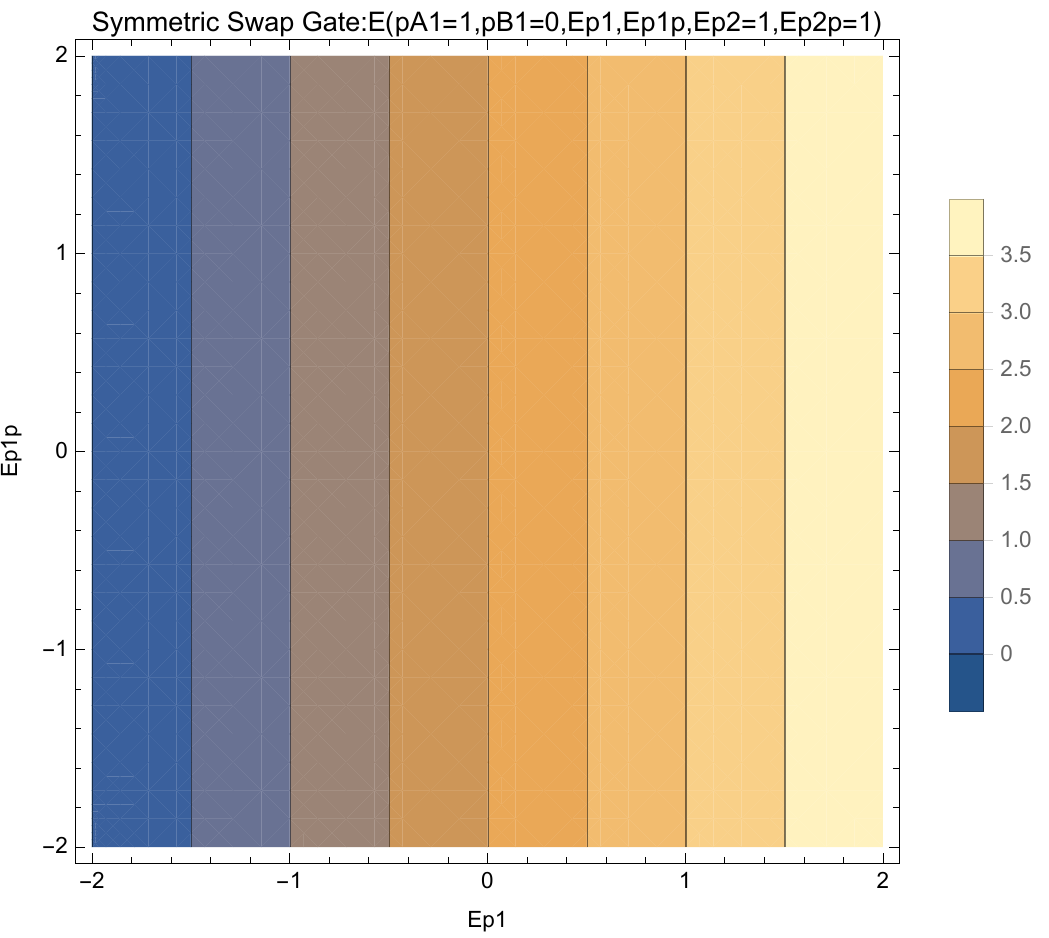}\includegraphics[scale=0.6]{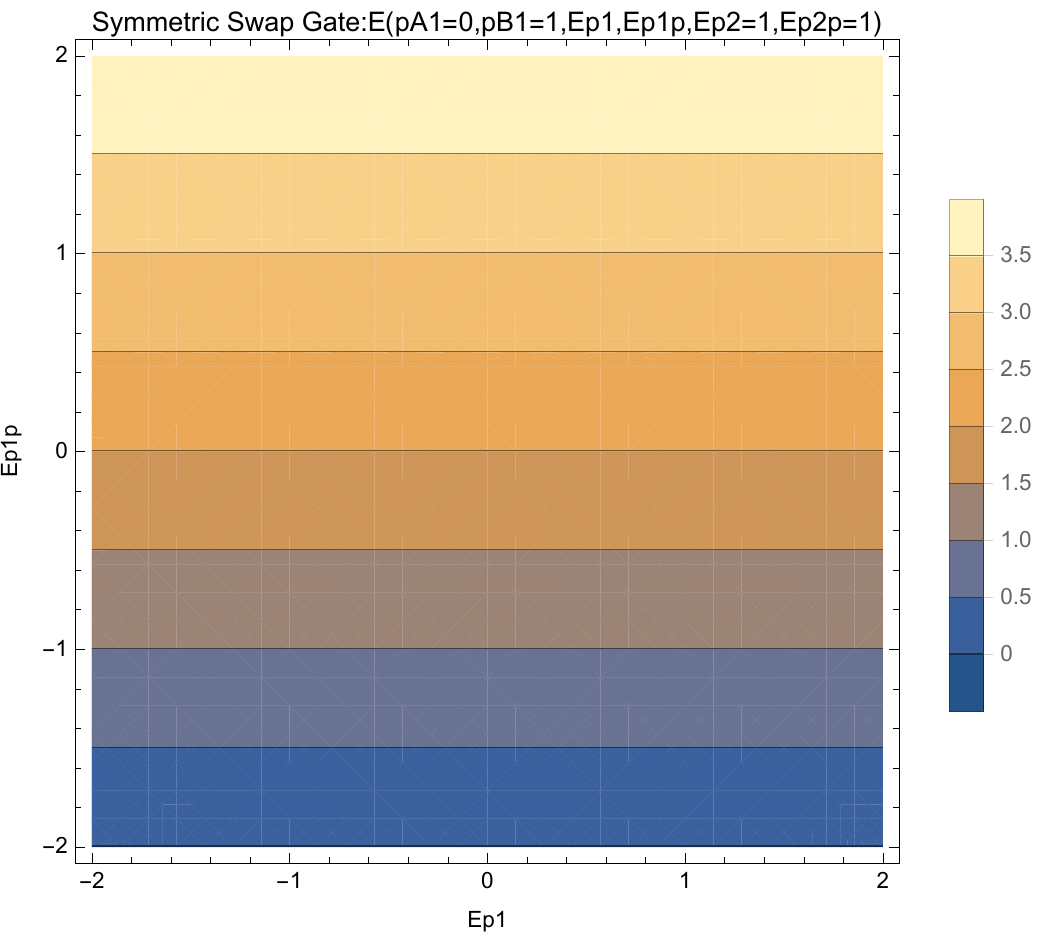}
 \includegraphics[scale=0.6]{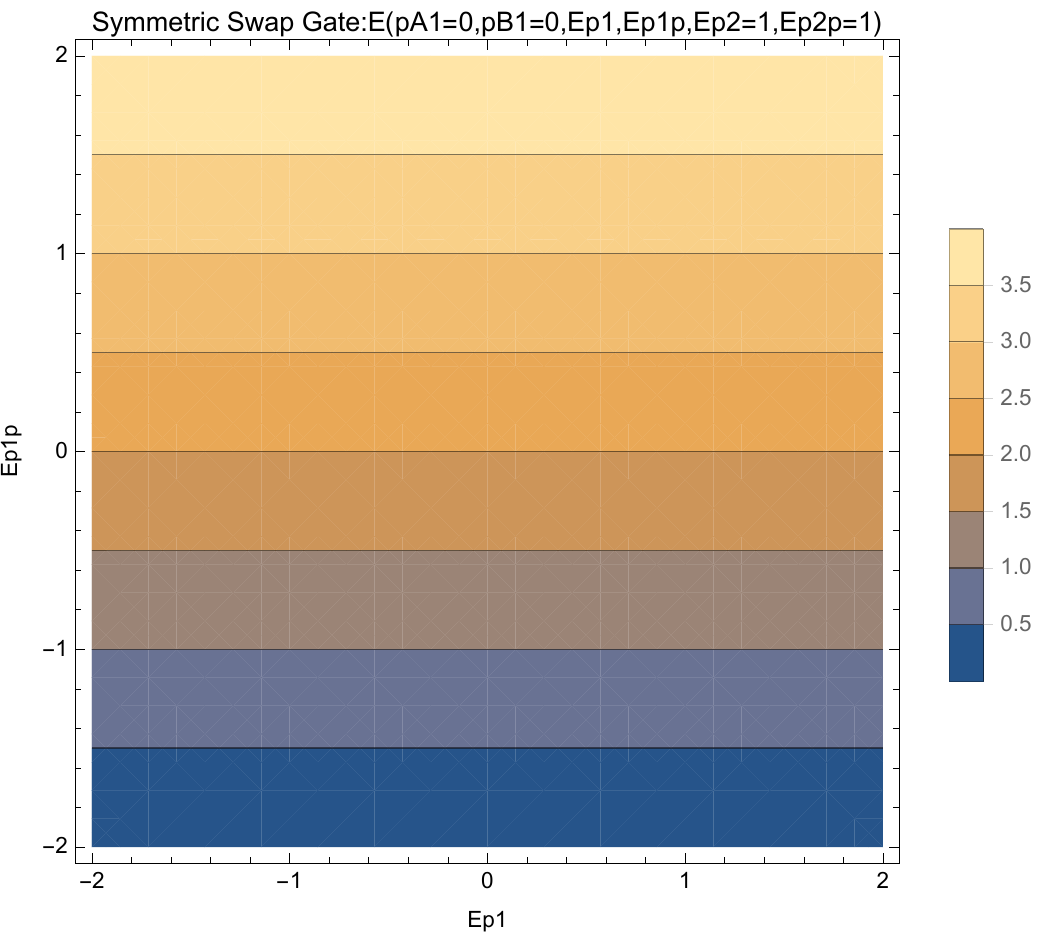}\includegraphics[scale=0.6]{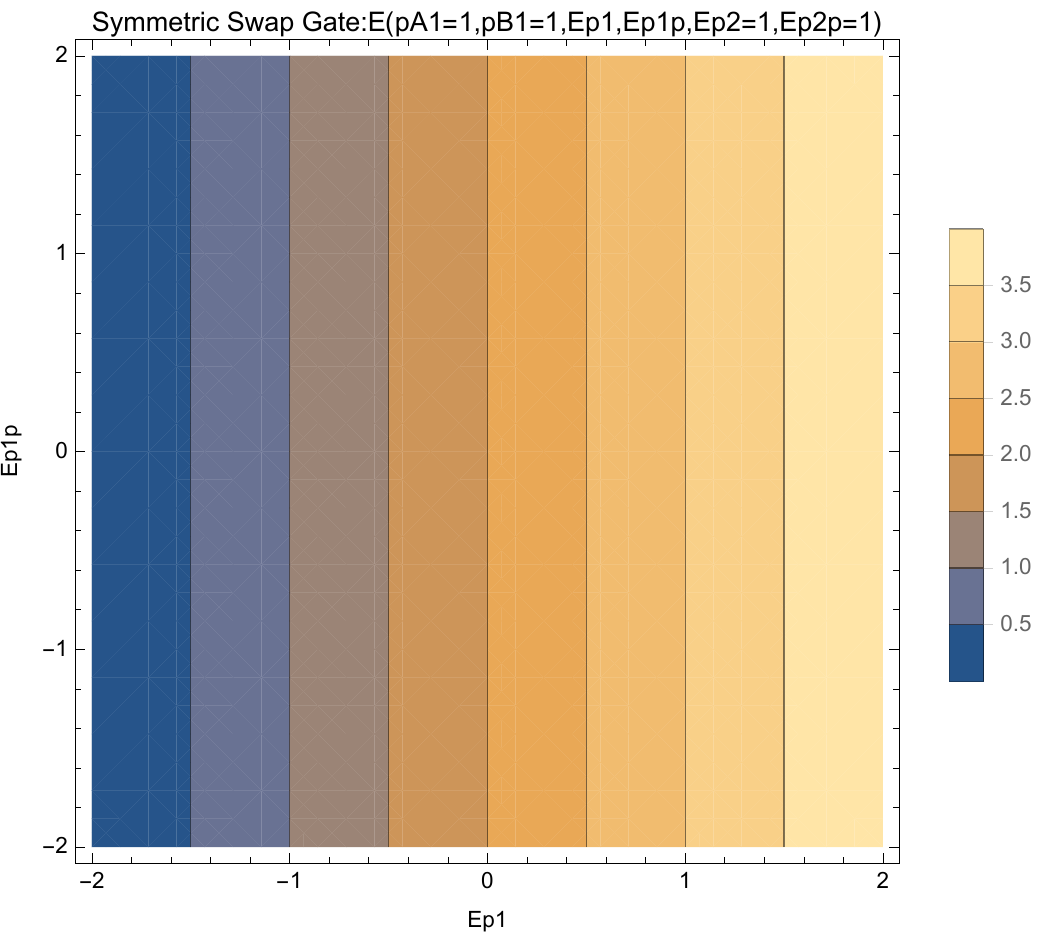}
 \caption{Dependence of quasi-classical Swap Gate Hamiltonian on polarizing potentials that are electrostatically controlled given by $E_{p1}$ and $E_{p1'}$ for fixed ($E_{p2}=1$,$E_{p2'}=1$).}
\label{qp1}
\end{figure}

\begin{figure}
\includegraphics[scale=0.6]{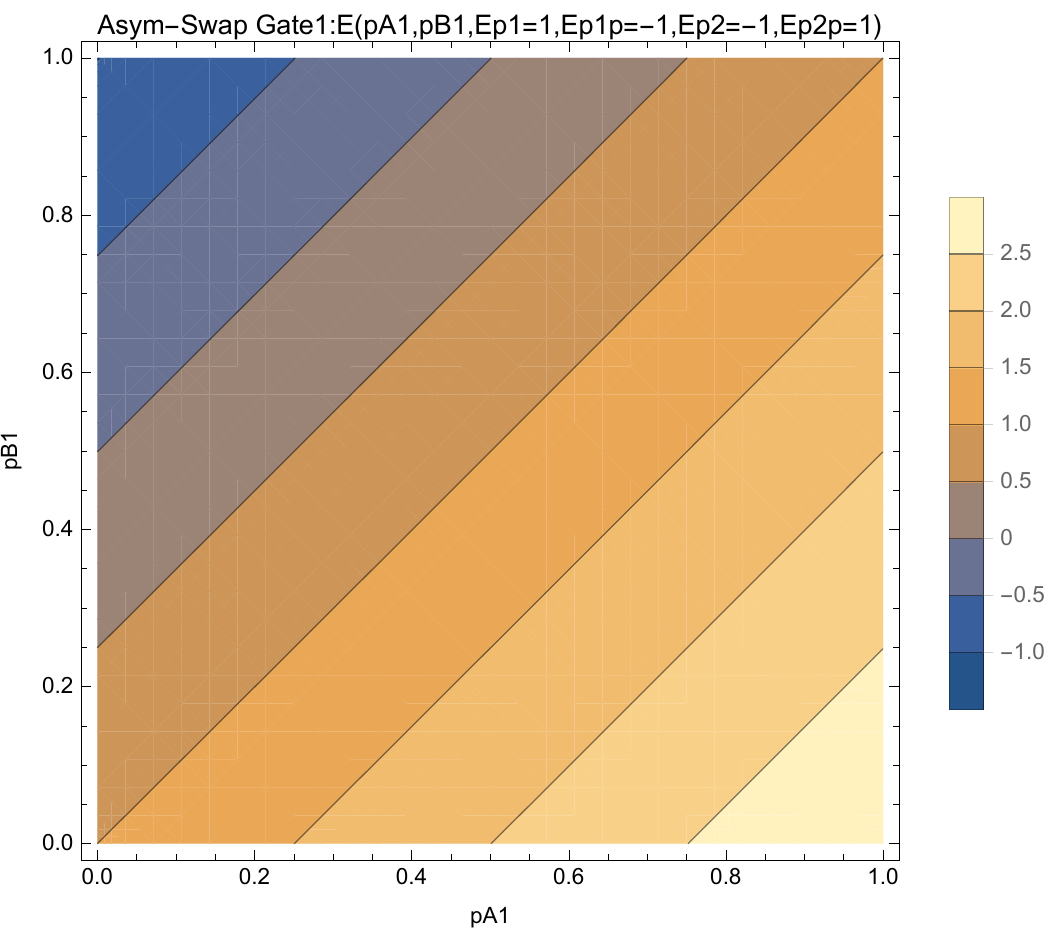}
\includegraphics[scale=0.6]{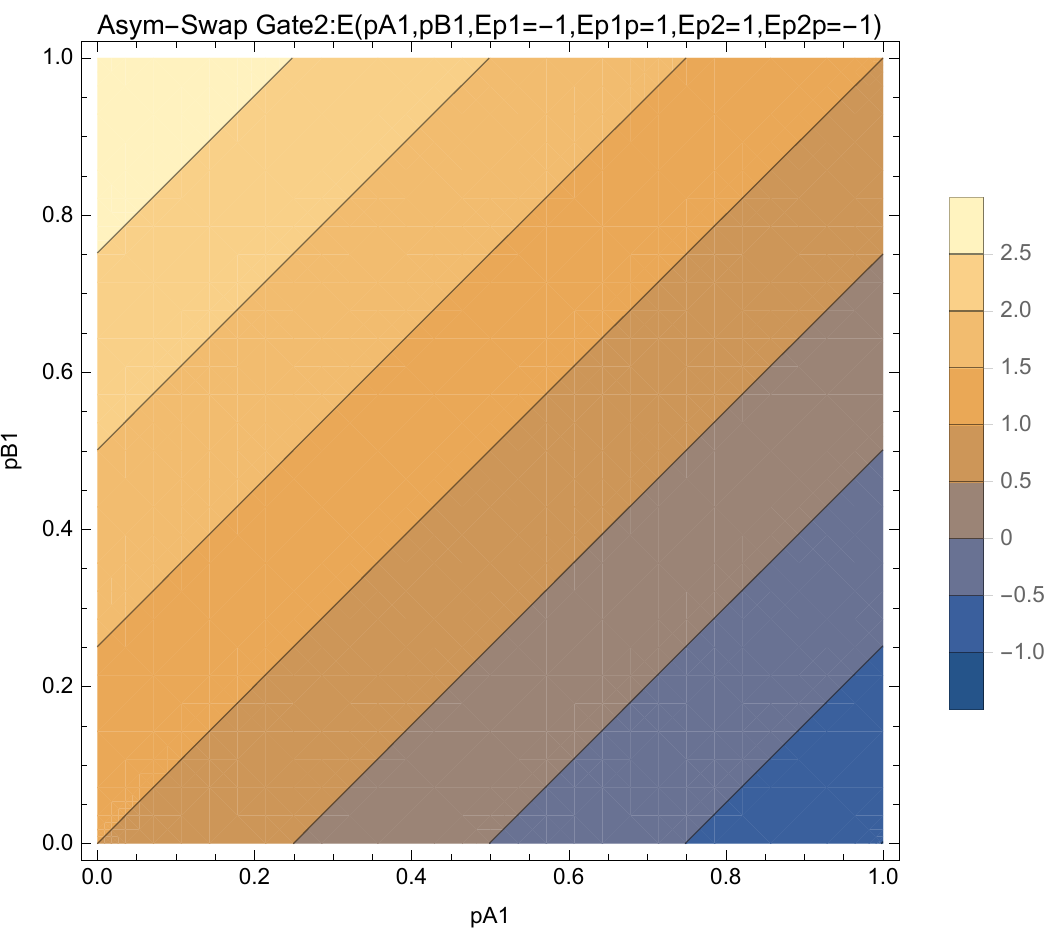}
\includegraphics[scale=0.6]{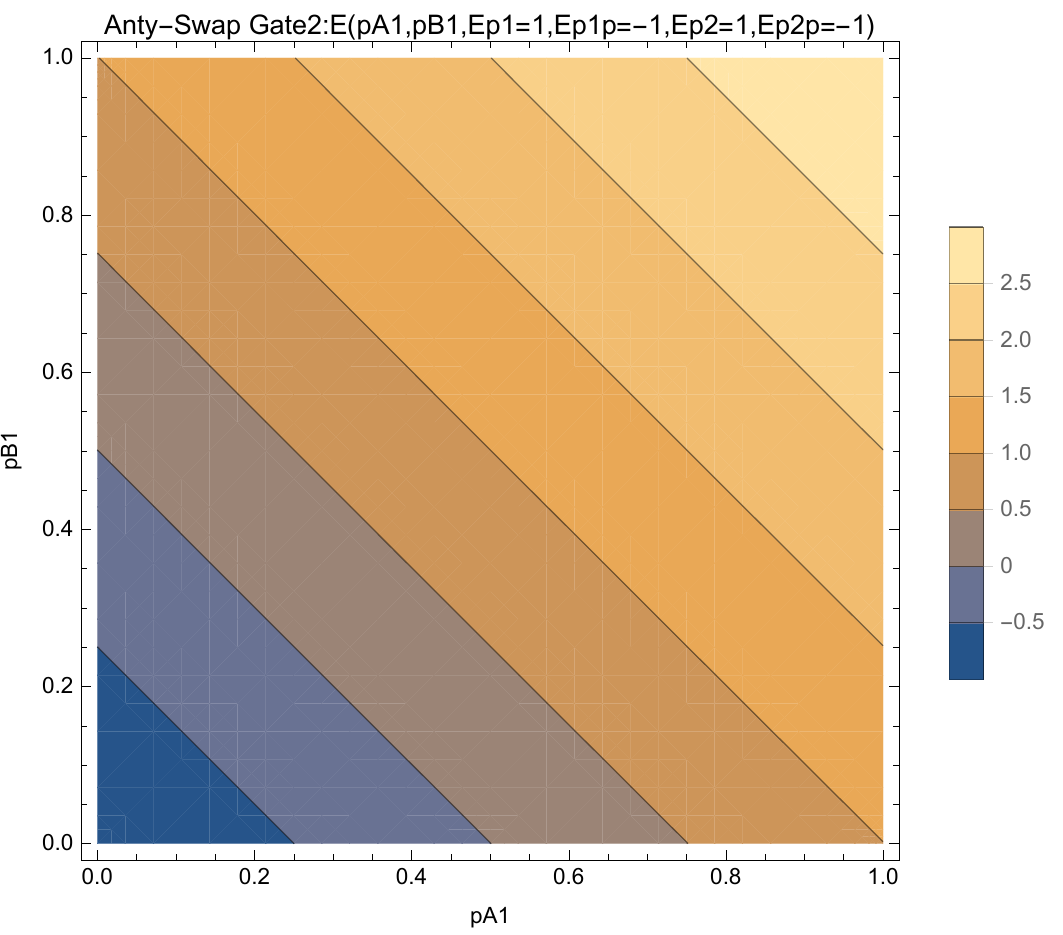}
\includegraphics[scale=0.6]{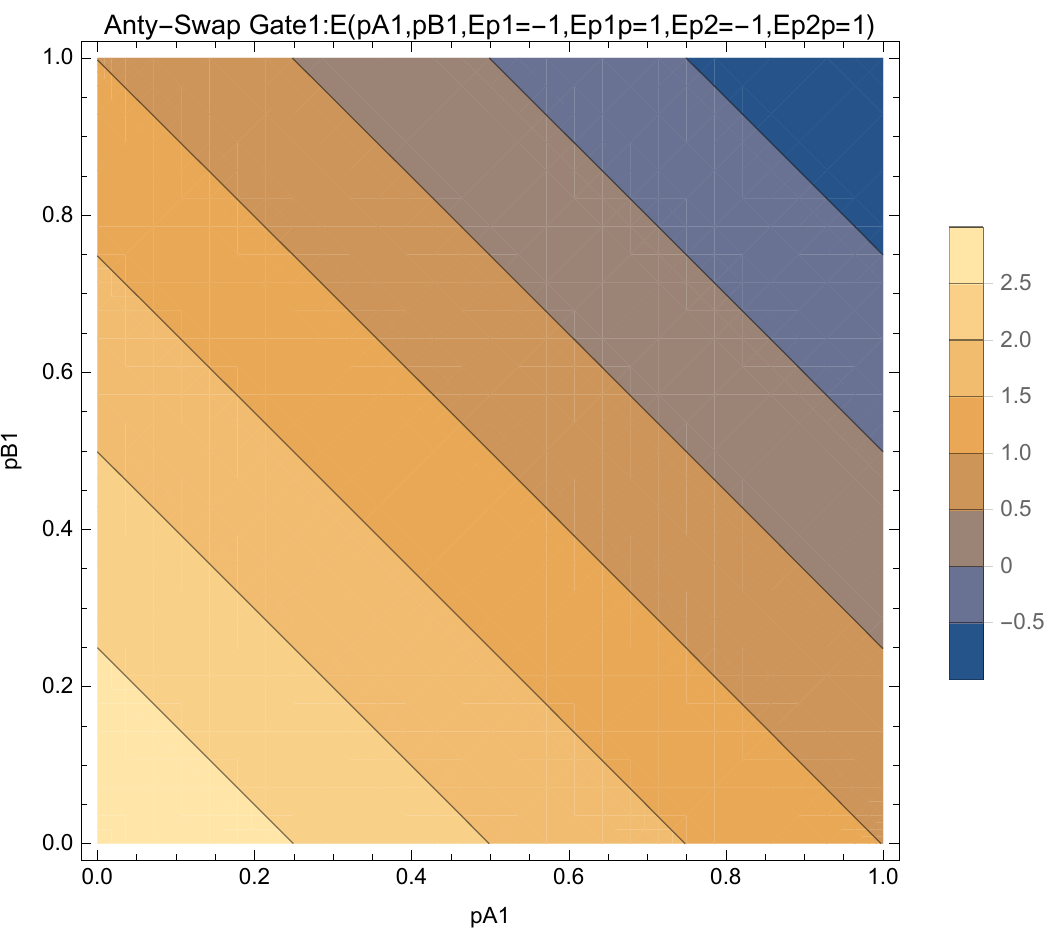}
\includegraphics[scale=0.6]{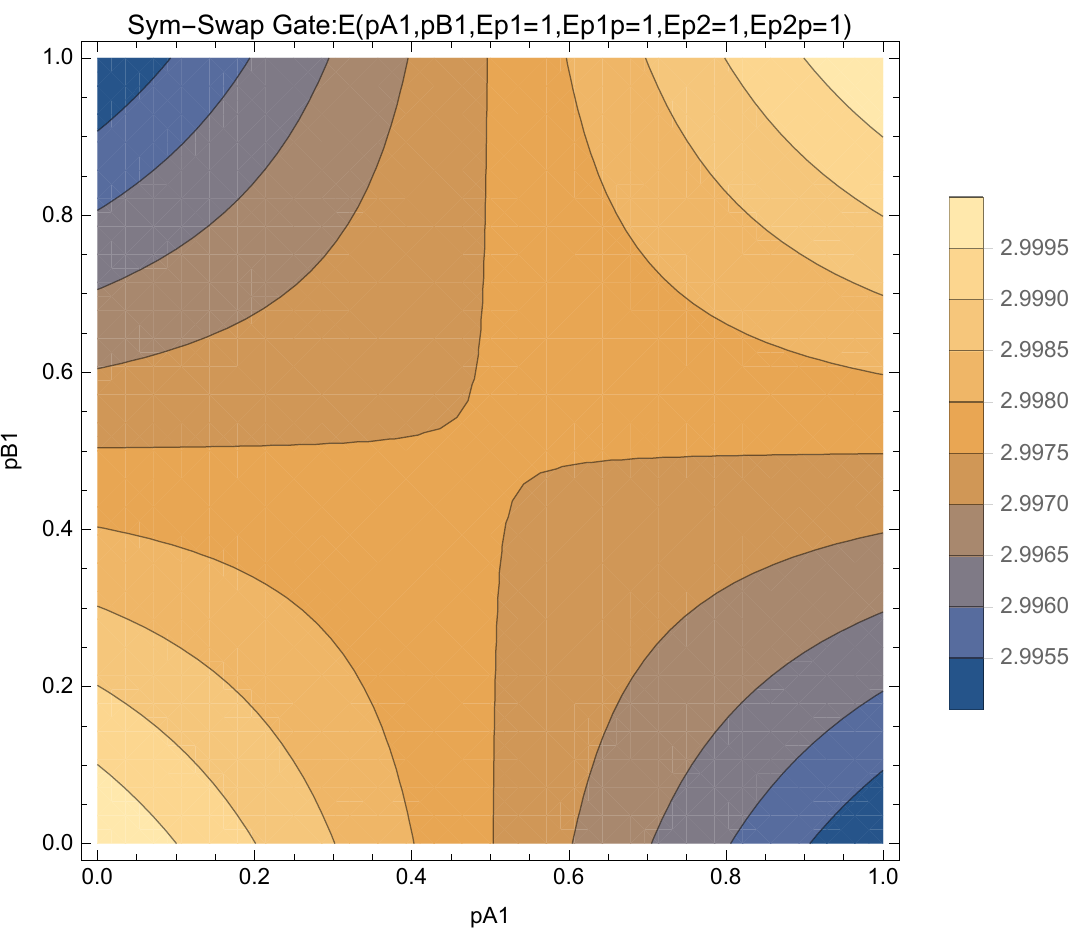}
\includegraphics[scale=0.6]{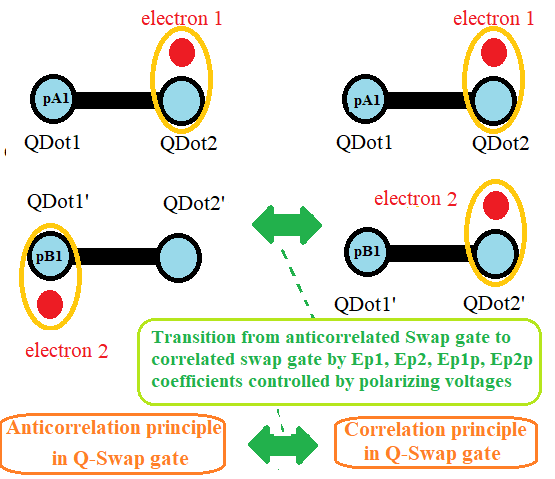}
 \caption{Role of polarizing voltages in transition from symmetric classical and quantum swap gate into asymmetric swap gate (upper middle and left) and into transition into anti-swap gate (bottom plots).}
 \label{qp2}
\end{figure}

\begin{figure}
\includegraphics[scale=0.5]{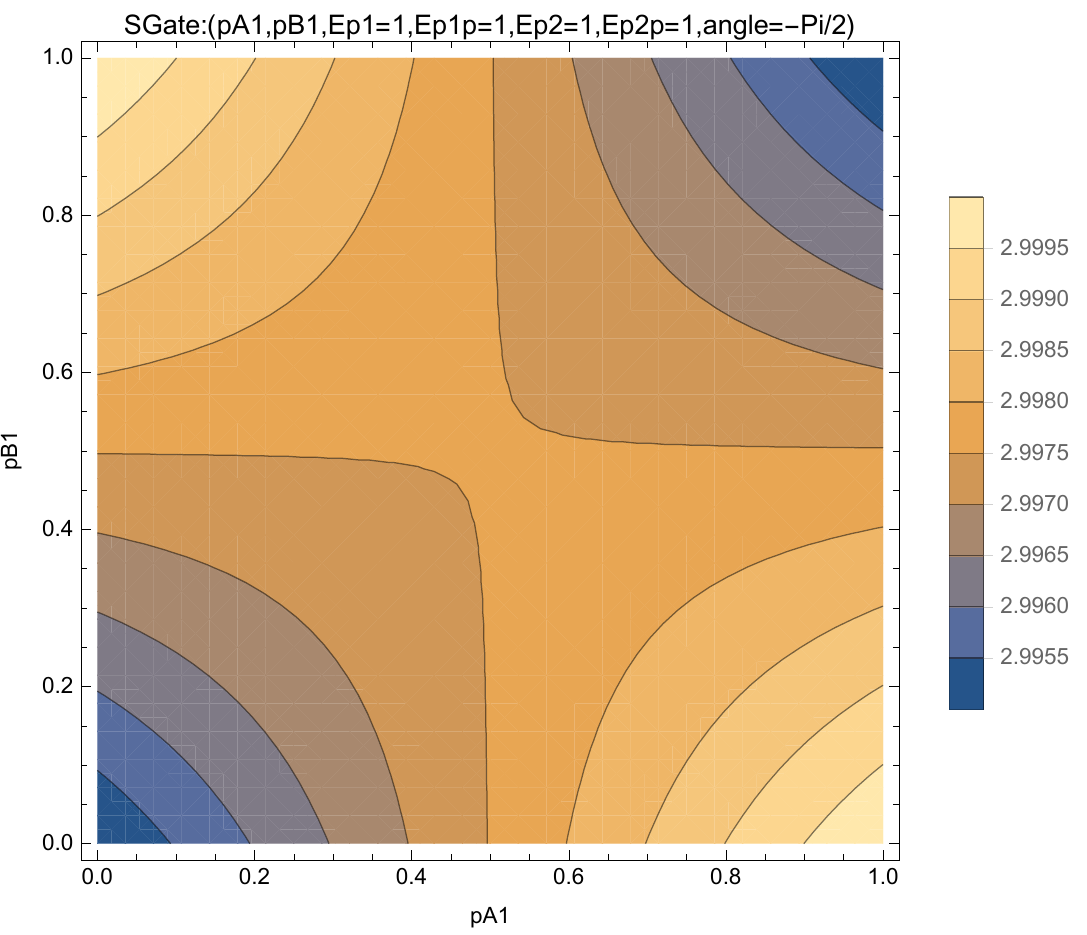}
\includegraphics[scale=0.4]{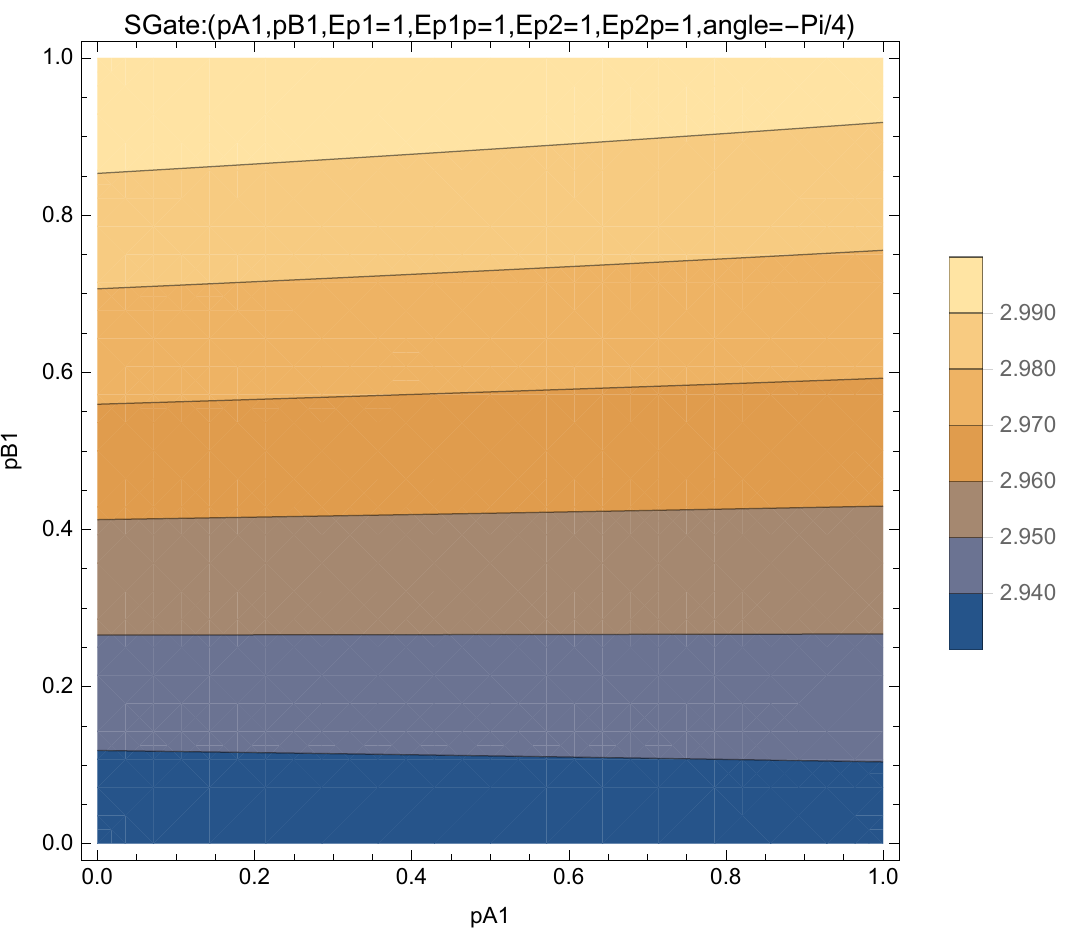}
\includegraphics[scale=0.4]{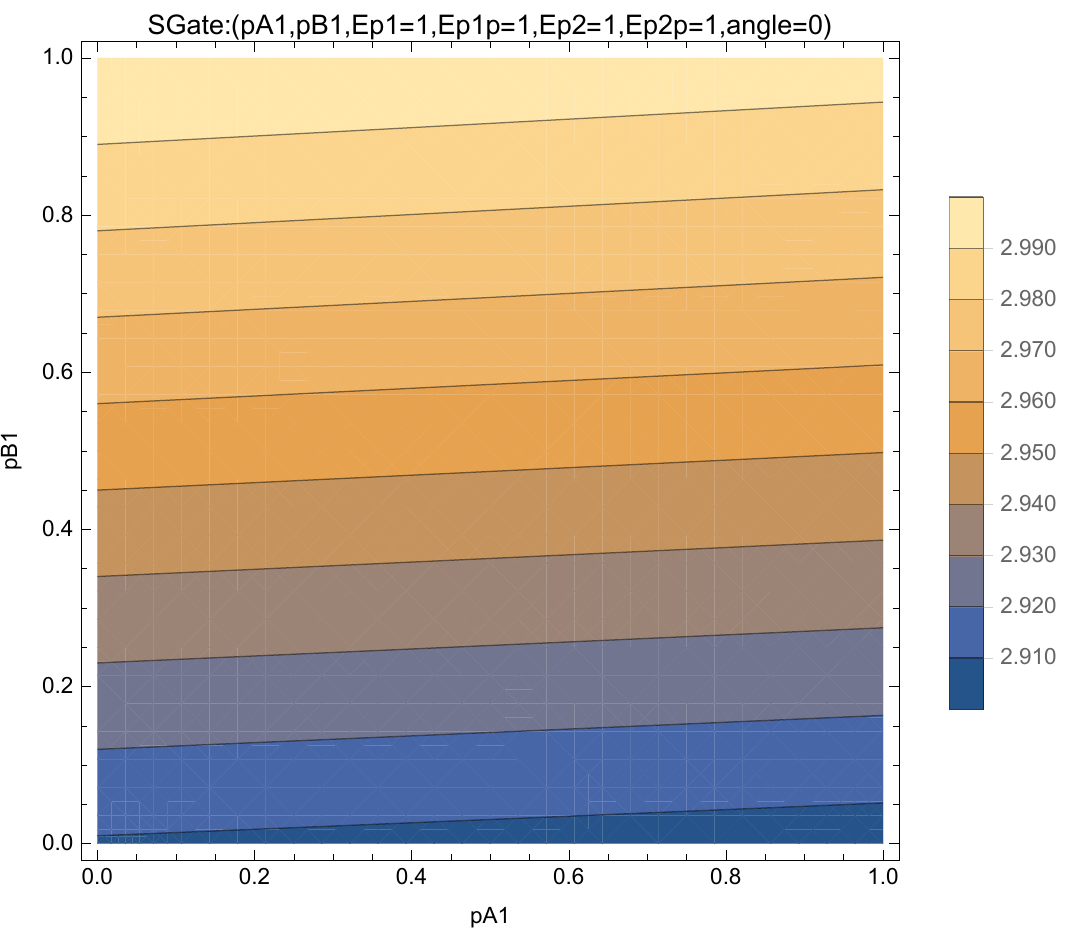}
\includegraphics[scale=0.4]{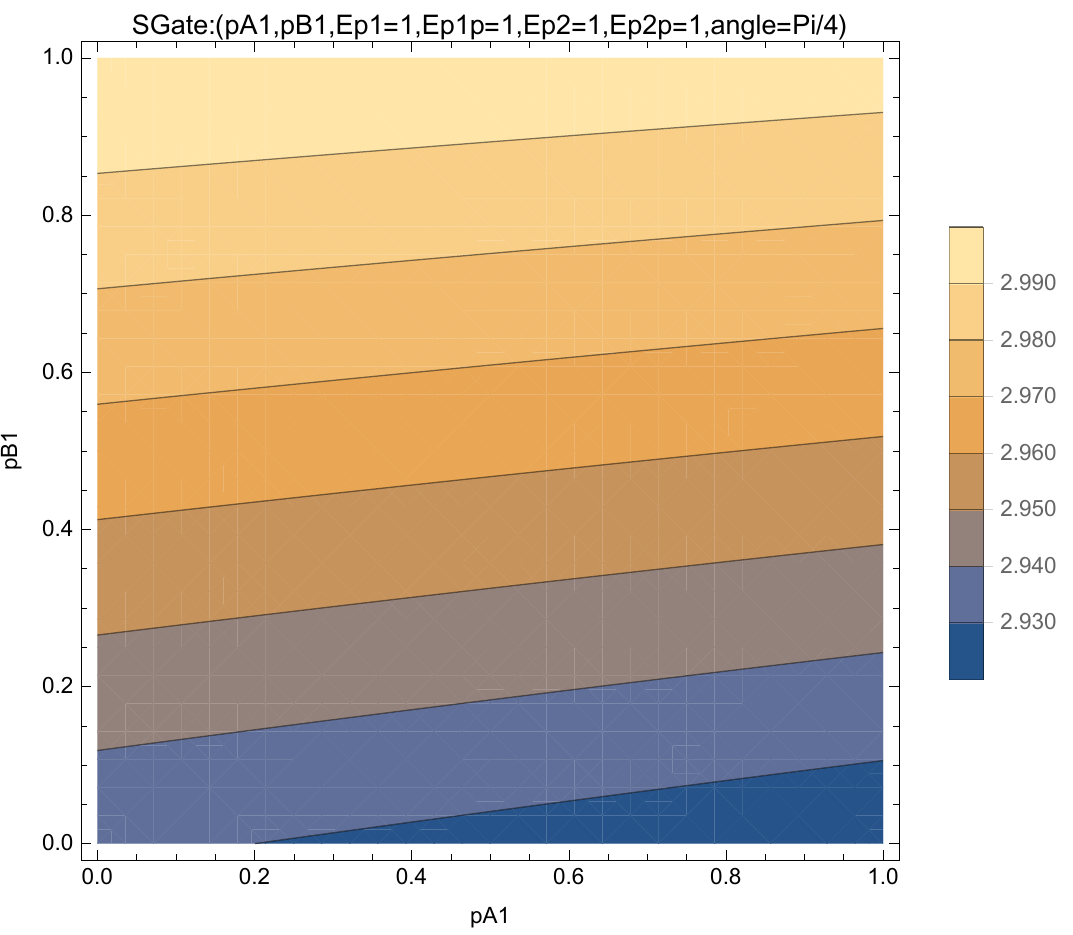}
\includegraphics[scale=0.5]{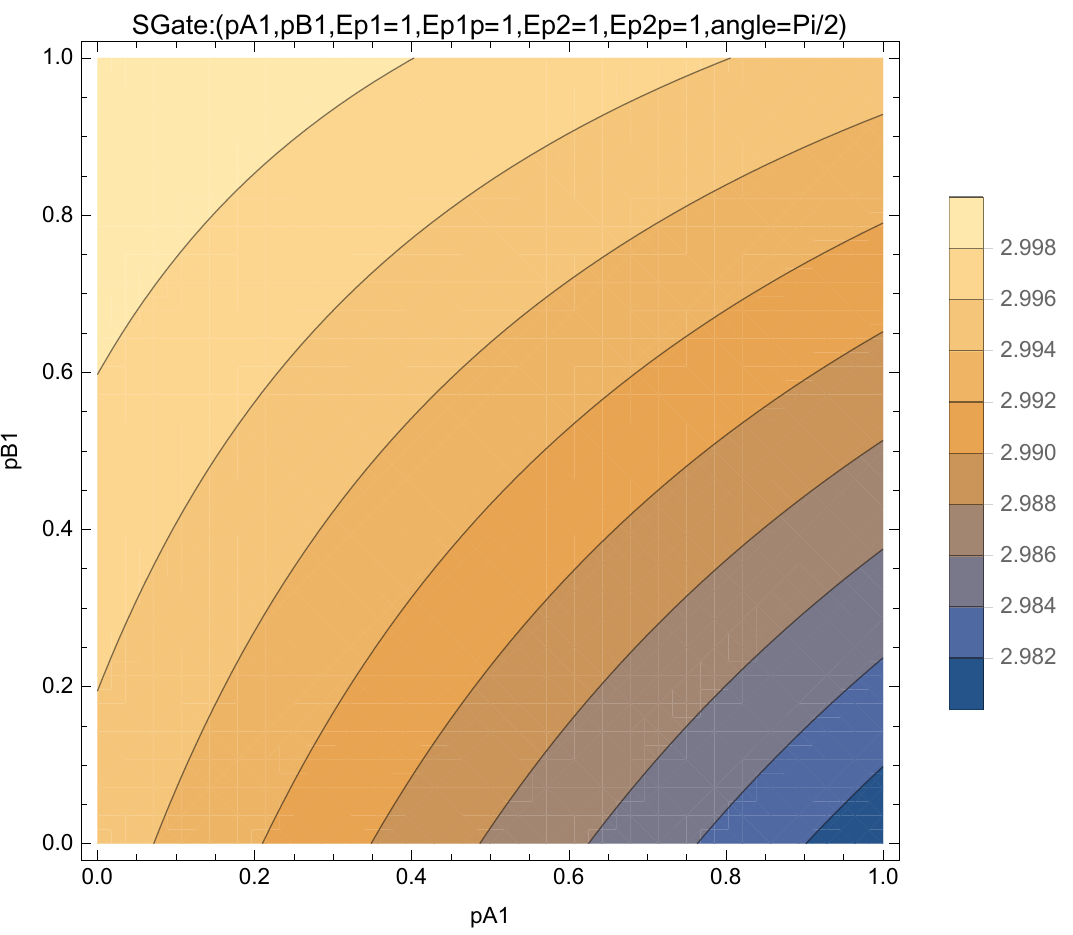}
\includegraphics[scale=0.3]{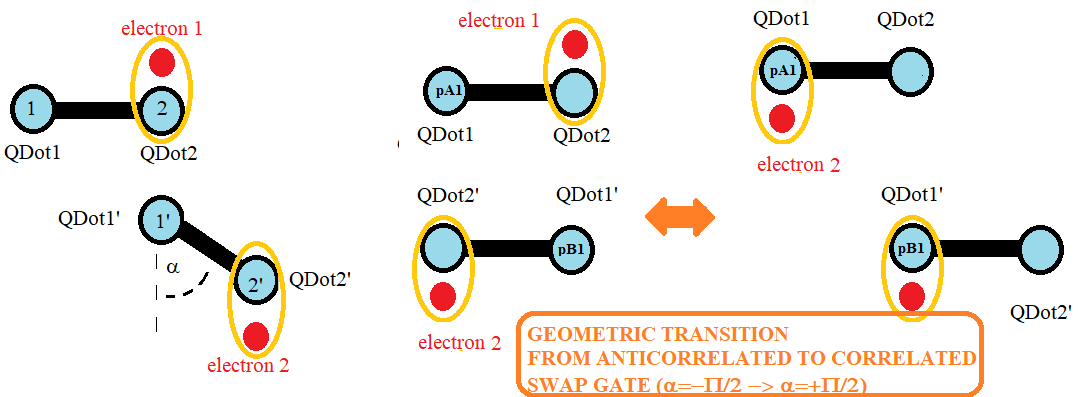}
\includegraphics[scale=0.4]{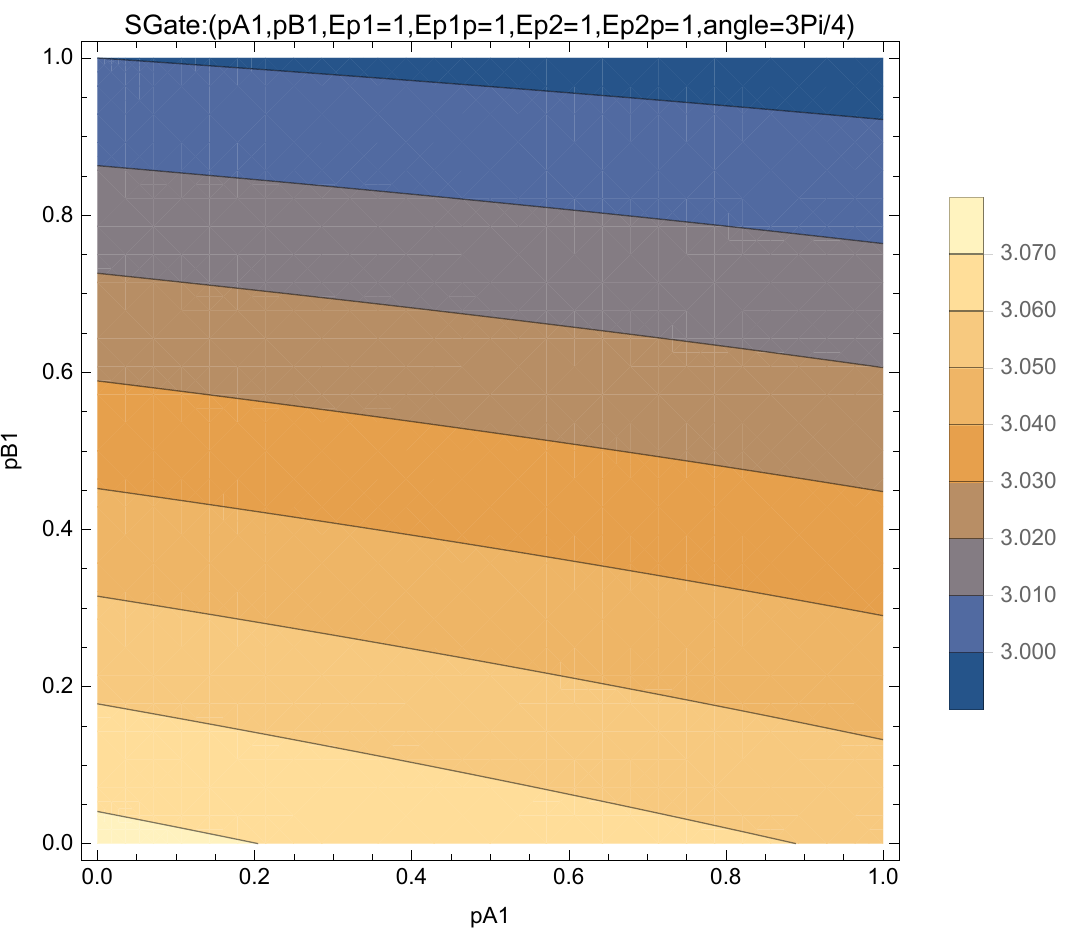}
\includegraphics[scale=0.4]{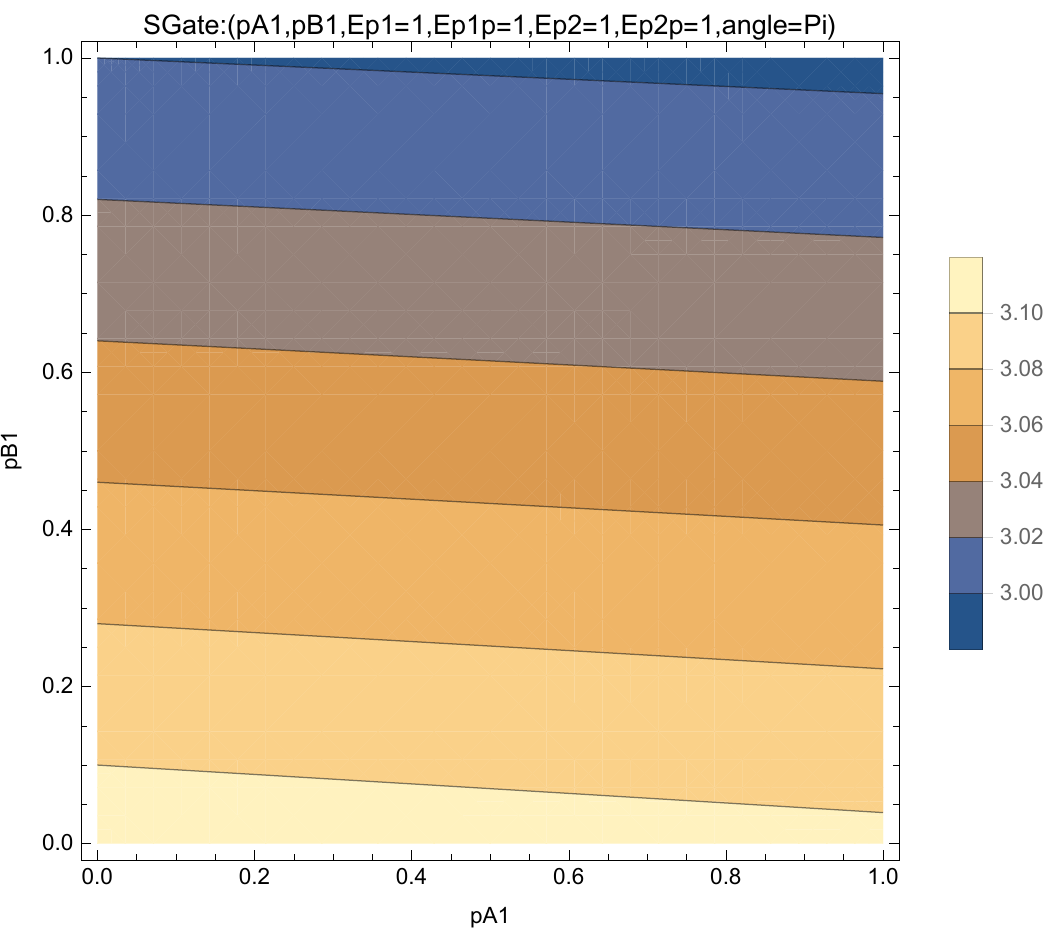}
\includegraphics[scale=0.4]{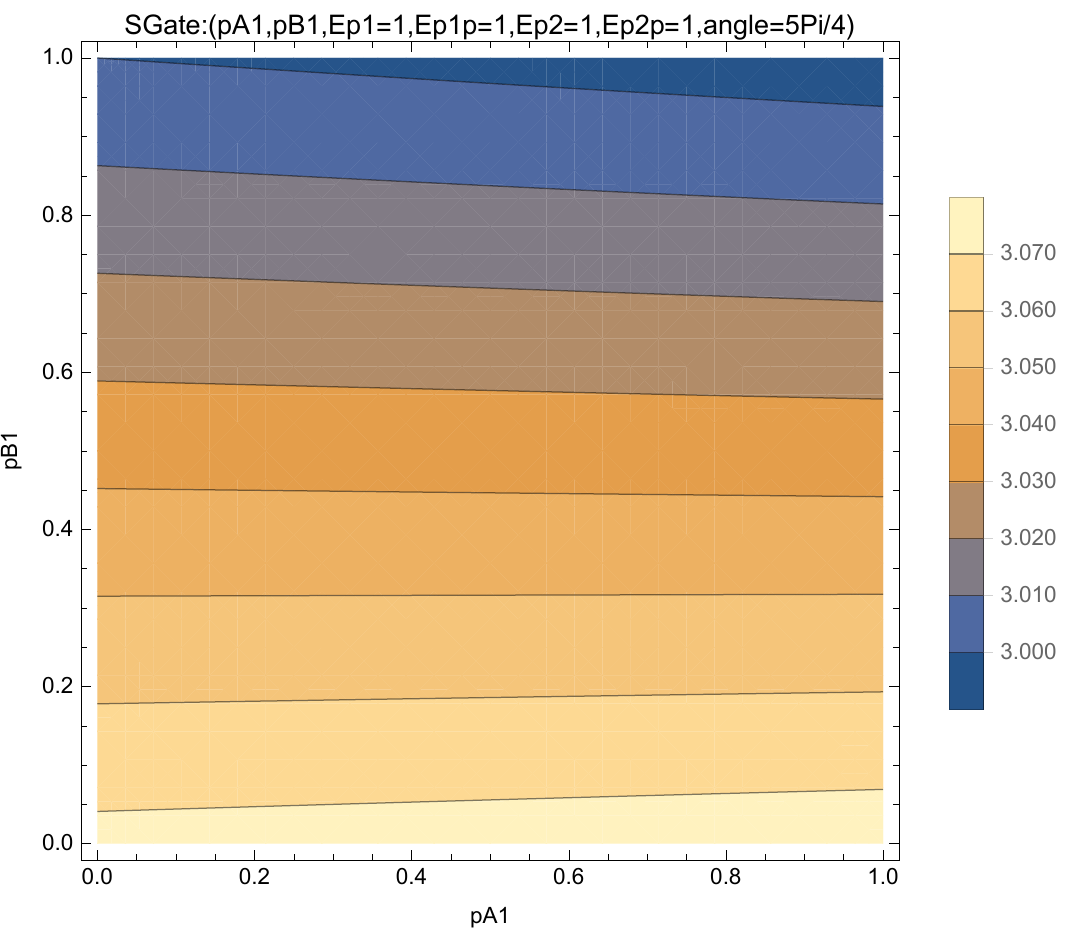}
\caption{Transition from classical/quantum swap gate into antiswap gate by changing the angle $\alpha$ from $-\Pi$ to $\Pi$. Minima in phenomenological Hamiltonian identifies the preferable logical state (states of ($p_{A1}=1,p_{B1}=0$) and ($p_{A1}=0,p_{B1}=1$) are preferred in the same way) into only one preferred state ($p_{A1}=1,p_{B1}=0$). Contour plots describing qubit-qubit Hamiltonian values in function of probabilities ($p_{A1}$,$p_{B1}$) of SWAP and ANTISWAP gates were magnified.}
\label{GeometricTransition}
\end{figure}

\begin{figure}
\centering
\includegraphics[scale=0.6]{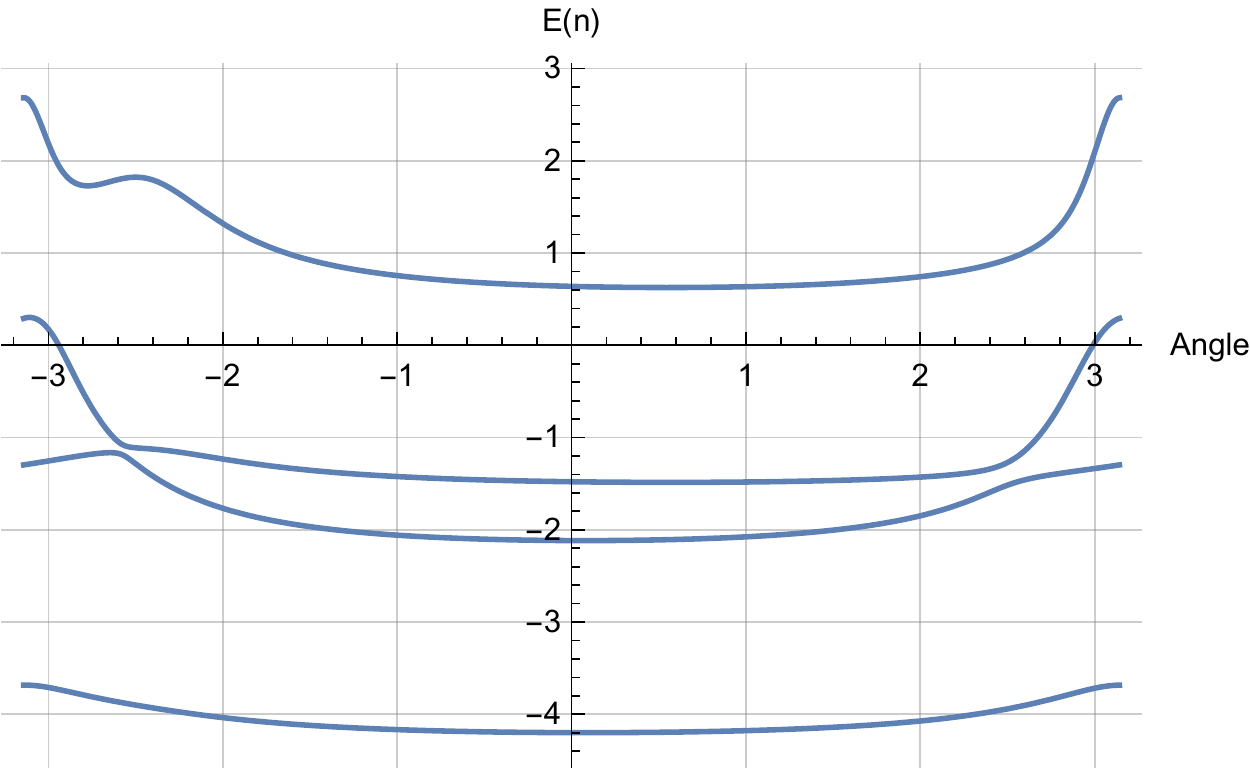}\includegraphics[scale=0.6]{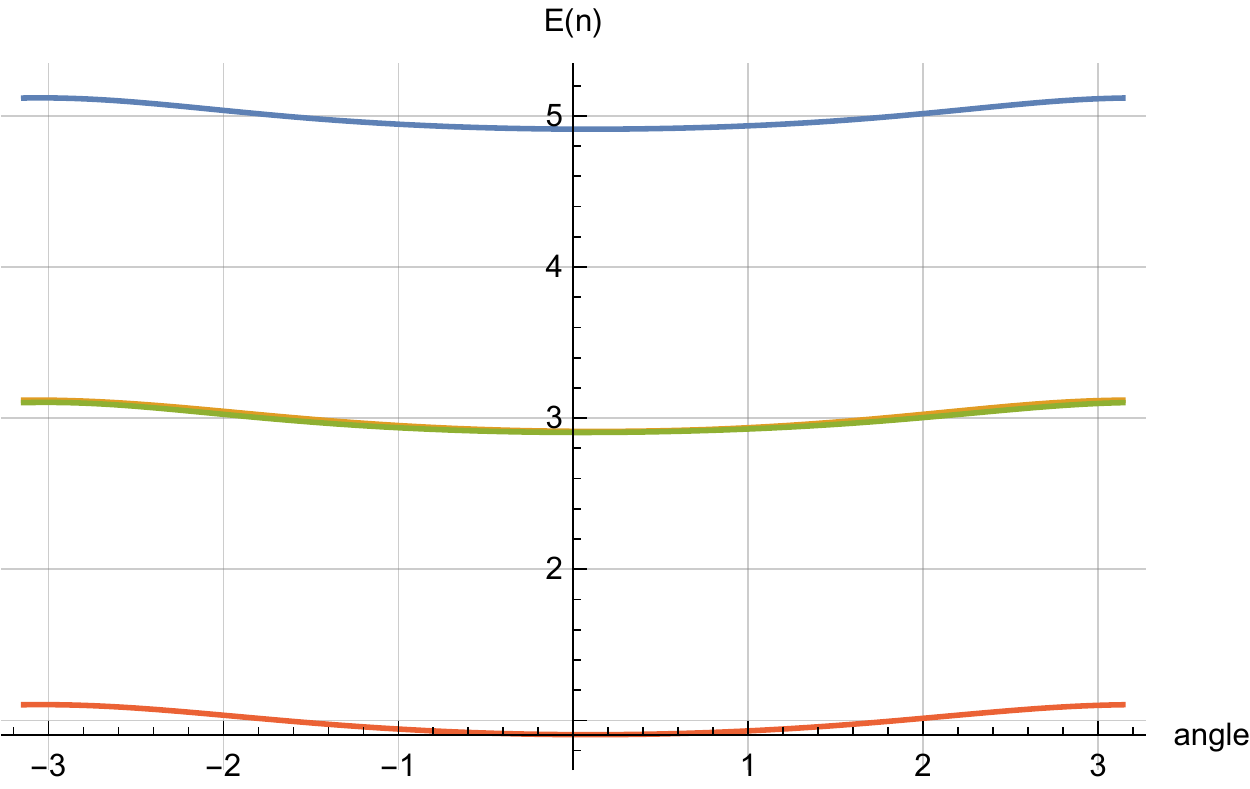}
\caption{Two main scenarios of dependence of eigenergies of electrostatically interacting qubit on angle. Left plot corresponds to parameters (d=1, a+b=0.8, q=1, $E_{p1}=1$,$E_{p2}=-1$,$E_{p1'}=-3$,$E_{p2'}= -2$, $t_{s12}=1$, $t_{s1'2'}=1$) and right plot has the same parameters except ( $E_{p1}=E_{p2}=E_{p1'}=E_{p2'}=1$, a+b=0.1). }
\label{qp4}
\end{figure}

\begin{figure}
\centering
\includegraphics[scale=0.38]{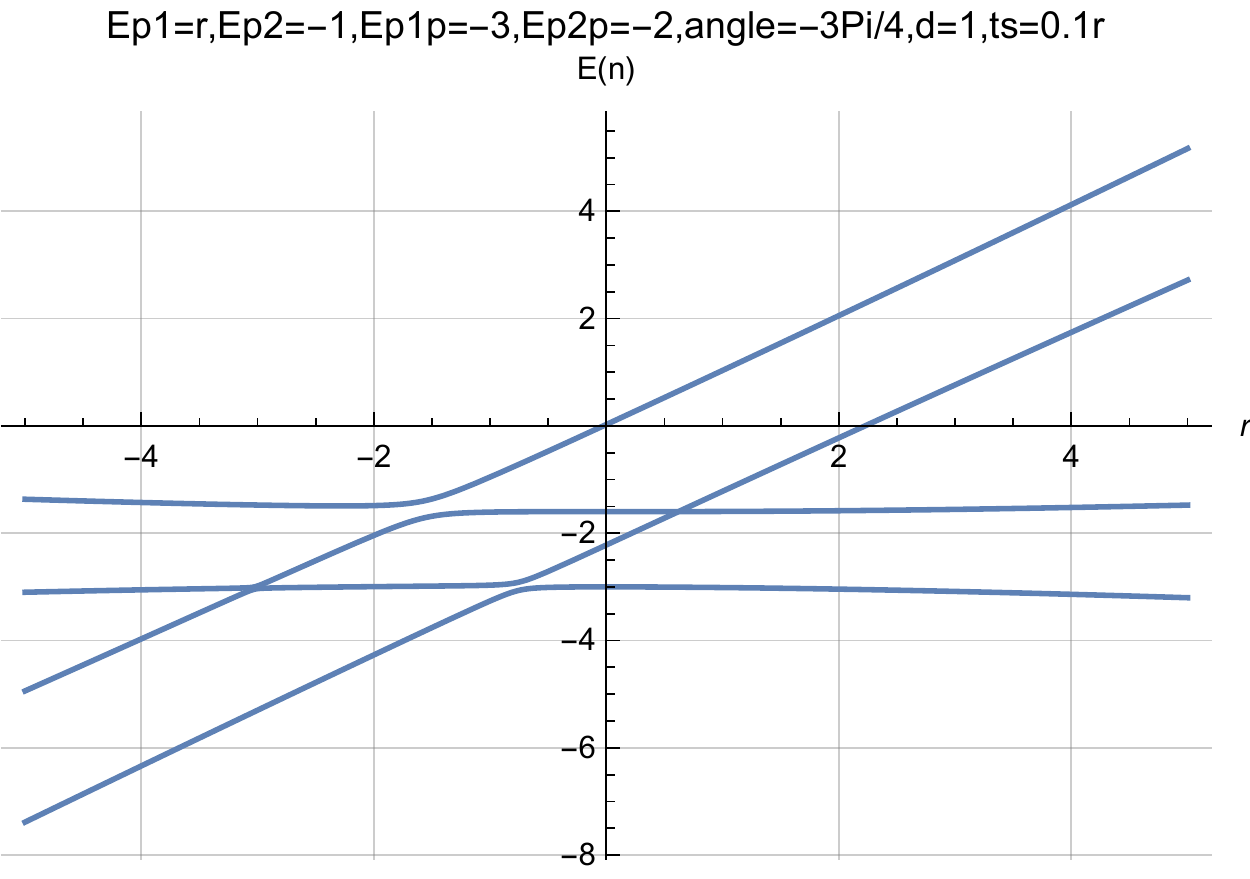}
\includegraphics[scale=0.38]{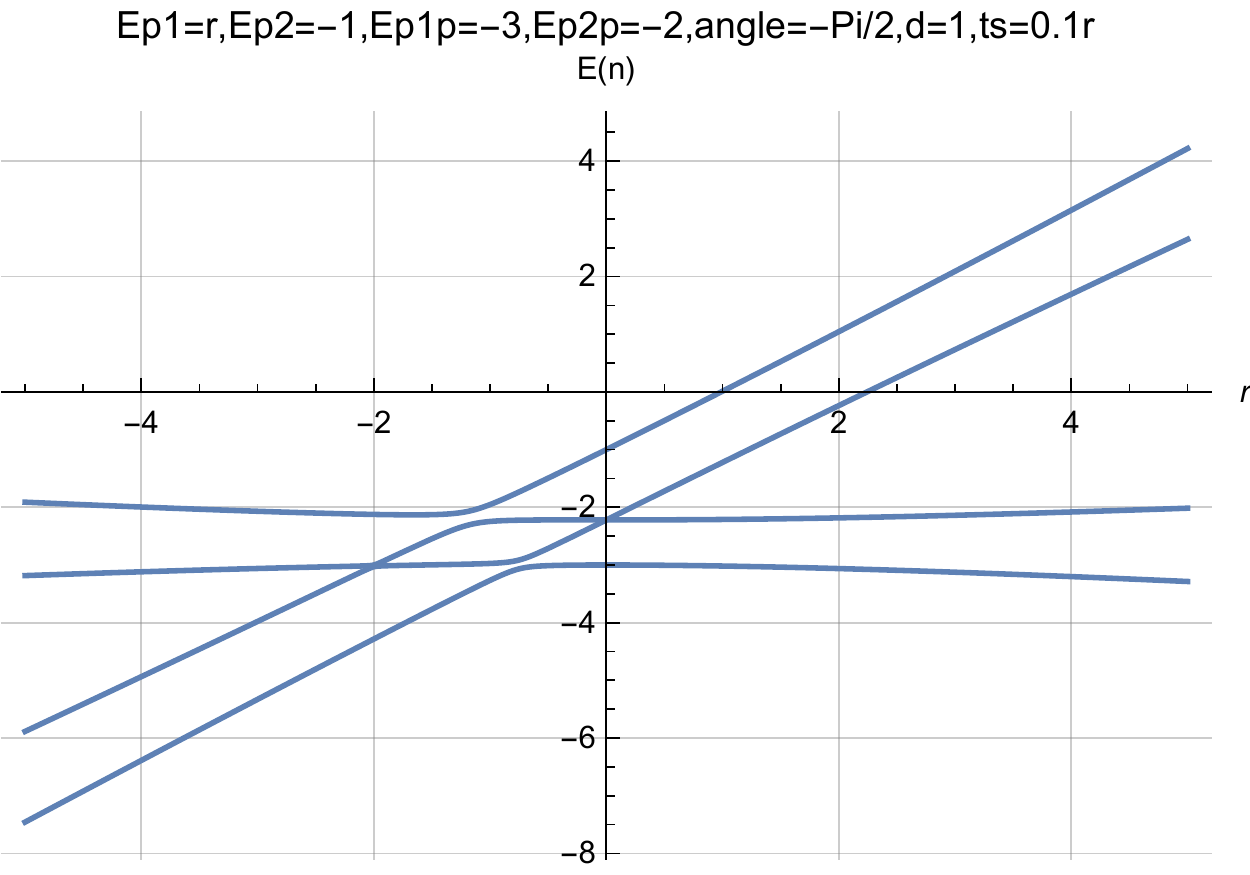}
\includegraphics[scale=0.38]{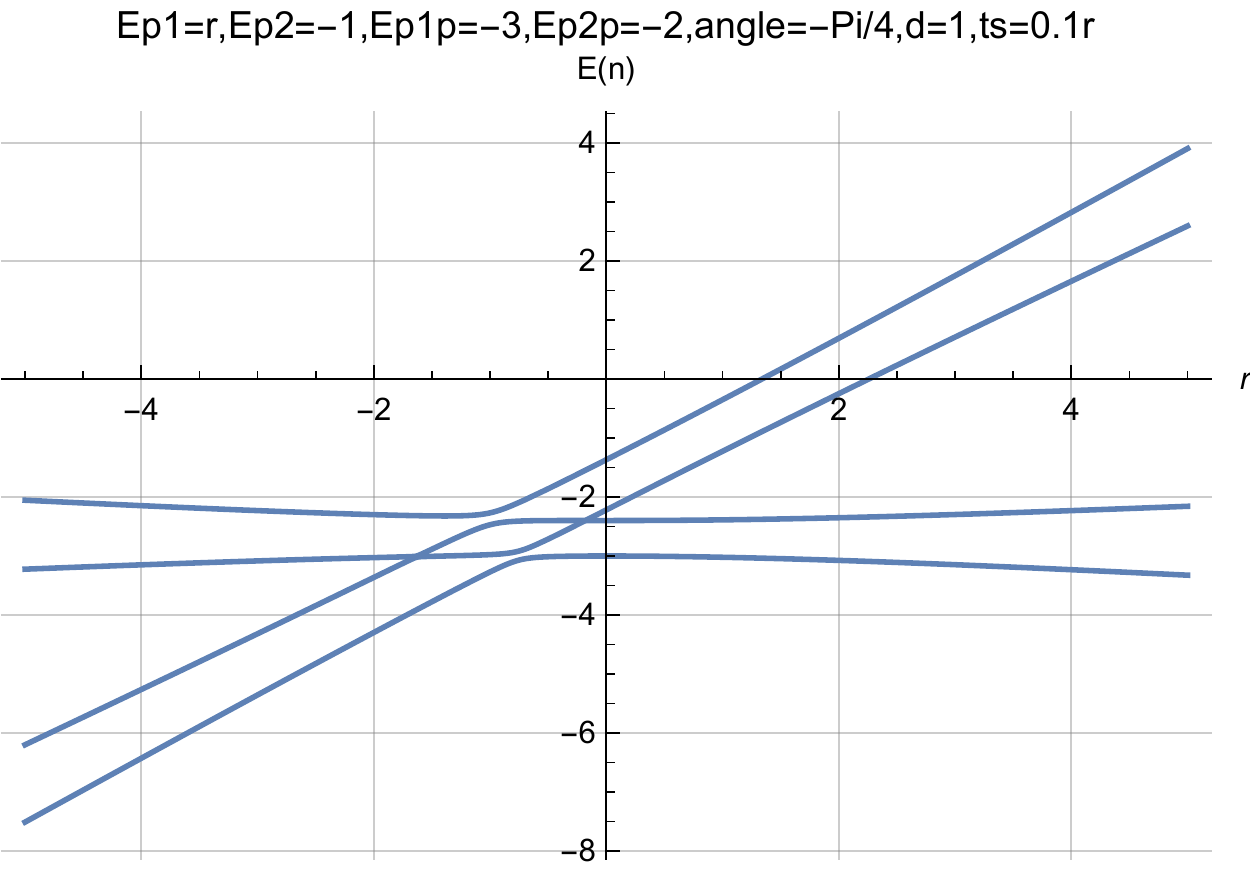}
\includegraphics[scale=0.38]{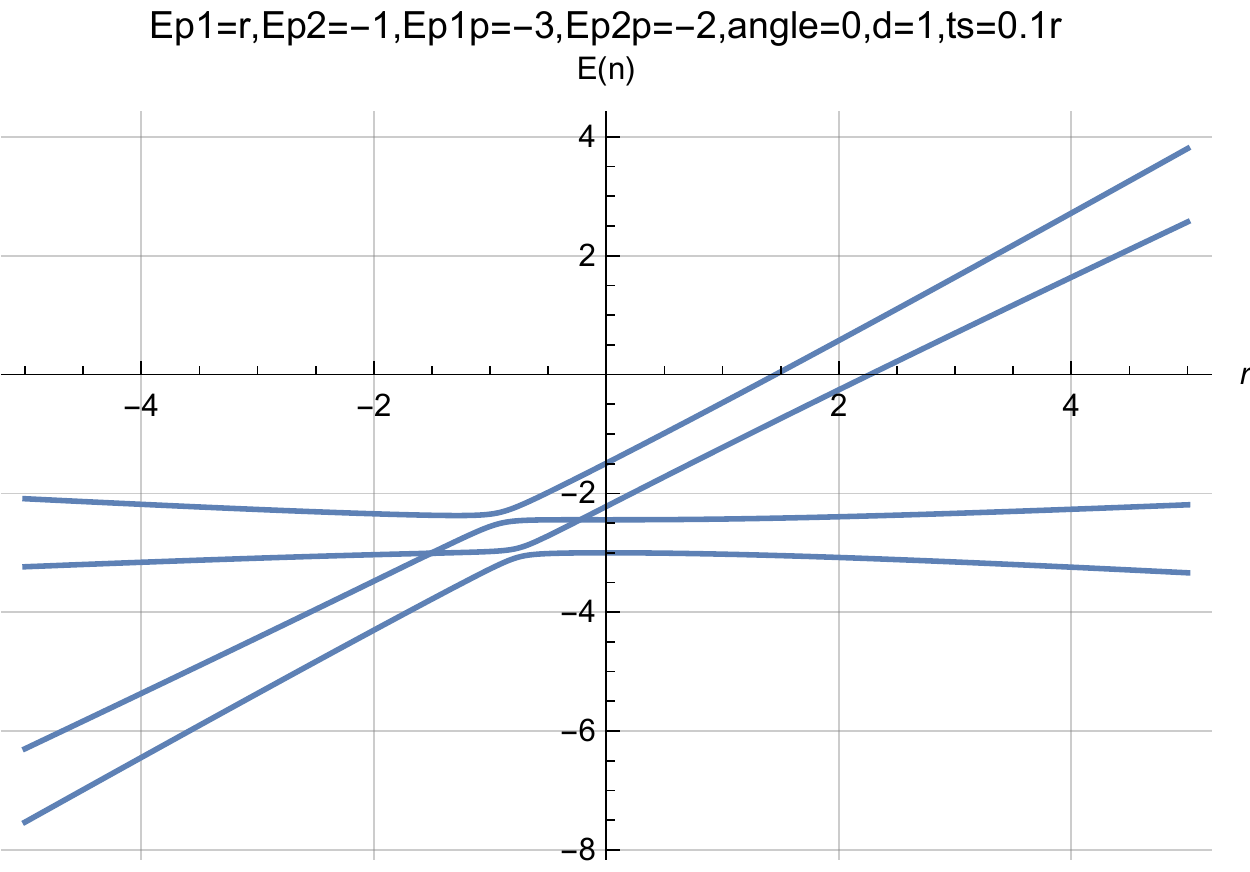}
\includegraphics[scale=0.38]{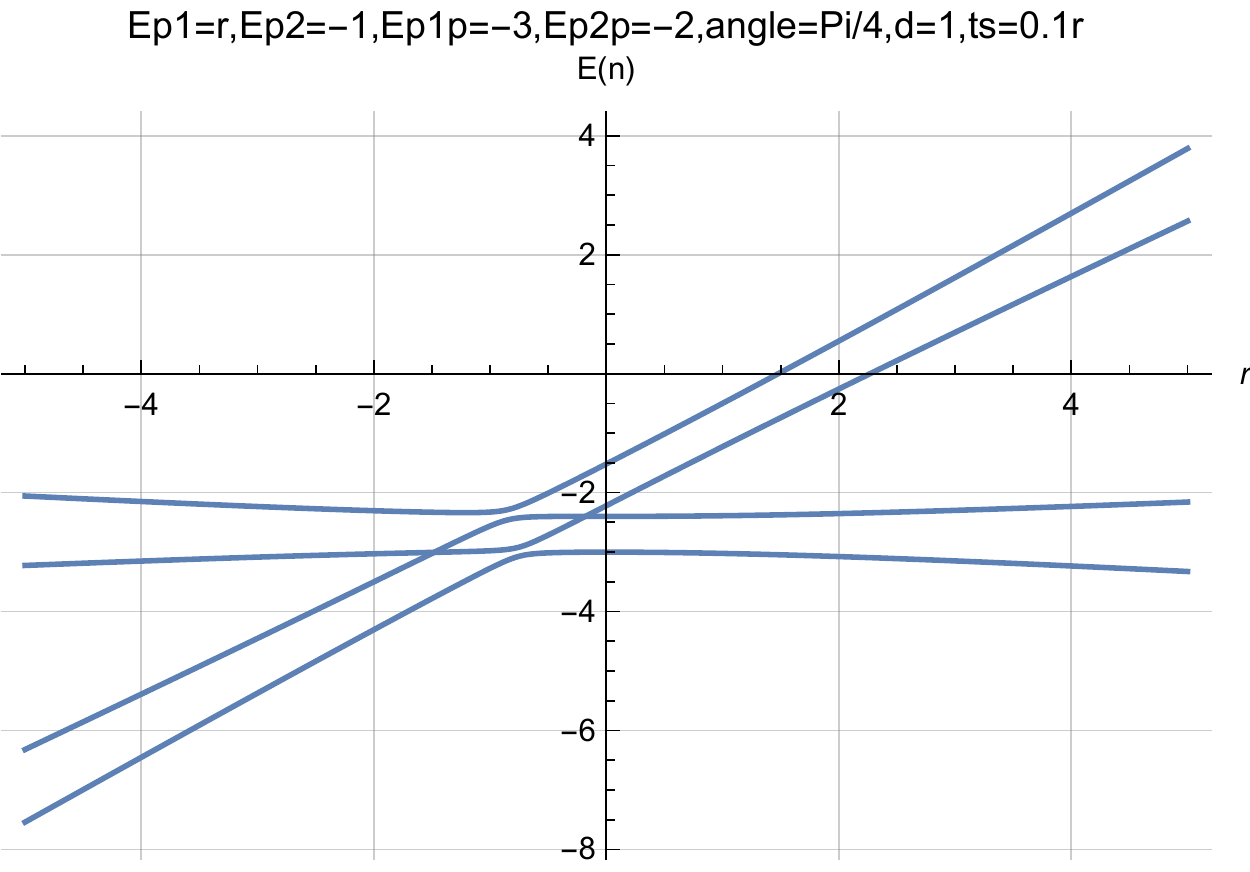}
\includegraphics[scale=0.38]{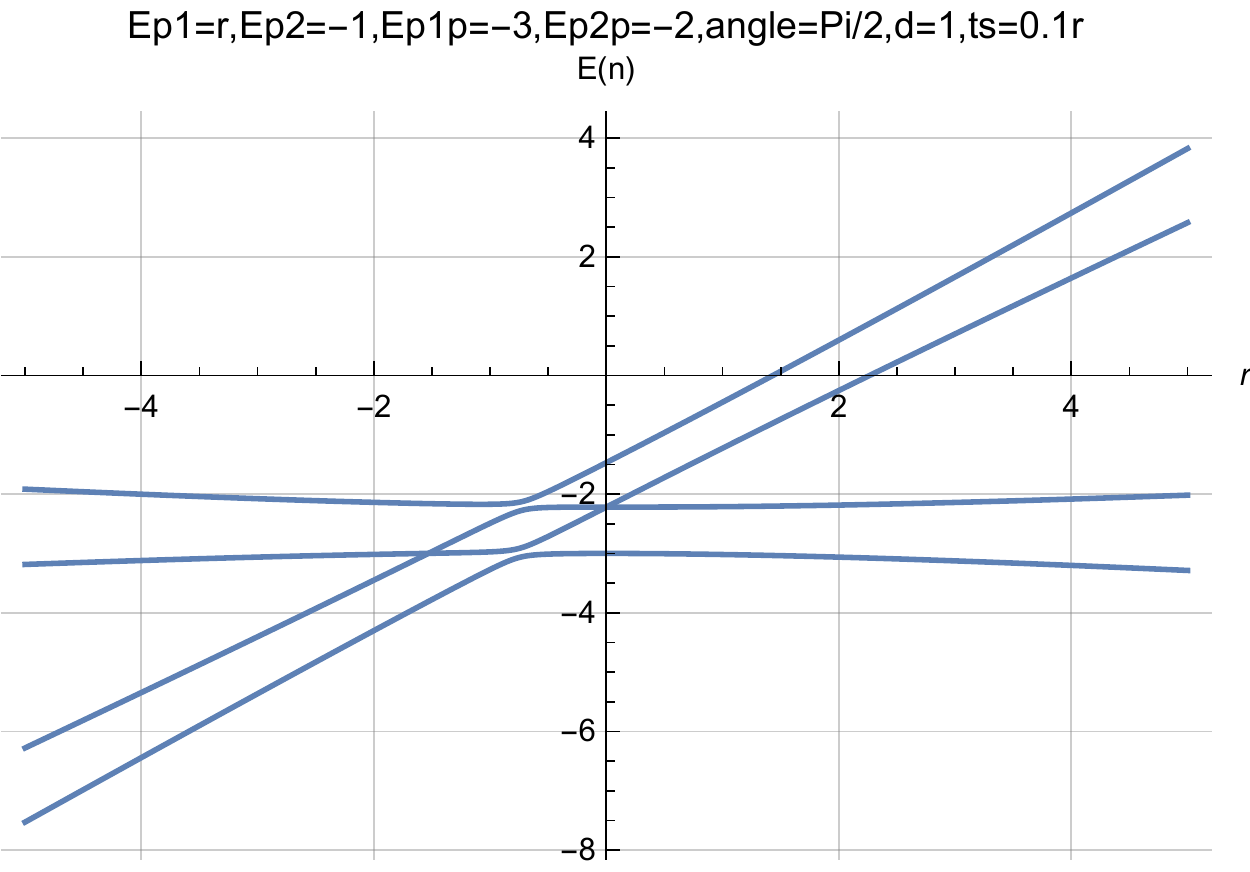}
\includegraphics[scale=0.38]{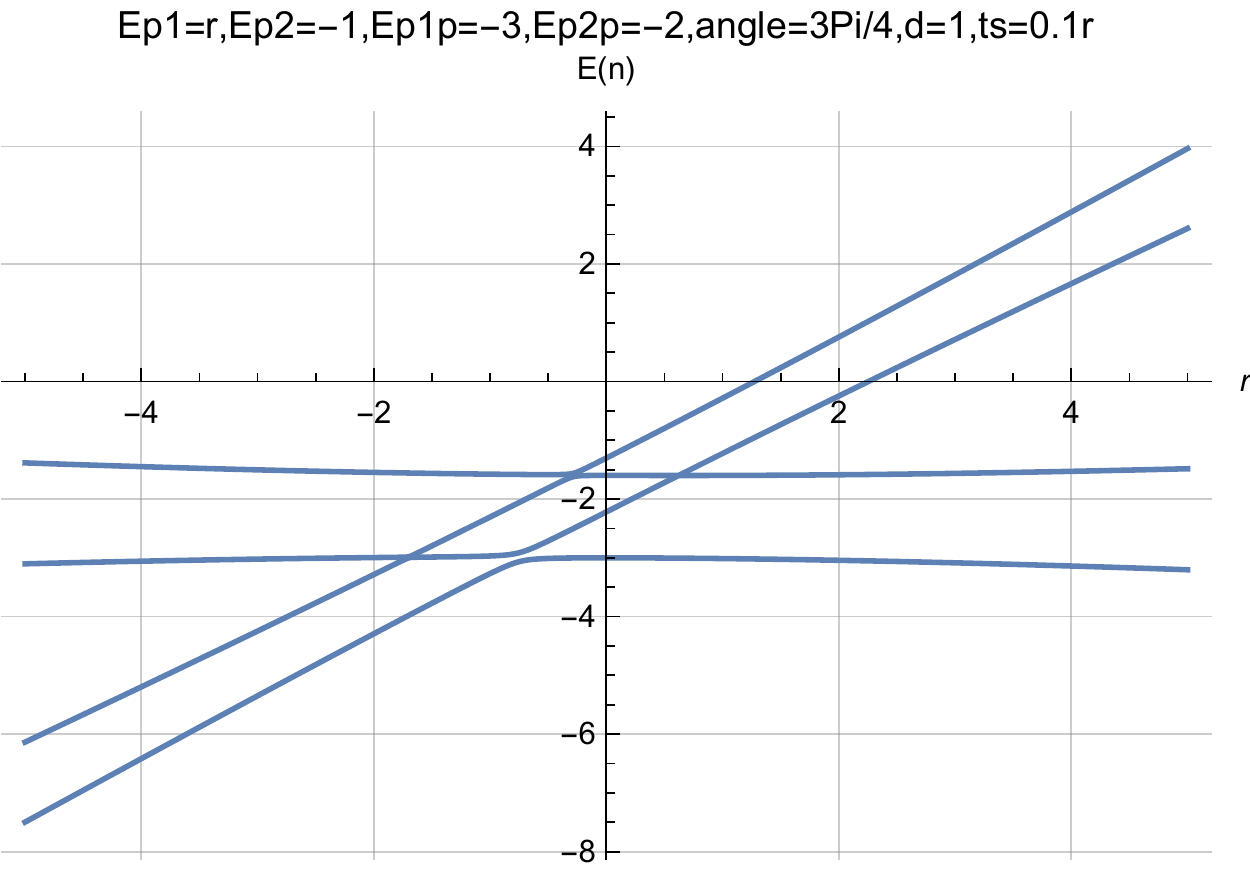}
\includegraphics[scale=0.38]{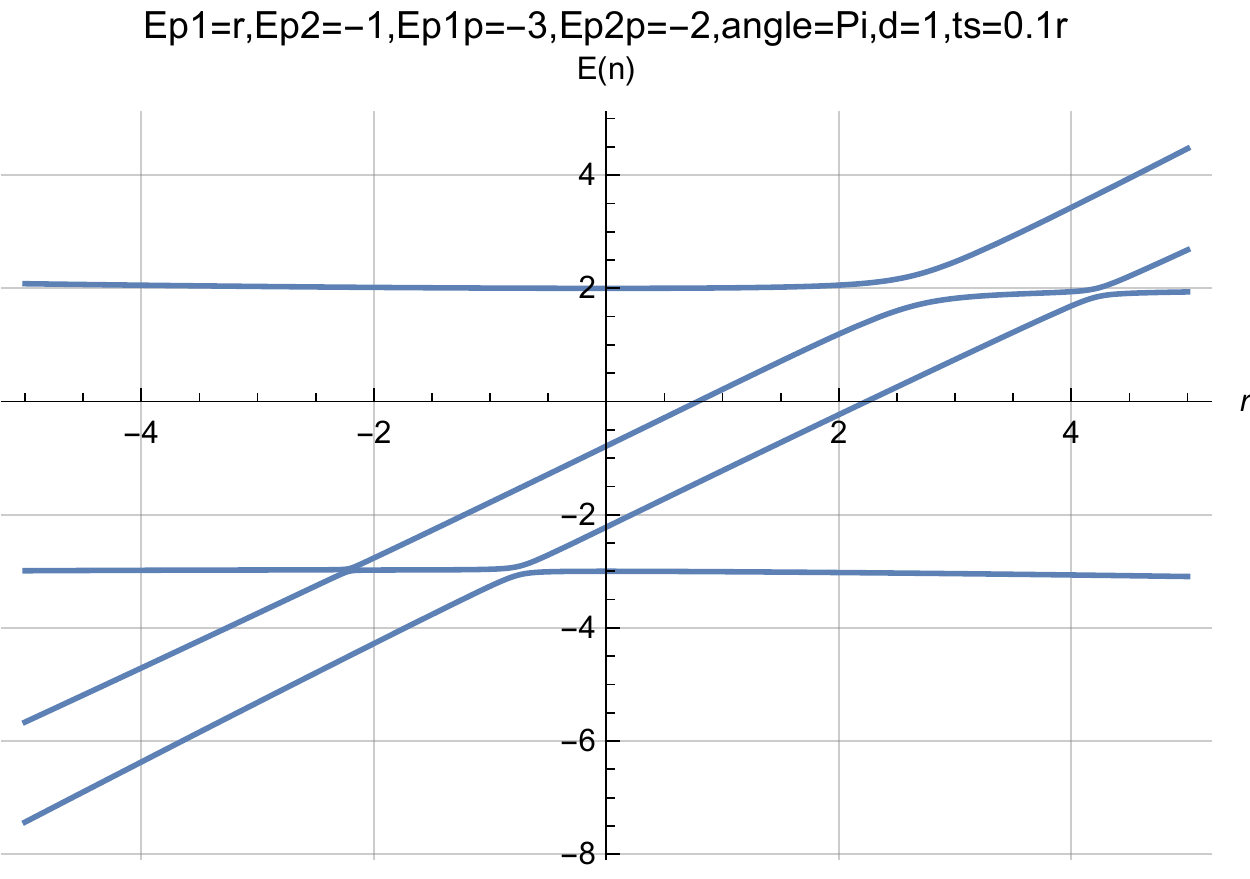}
\includegraphics[scale=0.38]{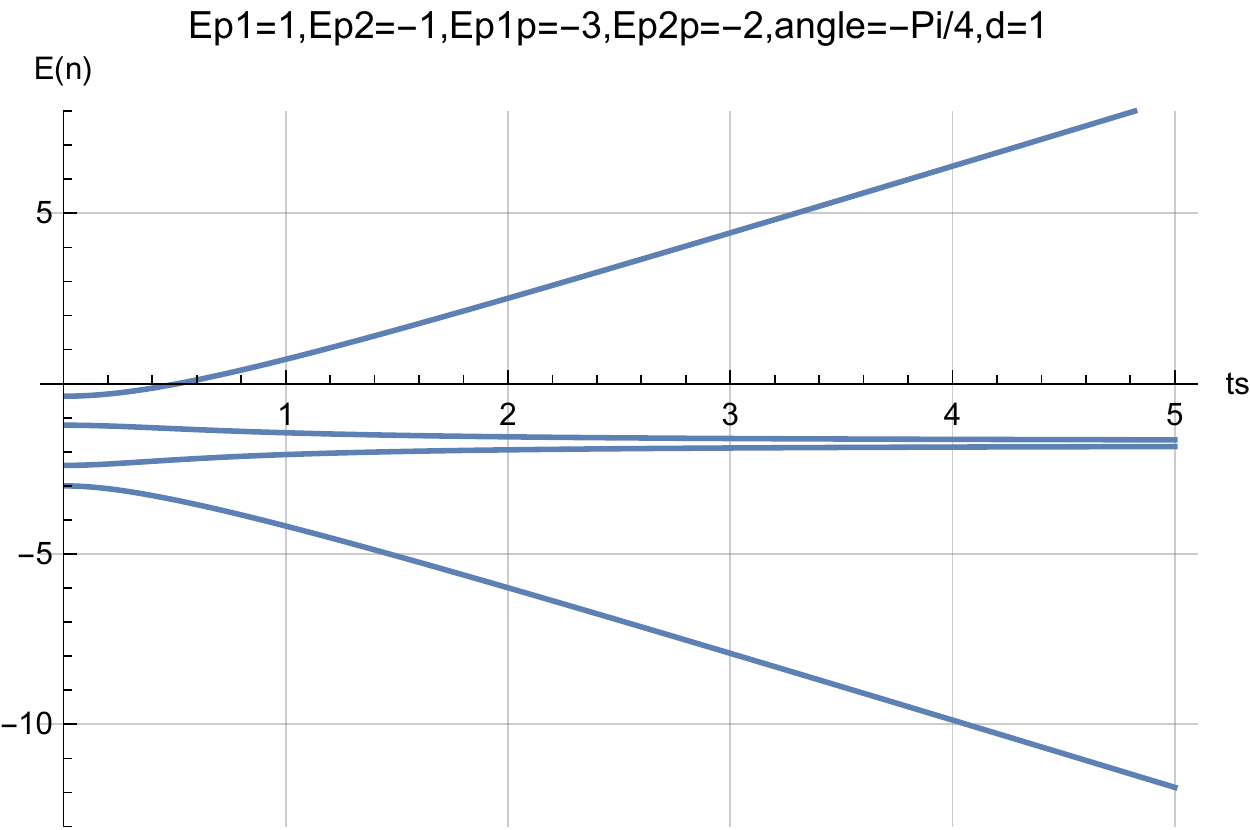}
\includegraphics[scale=0.38]{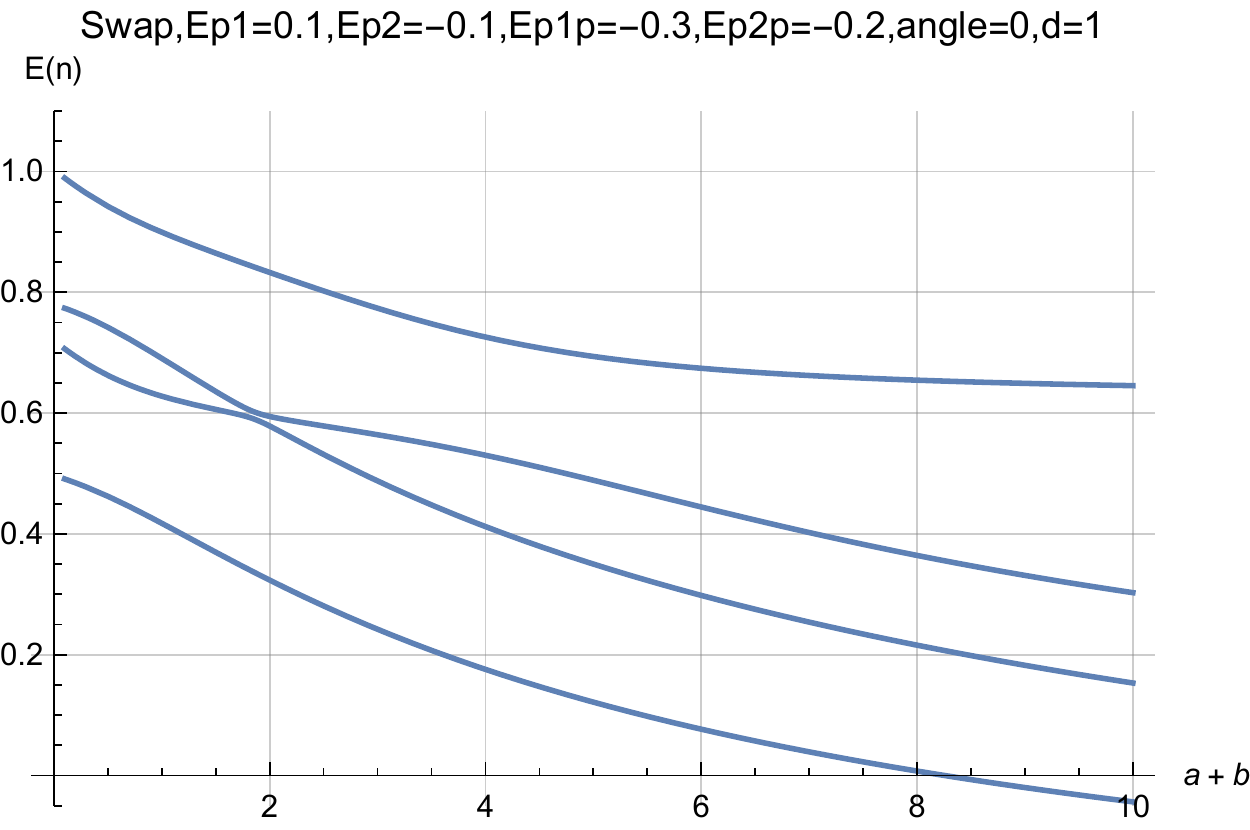}
\includegraphics[scale=0.38]{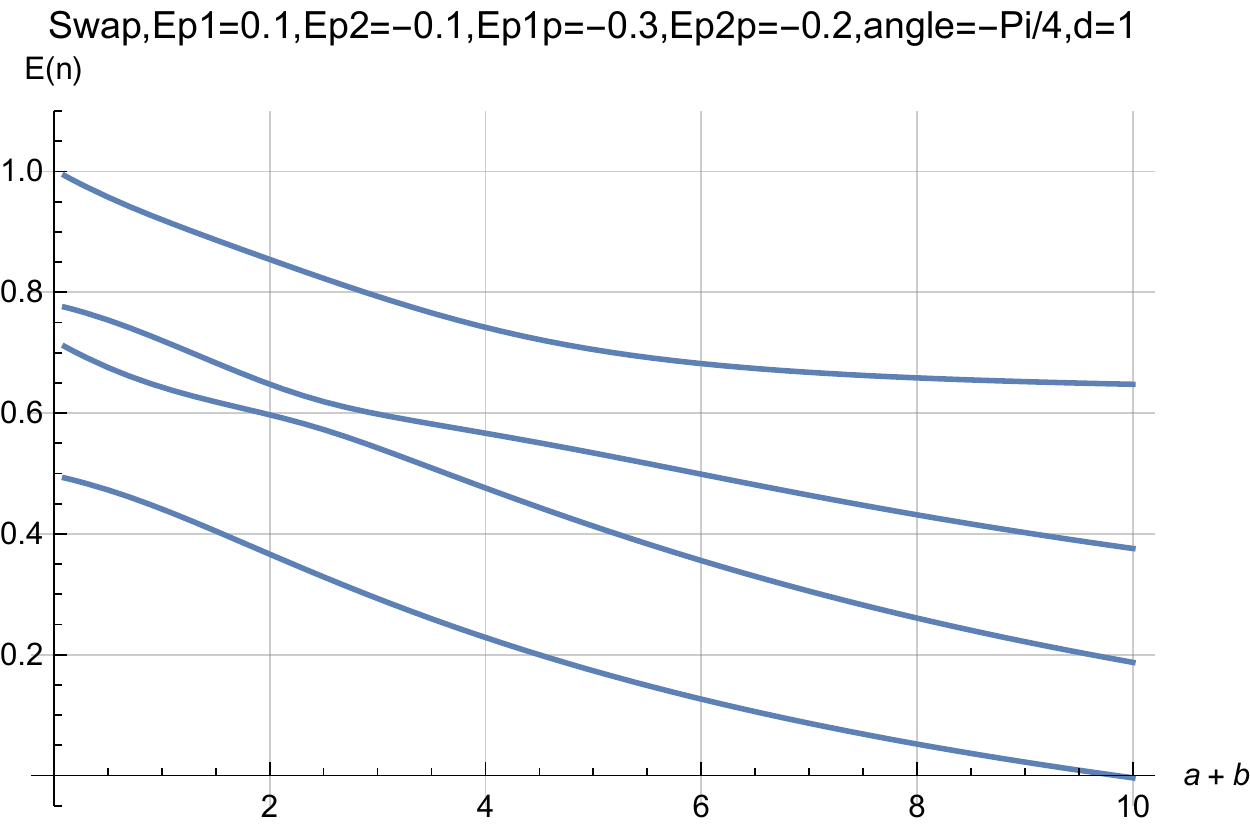}
\includegraphics[scale=0.38]{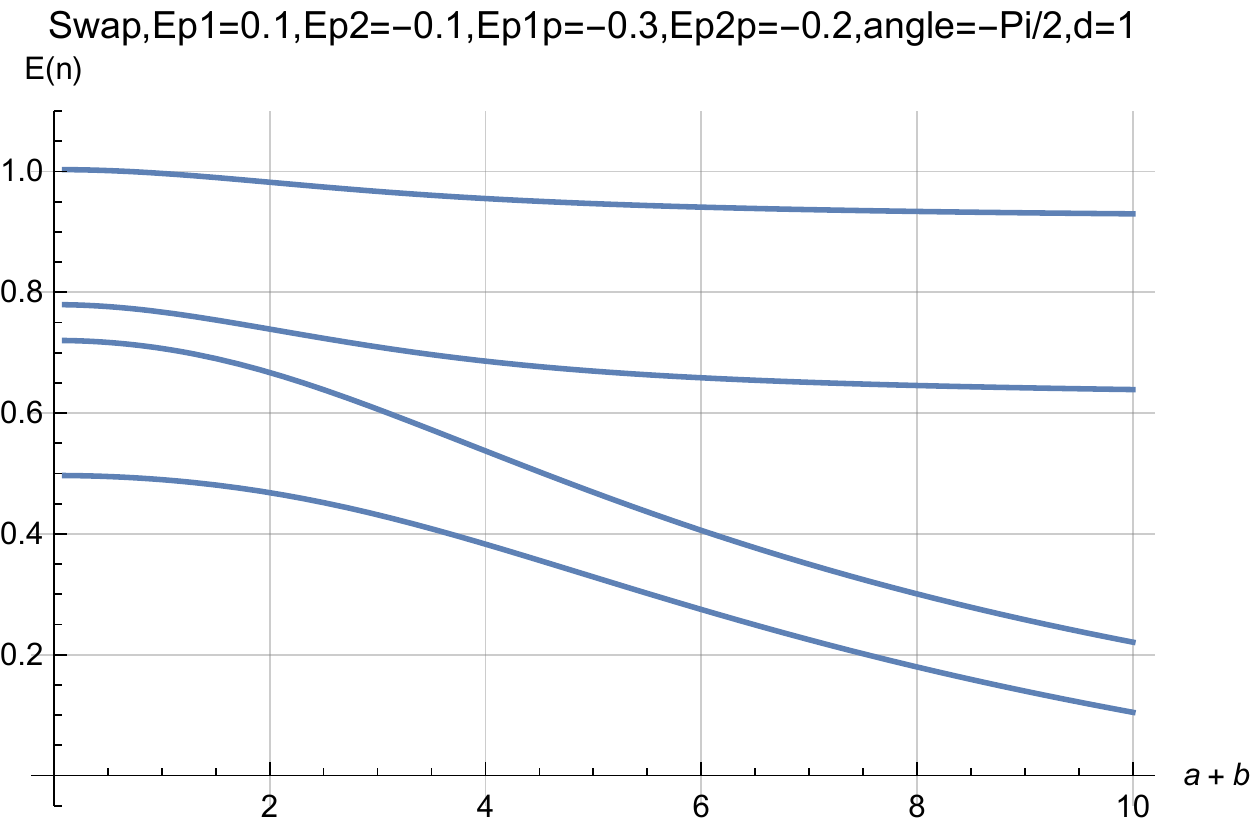}
\includegraphics[scale=0.38]{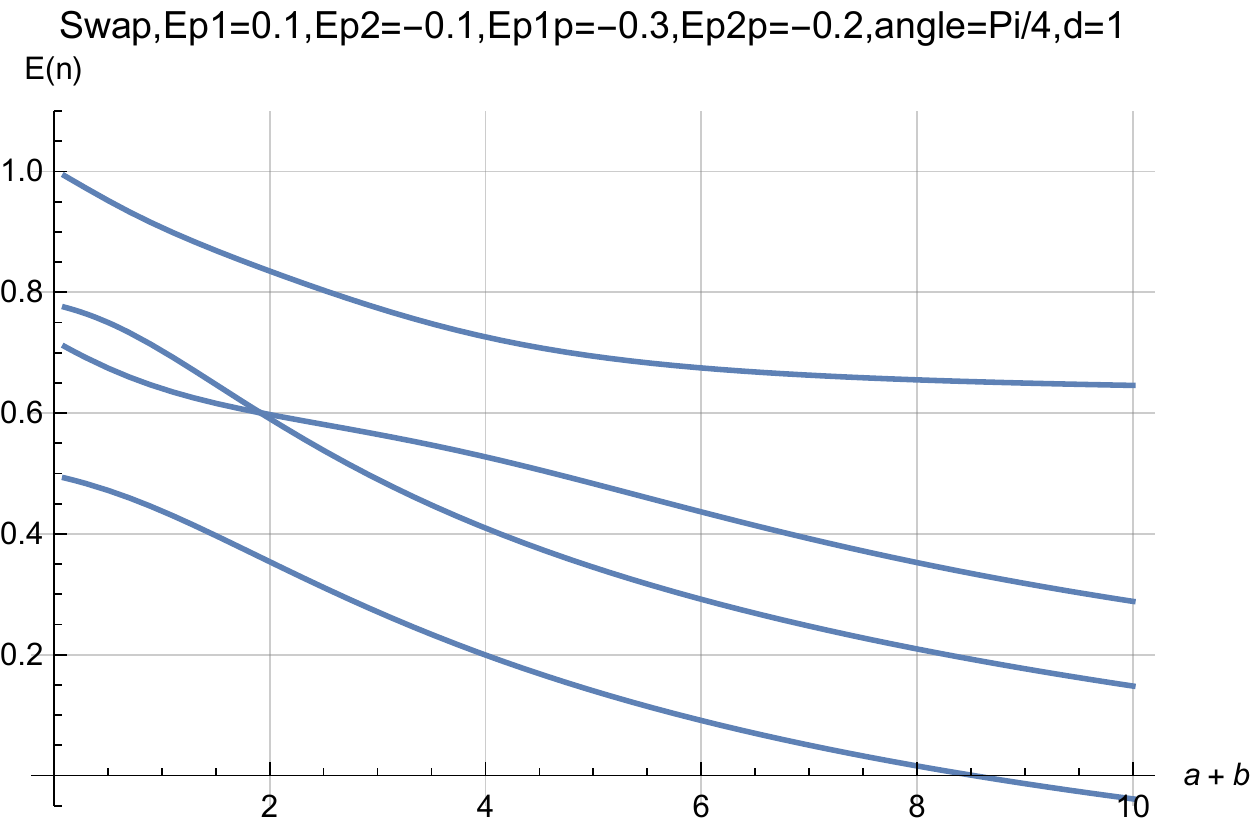}
\includegraphics[scale=0.38]{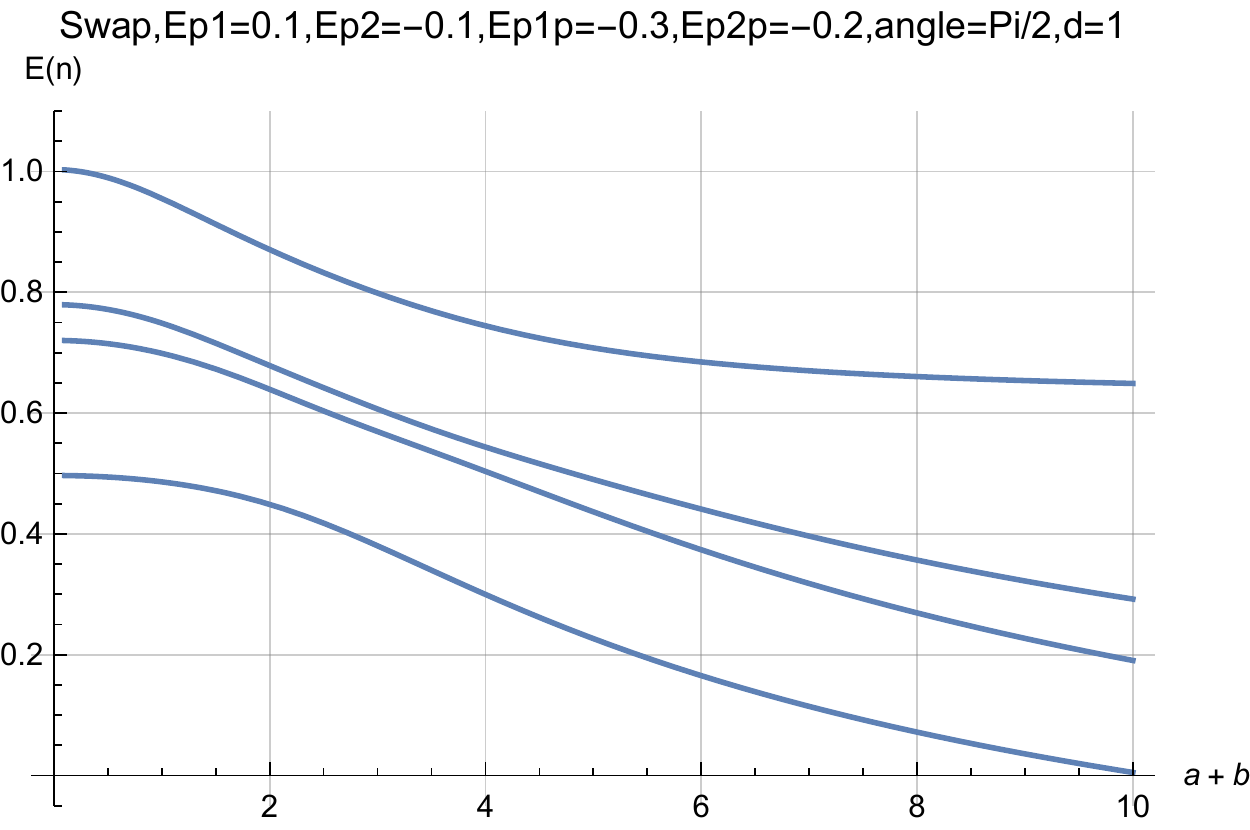}
\includegraphics[scale=0.38]{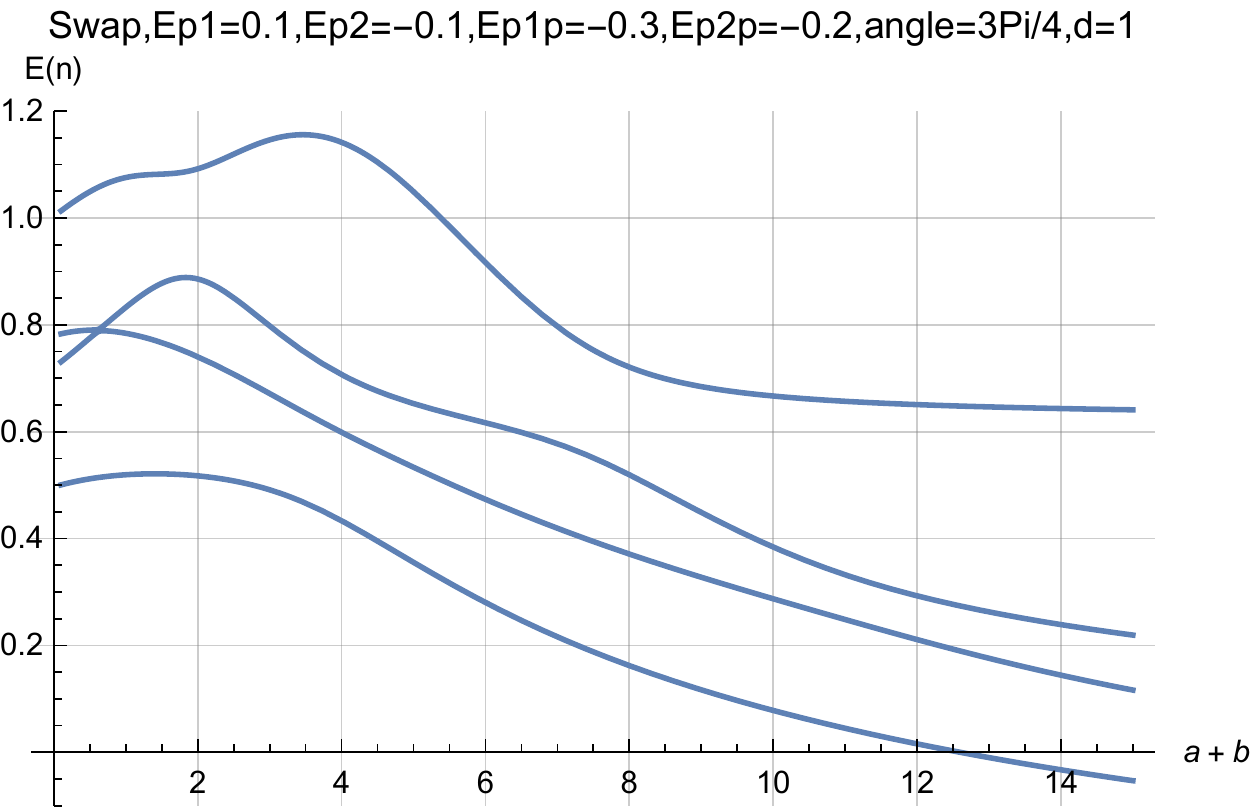}
\includegraphics[scale=0.38]{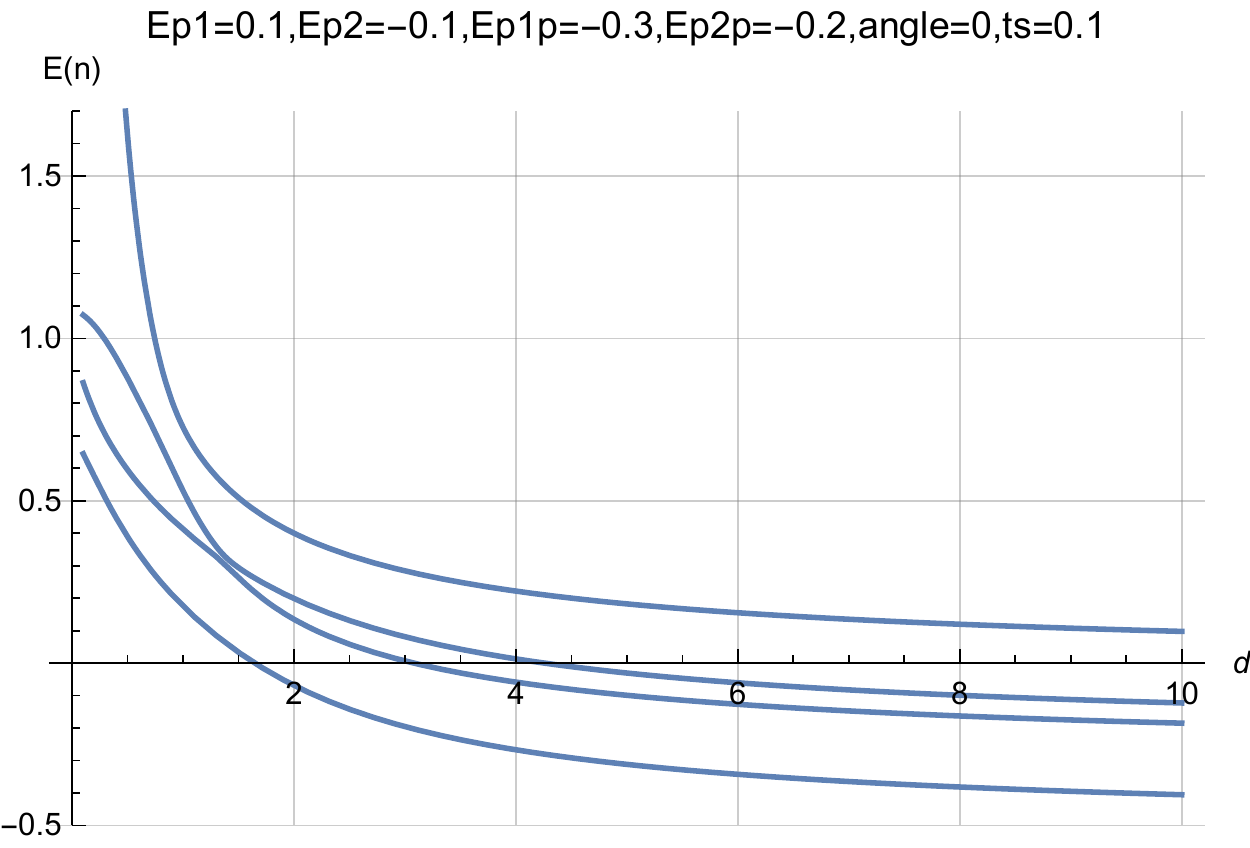}
\includegraphics[scale=0.38]{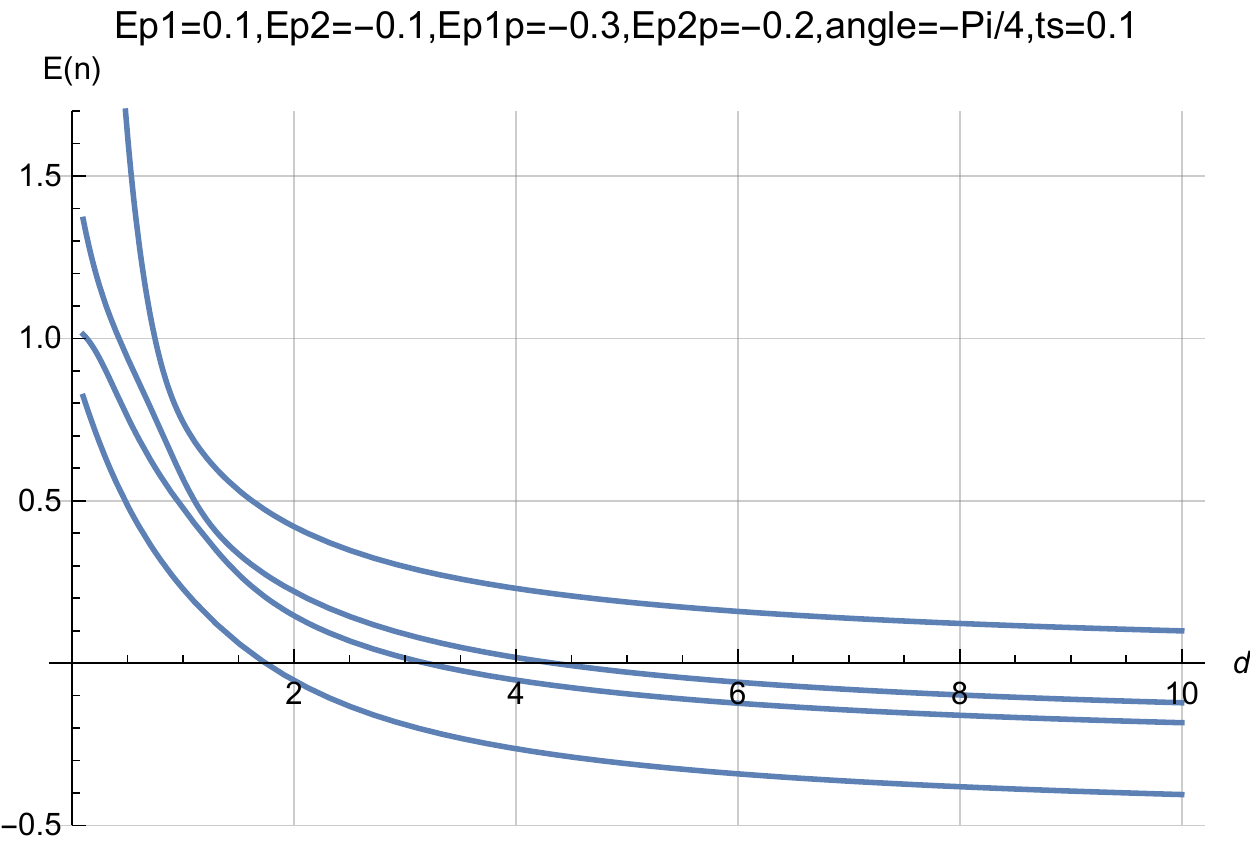}
\includegraphics[scale=0.38]{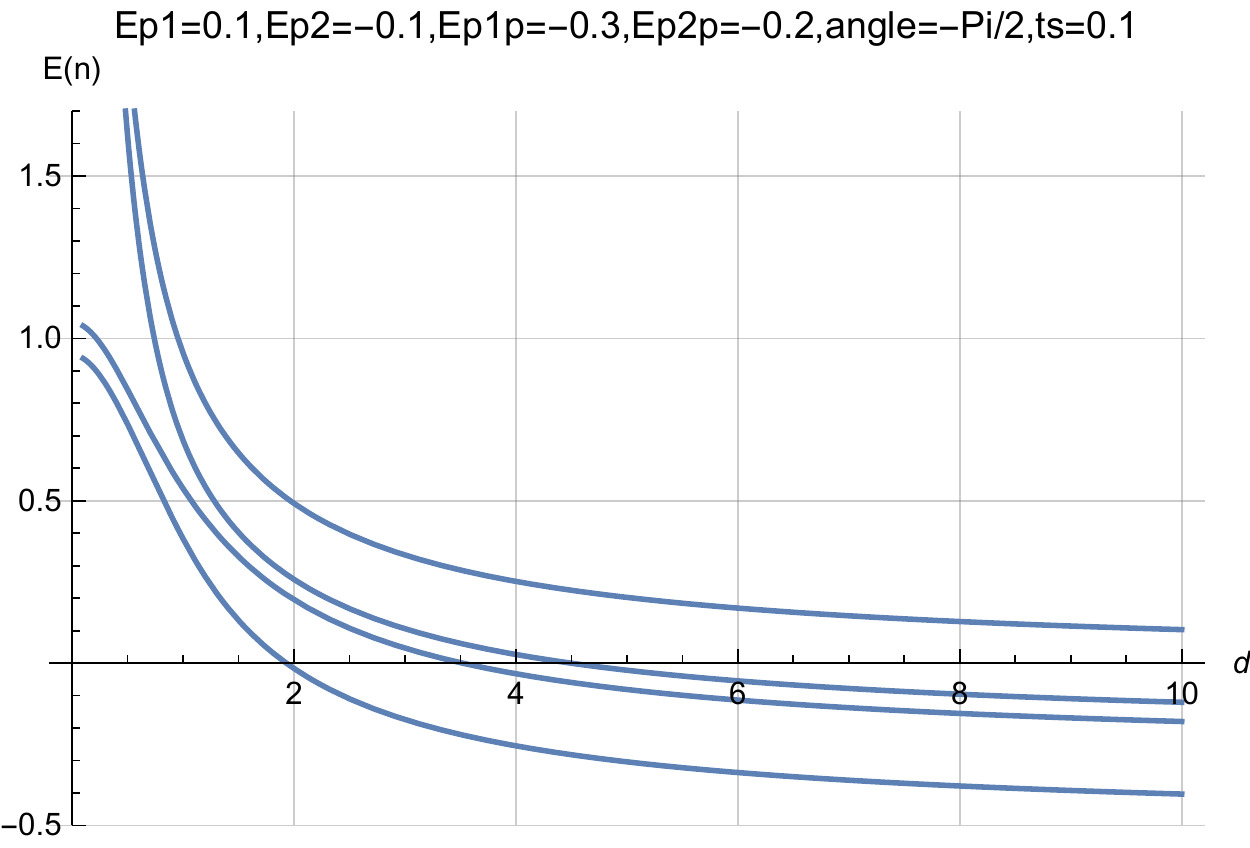}
\includegraphics[scale=0.38]{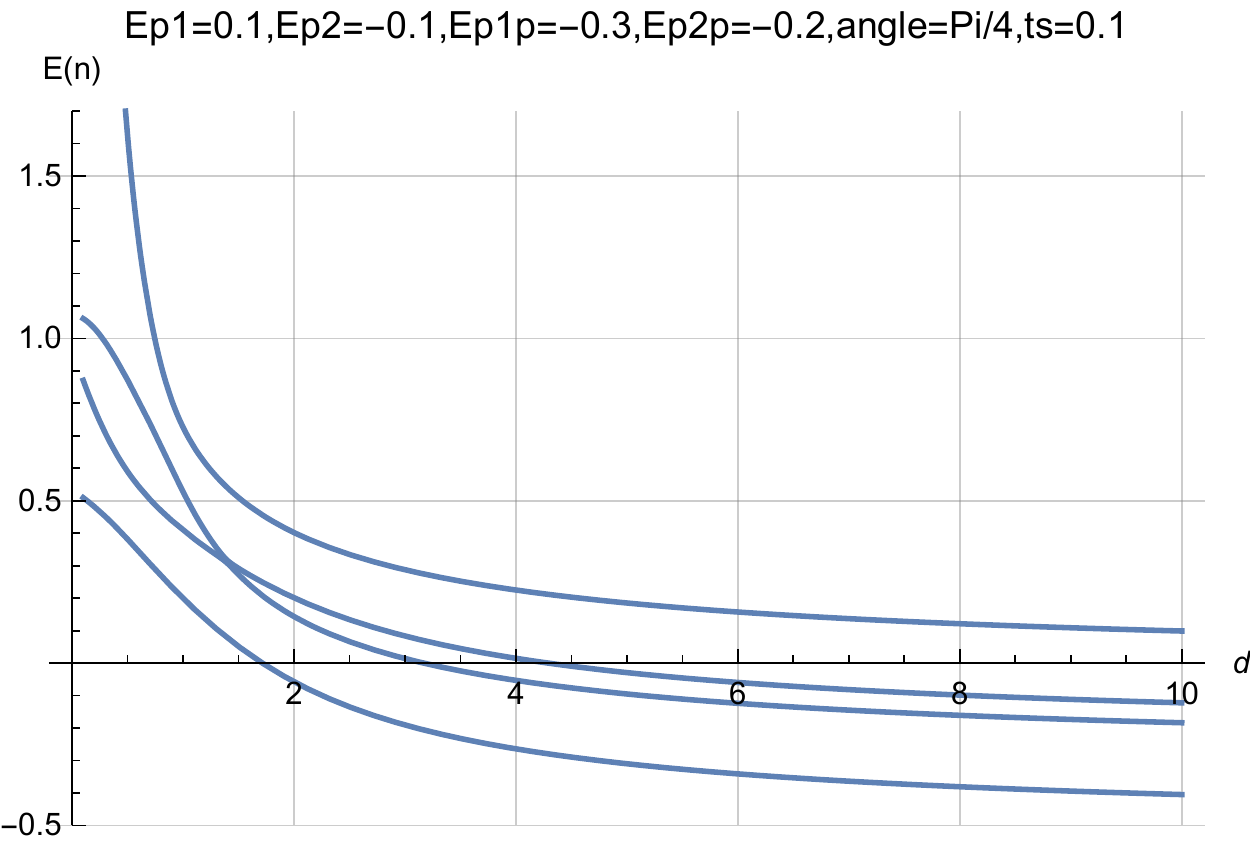}
\includegraphics[scale=0.38]{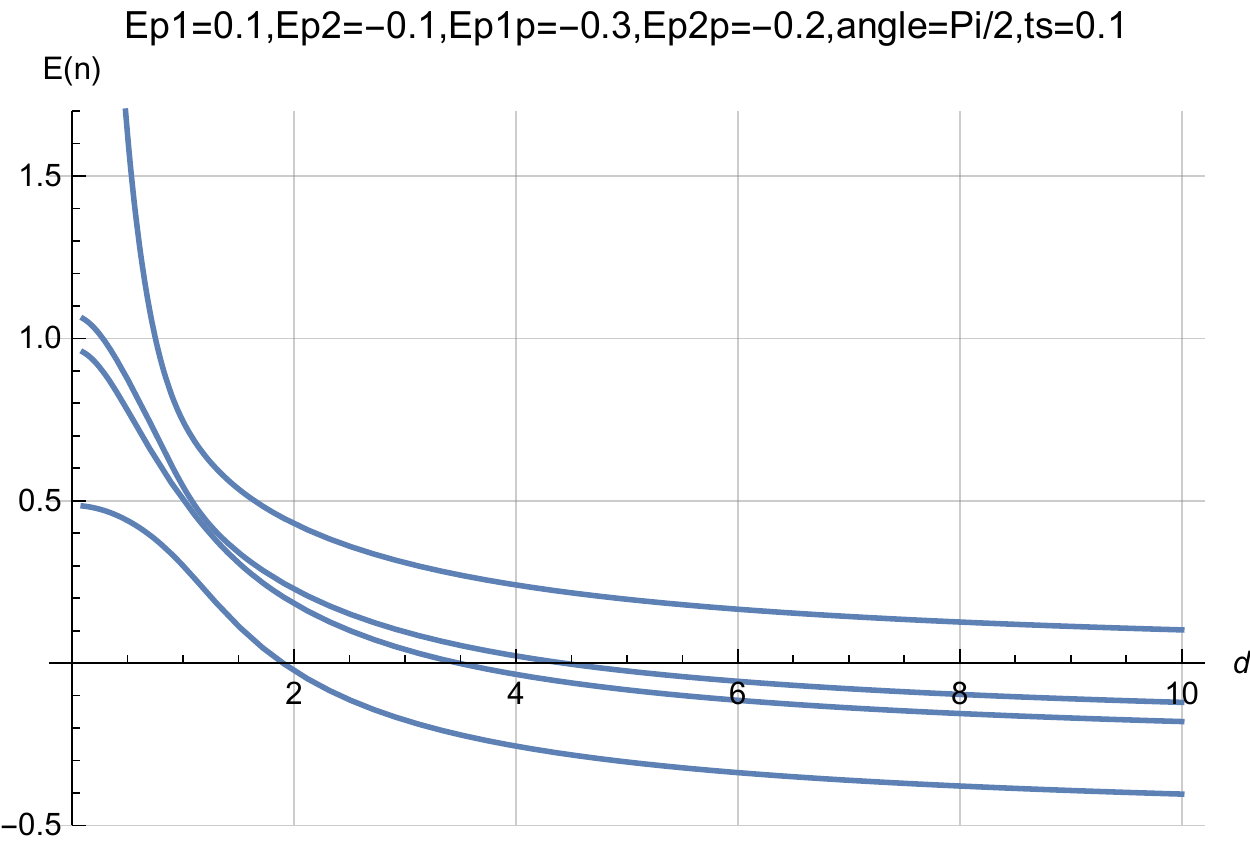}
\includegraphics[scale=0.38]{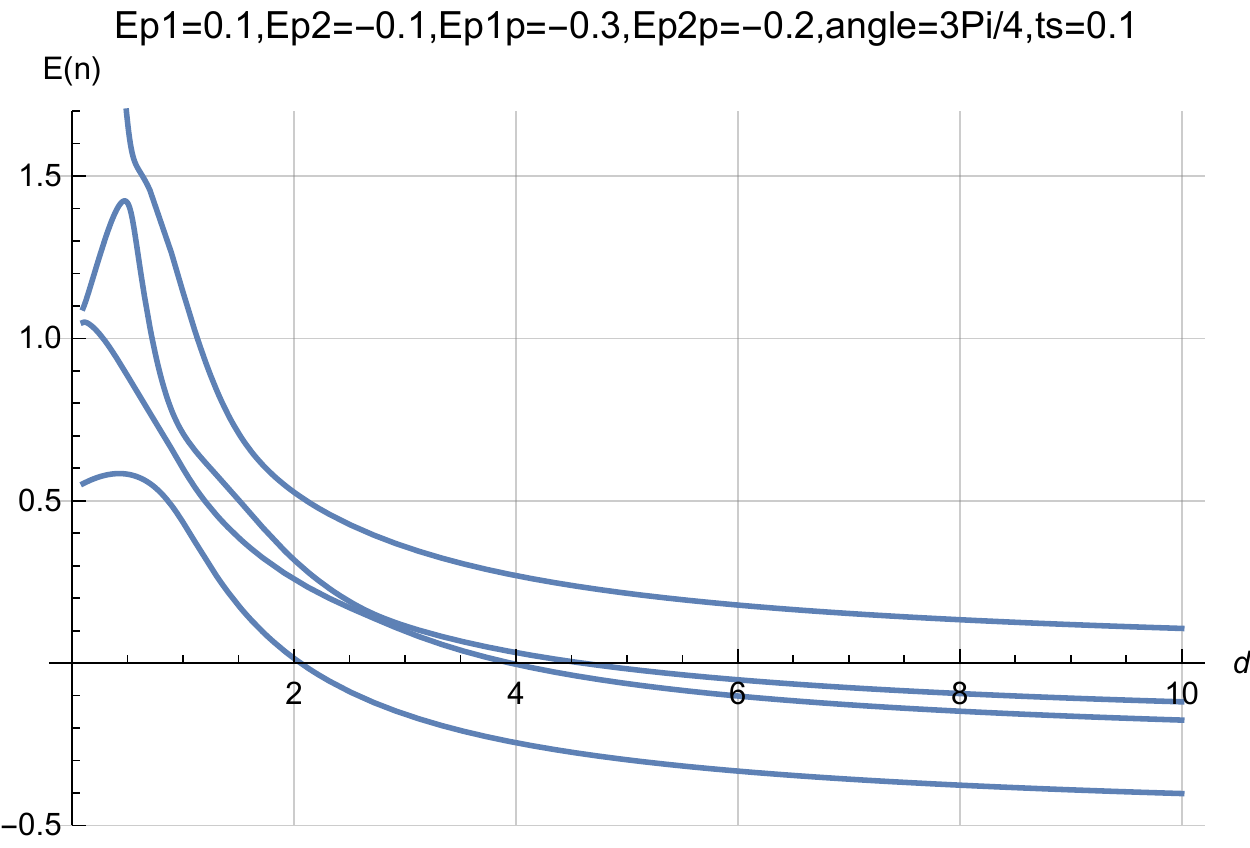}
	\caption{Energy spectrum of generalized electrostatic quantum gate in dependence on angle for fixed $E_{p1}$, $E_{p2}$, $E_{p1'}$, $E_{p2'}$, $t_{s12}=t_{s1'2'}=0.1$, $q=1$, $a+b=0.8$. No crossing of energy levels was reported.}
	\label{GQSwapGateSpectrum} 
\end{figure}

The results obtained by quasiclassical and quantum two body interaction Hamiltonian support the concept of quantum programmable matter as specfied by Fig. \ref{qp1}, \ref{qp2}, \ref{GeometricTransition} , \ref{qp4}, \ref{GQSwapGateSpectrum}.

Dynamics of two electrons confined by two local potentials is determined for various cases both in analytical and in numerical way in tight binding model [6]. In such way the system of two coupled electrostatic position based qubits can be used for the implementation of quantum swap or antiswap gate.
Broader picture is drawn by work [6,13,15].
The obtained results have its meaning in designing the proper operation of the quantum gates implemented in chain of coupled semiconductor quantum dots that are electrostatically controlled. It opens the path for implementation of CMOS quantum computer that is only controlled by voltages applied to CMOS transistors with no need of usage of magnetic field. It is nice alternative for implementation of quantum electronics other than by the usage of Josephson junctions [7,12]. The best way of detection of entanglement present between electrostatically interacting qubits is by measure of the correlation-anticorrelation function that is achievable in experimental way. Both formulas for von-Neumann entanglement entropy and anticorrelation functions are given in this work in analytical form.
\section{Acknowledgment}
I thank to Erik Staszewski from University College Dublin for his assistance in figures preparation.
\twocolumn



\end{document}